\documentclass[11pt,a4paper,british]{extarticle}
\usepackage[T1]{fontenc}
\usepackage[latin1]{inputenc}
\usepackage{amsmath}
\usepackage{amssymb}
\usepackage{setspace}
\usepackage{esint}
\usepackage{nameref}

\makeatletter

\usepackage{babel}
\usepackage{bbm}
\date{}
\allowdisplaybreaks
\numberwithin{equation}{section}

\makeatother

\usepackage{babel}
\begin{document}

\title{On the field-antifield (a)symmetry of the pure spinor superstring}

\author{Renann Lipinski Jusinskas\thanks{renannlj@fzu.cz}}

\maketitle

\begin{center}
Institute of Physics AS CR \\
Na Slovance 2, 182 21, Praha 8 \\
Prague - Czech Republic
\par\end{center}

\

\begin{abstract}
In this work, the DDF-like approach to the pure spinor cohomology
is extended to the next ghost number level, the so called antifields.
In a direct (supersymmetric) parallel to the bosonic string, some
properties of the ghost number two cohomology are derived with the
enlargement of the DDF algebra. Also, the DDF conjugates of the $b$
ghost zero mode emerge naturally from the extended algebra and the
physical state condition is discussed. Unlike the bosonic string case,
the cohomology analysis of the pure spinor $b$ ghost is restricted
to BRST-closed states.
\tableofcontents{}
\end{abstract}

\section{Introduction\label{sec:Introduction}}

\

When it comes to determining the physical content of a given theory,
it is clear that a thoroughly symmetry preserving description is not
always the best option. From this perspective, the light-cone has
played an important role since the first steps of string theory \cite{Goddard:1973qh}.
The starting point of this approach is a Lorentz breaking gauge fixing,
which enables the decoupling of unphysical degrees of freedom even
before quantisation. For the RNS superstring, for example, reparametrisation
and local (worldsheet) supersymmetry of the action are used to decouple
the light-cone components of $X^{m}$ and $\psi^{m}$, leaving only
the physical components, represented by $X^{i}$ and $\psi^{i}$,
the transverse $SO\left(8\right)$ directions. 

In the pure spinor superstring \cite{Berkovits:2000fe}, the covariant
formulation is not suitable to investigate the physical degrees of
freedom. While the massless cohomology is elegantly given in terms
of the Yang-Mills superfield \cite{Siegel:1978yi}, any attempt to
fully describe the massive spectrum is practically hopeless due to
the introduction of extra auxiliary superfields and gauge transformations.
To date, only the first massive level has a covariant superfield description
\cite{Berkovits:2002qx}. 

The first approaches to determine the pure spinor cohomology \cite{Berkovits:2000nn,Berkovits:2004tw}
involved a series of indirect methods to produce a light-cone version
of its BRST-like charge, but nothing as clear and plain as for the
other formalisms. A full ``ungauged'' description has been presented
in \cite{Berkovits:2014aia,Berkovits:2015yra}, involving a twistor-like
symmetry. A master action was proposed and shown to give rise to the
usual pure spinor BRST symmetry or to the Green-Schwarz $\kappa$-symmetry
\cite{Green:1983wt}, depending on the chosen gauge fixing. This work
has established a much better understanding on the origin of the formalism
and has yet to be explored, but the known light-cone gauge fixing
from bosonic string theory or even the RNS and Green-Schwarz superstrings
is still lacking. 

Recently, a DDF-like construction of the massless vertices was proposed
\cite{Jusinskas:2014vqa}, introducing a supersymmetric creation-annihilation
algebra that can be used to span the pure spinor cohomology \cite{Berkovits:2014bra}.
It enables an $SO\left(8\right)$-covariant superfield description
of the spectrum in a systematic way, relying solely on the well established
massless spectrum.

It is interesting to point out that physical states in the pure spinor
superstring are defined to be at the ghost number one cohomology.
Although supported by successful comparisons with the other formalisms
\cite{Berkovits:2000ph,Berkovits:2001us,Berkovits:2005ng} and even
providing some impressive new results, for example \cite{Gomez:2013sla},
this definition seems rather \emph{ad-hoc} and illustrates an incomplete
understanding of some fundamental aspects of the formalism. Taking
the bosonic string as an example, the cohomology at ghost number two
(antifields) has a clear kinematic property that distinguishes it
from the cohomology at ghost number one (fields), namely BRST-closedness
does not impose the mass-shell condition. In a more fundamental level,
it can be shown that unitarity of the scattering amplitudes projects
out the ghost number two states. This leads to the physical state
condition, known as Siegel gauge: any element of the BRST cohomology
annihilated by the $b$ ghost zero mode, $b_{0}$, is defined to be
a \emph{physical state}.

In \cite{Mikhailov:2012uh,Mikhailov:2014qka}, the massless sector
of the pure spinor cohomology at ghost number higher than one is discussed
in detail. But it seems that a complete understanding of the spectrum
beyond that was never achieved, especially when it comes to the antifields.
The doubling of the degrees of freedom was expected, although through
a nontrivial realisation. For example, when comparing the massless
vertices $U=\lambda^{\alpha}A_{\alpha}\left(X,\theta\right)$ and
$U^{*}=\lambda^{\alpha}\lambda^{\beta}A_{\alpha\beta}\left(X,\theta\right)$
\cite{Berkovits:2001rb}, it is far from obvious that the number of
degrees of freedom matches at different ghost numbers. Unlike in the
bosonic string, where the zero mode of the $c$ ghost introduces a
sort of degeneracy of the ground state, the pure spinor variable makes
the doubling of the cohomology much more intricate. In this sense,
the DDF construction of \cite{Jusinskas:2014vqa} seems to be a good
way to approach the problem. The creation/annihilation algebra is
already determined, and it remains to introduce a ground state associated
to the antifields. 

For the fields, the DDF ground state is defined through the state-operator
map of the vertex $U$ in the frame $P^{+}=0$. While one of the $SO\left(8\right)$
chiralities of $A_{\alpha}$ can be gauged to zero ($A_{a}=0$), the
other, $\bar{A}_{\dot{a}}$, is set to depend on only half of
the $\theta^{\alpha}$'s. This solution was first presented in \cite{Brink:1983pf}
and is discussed in the review section \ref{sec:purespinor}. The
ghost number two massless vertex can be analogously gauged to $U^{*}=\bar{\lambda}_{\dot{a}}\bar{\lambda}_{\dot{b}}\bar{A}_{\dot{a}\dot{b}}$,
where $\bar{A}_{\dot{a}\dot{b}}$ represents the nonvanishing
components of $A_{\alpha\beta}$ and is simply given in terms of the
superfield $\bar{A}_{\dot{a}}$. This immediately leads to the
definition of the antifields ground state and the spectrum is build
in terms of the DDF creation operators.

The introduction of the integrated vertex associated to $U^{*}$ is
a natural step and it is straightforward to build. The first implication
is the extension of the DDF algebra. Denoting the integrated vertices
by $V_{\textrm{L.C.}}(k;a_{i},\bar{\xi}_{\dot{a}})$ and $V_{\textrm{L.C.}}^{*}(k;a_{i}^{*},\bar{\xi}_{\dot{a}}^{*})$
for the massless field and antifield respectively, the algebra can
be cast as\begin{subequations}\label{eq:extendedDDFalgebra}
\begin{eqnarray}
[V_{\textrm{L.C.}}(k;a_{i},\bar{\xi}_{\dot{a}}),V_{\textrm{L.C.}}(p;b_{i},\bar{\chi}_{\dot{a}})] & = & \delta_{k+p}\sqrt{2}\{ka_{j}b_{j}+i\bar{\xi}_{\dot{a}}\bar{\chi}_{\dot{a}}\}P^{+},\\
{}[V_{\textrm{L.C.}}(k;a_{i},\bar{\xi}_{\dot{a}}),V_{\textrm{L.C.}}^{*}(p;a_{i}^{*},\bar{\xi}_{\dot{a}}^{*})] & \approx & 2k\delta_{p+k}\{ka_{j}a_{j}^{*}+i\bar{\xi}_{\dot{a}}\bar{\xi}_{\dot{a}}^{*}\}c_{0}^{+}\nonumber \\
 &  & -ik\delta_{p+k}\{a_{j}(\sigma^{j}\bar{\xi}^{*})_{a}-a_{j}^{*}(\sigma^{j}\bar{\xi})_{a}\}W_{a}^{*}\left(0\right),\\
{}[V_{\textrm{L.C.}}^{*}(k;a_{i}^{*},\bar{\xi}_{\dot{a}}^{*}),V_{\textrm{L.C.}}^{*}(p;b_{i}^{*},\bar{\chi}_{\dot{a}}^{*})] & \approx & -4k\delta_{k+p}\{ka_{j}^{*}b_{j}^{*}+i\bar{\xi}_{\dot{a}}^{*}\bar{\chi}_{\dot{a}}^{*}\}M.
\end{eqnarray}
\end{subequations}Here, $k$ and $p$ parametrise the $P^{-}\neq0$
momenta, and the remaining arguments of $V_{\textrm{L.C.}}$ and $V_{\textrm{L.C.}}^{*}$
represent the $SO\left(8\right)$ polarisations of the massless field
and antifield. The first equation is exact, the usual creation/annihilation
algebra, while the last ones hold up to BRST-trivial terms. Just like
the operator $P^{+}$ is an element of the zero-momentum cohomology,
so are $W_{a}^{*}\left(0\right)$, $c_{0}^{+}$ and $M$. The latter,
for example, is the integrated form of the pure spinor measure of
integration.

The algebra \eqref{eq:extendedDDFalgebra} is a simple supersymmetric
extension of the bosonic string one, given by\begin{subequations}
\begin{eqnarray}
[V_{\textrm{L.C.}}(k;a_{i}),V_{\textrm{L.C.}}(p;b_{i})] & = & \sqrt{2}\delta_{k+p}ka_{j}b_{j}P^{+},\\
{}[V_{\textrm{L.C.}}(k;a_{i}),V_{\textrm{L.C.}}^{*}(p;a_{i}^{*})] & \approx & 2\delta_{p+k}k^{2}a_{j}a_{j}^{*}c_{0}^{+},\\
{}[V_{\textrm{L.C.}}^{*}(k;a_{i}^{*}),V_{\textrm{L.C.}}^{*}(p;b_{i}^{*})] & \approx & -4\delta_{k+p}k^{2}a_{j}^{*}b_{j}^{*}M_{bos},
\end{eqnarray}
\end{subequations}where $c_{0}^{+}\equiv-\sqrt{2}\ointctrclockwise\partial c\partial X^{+}$
and $M_{bos}\equiv\ointctrclockwise\left(\partial c\partial^{2}c\right)$.
The operator $c_{0}^{+}$ is of particular interest. Observe that
it satisfies
\begin{equation}
i\sqrt{2}P^{+}=\{b_{0},c_{0}^{+}\},\label{eq:DDFbosbc}
\end{equation}
where $b_{0}$ is the zero mode of the fundamental $b$ ghost. This
means that $c_{0}^{+}$ acts as a DDF conjugate of $b_{0}$ in the
subspace $P^{+}\neq0$. There is, of course, the $P^{-}\neq0$ analogous
of this operator, defined by $c_{0}^{-}\equiv-\sqrt{2}\ointctrclockwise\partial c\partial X^{-}$.

Because the cohomology of the bosonic string $b_{0}$ is trivial,
equation \eqref{eq:DDFbosbc} brings no new information. However,
this is potentially interesting in the pure spinor formalism. The
operators $c_{0}^{\pm}$ indeed act as a composite $c$ ghost, taking
any element of the ghost number one cohomology to its related antifield.
The explicit expression for $c_{0}^{+}$, for example, is 
\begin{equation}
c_{0}^{+}=\frac{1}{6}\ointctrclockwise\left\{ \Pi_{i}\theta_{ij}\Lambda_{j}-\frac{3}{\sqrt{2}}\bar{N}_{i}\Lambda_{i}-\frac{3}{2}\left(\theta\sigma_{i}\bar{d}\right)\Lambda_{i}-\frac{1}{2}\left(\theta\sigma_{i}\partial\bar{\theta}\right)\theta_{ij}\Lambda_{j}\right\} .
\end{equation}
The $SO\left(8\right)$ notation used above is discussed in detail
in subsection \ref{sub:DDF-like-operators}. By investigating the
role of the composite $b$ ghost, the extended DDF algebra implies
that
\begin{equation}
i\sqrt{2}P^{\pm}=\left\{ b_{0},c_{0}^{\pm}\right\} +\left\{ Q,\xi^{\pm}\right\} ,\label{eq:b0c0PS}
\end{equation}
which differs from the bosonic string analogous by a BRST-exact piece.
The operators $\xi^{\pm}$ depend on the specific form of the $b$
ghost (recall that the pure spinor $b$ ghost is not uniquely defined).
In this sense, there is a hidden $\left(b,c\right)$ structure in
the spectrum.

In spite of the more restricted result of \eqref{eq:b0c0PS}, it is
possible to extract some information about the cohomology of the pure
spinor $b_{0}$. Assuming that the DDF states span the ghost number
two cohomology, there is an interesting consequence concerning the
spectrum. The antifields have a singular kinematic property, in a
direct parallel to the bosonic string, and the definition of a physical
state condition in terms of the action of $b_{0}$ is roughly a consequence
of this analysis.

This work is organised as follows. Section \ref{sec:purespinor} is
a review on some basic aspects of the pure spinor cohomology, including
the DDF-like construction that is extensively used next. Section \ref{sec:Antifields}
presents the analysis of the ghost number two cohomology restricted
to \emph{conformal primary operators}, with an extension of the DDF
algebra and the map from the physical states to the antifields with
the introduction of the operators $c_{0}^{\pm}$, which behave as
zero modes of a $c$ ghost within the DDF construction. Section \ref{sec:DDFandb}
discusses the physical state condition known as the Siegel gauge and
the $b$ ghost cohomology is analysed in the subspace of BRST-closed
operators. Section \ref{sec:Summary} summarises the main results
of this work:
\begin{itemize}
\item $SO\left(8\right)$-covariant construction of the super Yang-Mills
antifields;
\item $SO\left(8\right)$-covariant construction of the massless integrated
vertex for the antifields;
\item extension of the DDF algebra;
\item doubling of the pure spinor cohomology at ghost number two and its
singular BRST-exact feature;
\item definition of the operators $c_{0}^{\pm}$ and their role as DDF conjugates
of $b_{0}$;
\item analysis of the physical state condition (Siegel gauge) on the BRST-cohomology:
ghost number one states are $b_{0}$-exact up to a gauge transformation;
ghost number two states do not belong to the cohomology of $b_{0}$.
\end{itemize}
Appendix \ref{sec:computations} includes several computations that
have been skipped in the main text, \emph{e.g.} the $SO\left(8\right)$
decomposition of the pure spinor measure of integration and the extended
DDF algebra. In order to present a more familiar ground, appendix
\ref{sec:bosonic} contains a short review on the DDF operators in
bosonic string theory. Most of the results of this work have a simple
analogous there and understanding the bosonic picture will make the
pure spinor case clearer.

\section{Review of the pure spinor cohomology\label{sec:purespinor}}

\

The pure spinor (left-moving) BRST-charge is given by
\begin{equation}
Q=\ointctrclockwise\left(\lambda^{\alpha}d_{\alpha}\right),
\end{equation}
where $\lambda^{\alpha}$ is a pure spinor variable, and
\begin{equation}
d_{\alpha}=p_{\alpha}-\frac{1}{2}\partial X^{m}\left(\theta\gamma_{m}\right)_{\alpha}-\frac{1}{8}\left(\theta\gamma^{m}\partial\theta\right)\left(\theta\gamma_{m}\right)_{\alpha}
\end{equation}
is the field realization of the supersymmetric derivative
\begin{equation}
D_{\alpha}=\frac{\partial}{\partial\theta^{\alpha}}-\frac{1}{2}\left(\gamma^{m}\theta\right)_{\alpha}\frac{\partial}{\partial X^{m}}.\label{eq:SUSYderivative}
\end{equation}
Note that
\begin{equation}
d_{\alpha}\left(z\right)F\left(X,\theta;y\right)\sim\frac{D_{\alpha}F}{\left(z-y\right)},
\end{equation}
with $F$ being a superfield of the variables $X^{m}$ and $\theta^{\alpha}$.
It is straightforward to check that
\begin{equation}
d_{\alpha}\left(z\right)d_{\beta}\left(y\right)\sim-\frac{\gamma_{\alpha\beta}^{m}\Pi_{m}}{\left(z-y\right)},
\end{equation}
with $\Pi^{m}$ the supersymmetric momentum operator, defined by
\begin{equation}
\Pi^{m}=\partial X^{m}+\frac{1}{2}\left(\theta\gamma^{m}\partial\theta\right)
\end{equation}
and satisfying
\begin{eqnarray}
\Pi^{m}\left(z\right)d_{\alpha}\left(y\right) & \sim & -\frac{\gamma_{\alpha\beta}^{m}\partial\theta^{\beta}}{\left(z-y\right)},\\
\Pi^{m}\left(z\right)\Pi^{n}\left(y\right) & \sim & -\frac{\eta^{mn}}{\left(z-y\right)^{2}}.
\end{eqnarray}

Both $\Pi^{m}$ and $d_{\alpha}$ are invariant under the action of
the supersymmetry charge
\begin{equation}
Q_{\alpha}=\ointctrclockwise\left\{ p_{\alpha}+\frac{1}{2}\partial X^{m}\left(\theta\gamma_{m}\right)_{\alpha}+\frac{1}{24}\left(\theta\gamma^{m}\partial\theta\right)\left(\theta\gamma_{m}\right)_{\alpha}\right\} ,\label{eq:SUSYcharge}
\end{equation}
with algebra $\left\{ Q_{\alpha},Q_{\beta}\right\} =-i\gamma_{\alpha\beta}^{m}P_{m}$,
where $P^{m}=i\ointctrclockwise\partial X^{m}$. The matter energy-momentum
tensor can be written as
\begin{equation}
T_{\textrm{matter}}=-\frac{1}{2}\Pi^{m}\Pi_{m}-d_{\alpha}\partial\theta^{\alpha},
\end{equation}
and supersymmetry is explicit.

\subsection{Massless cohomology\label{sub:Masslesscohomology}}

\

The pure spinor constraint,
\begin{equation}
\lambda\gamma^{m}\lambda=0,\label{eq:psconstraint}
\end{equation}
is essential for the nilpotency of the BRST charge, as
\begin{equation}
Q^{2}=-\frac{1}{2}\ointctrclockwise\left(\lambda\gamma^{m}\lambda\right)\Pi_{m},
\end{equation}
and it clearly plays a fundamental role in determining its cohomology.
Perhaps the easiest way to see this is through the zero momentum states,
which can be cast as
\[
\left\{ \begin{array}{cccc}
\mathbbm{1}, & \begin{array}{c}
\left(\lambda\gamma^{m}\theta\right),\\
\left(\lambda\gamma^{m}\theta\right)\left(\gamma_{m}\theta\right)_{\alpha},
\end{array} & \begin{array}{c}
\left(\lambda\gamma_{m}\theta\right)\left(\lambda\gamma_{n}\theta\right)\left(\gamma^{mn}\theta\right)^{\alpha},\\
\left(\lambda\gamma_{n}\theta\right)\left(\lambda\gamma_{p}\theta\right)\left(\theta\gamma^{mnp}\theta\right),
\end{array} & \left(\lambda\gamma^{m}\theta\right)\left(\lambda\gamma^{n}\theta\right)\left(\lambda\gamma^{p}\theta\right)\left(\theta\gamma_{mnp}\theta\right)\end{array}\right\} .
\]
The above set is organized according to the ghost number charge defined
by the current $J=-\omega_{\alpha}\lambda^{\alpha}$, where $\omega_{\alpha}$
is the conjugate of the pure spinor, such that
\begin{equation}
J\left(z\right)\lambda^{\alpha}\left(y\right)\sim\frac{\lambda^{\alpha}}{\left(z-y\right)}.
\end{equation}

The unit operator is the only element at ghost number zero, as there
is no nontrivial structure associated to the constraint \eqref{eq:psconstraint}.
The ghost number one states correspond to the unintegrated vertices
of the super-Poincar\'e generators $P^{m}$ and $Q_{\alpha}$, \emph{cf.}
equation \eqref{eq:SUSYcharge}. Note also that the higher ghost number
elements can all be composed from the ghost number one states:\begin{subequations}
\begin{eqnarray}
\left(\lambda\gamma_{m}\theta\right)\left(\lambda\gamma_{n}\theta\right)\left(\gamma^{mn}\theta\right)^{\alpha} & = & [\left(\lambda\gamma^{m}\theta\right)][\left(\lambda\gamma^{n}\theta\right)\left(\gamma_{n}\theta\right)_{\beta}]\gamma_{m}^{\alpha\beta},\\
\left(\lambda\gamma_{n}\theta\right)\left(\lambda\gamma_{p}\theta\right)\left(\theta\gamma^{mnp}\theta\right) & = & [\left(\lambda\gamma^{m}\theta\right)\left(\gamma_{m}\theta\right)_{\alpha}][\left(\lambda\gamma^{n}\theta\right)\left(\gamma_{n}\theta\right)_{\beta}]\gamma_{p}^{\alpha\beta},\\
\left(\lambda\gamma^{m}\theta\right)\left(\lambda\gamma^{n}\theta\right)\left(\lambda\gamma^{p}\theta\right)\left(\theta\gamma_{mnp}\theta\right) & = & [\left(\lambda\gamma_{m}\theta\right)][\left(\lambda\gamma_{n}\theta\right)\left(\lambda\gamma_{p}\theta\right)\left(\theta\gamma^{npq}\theta\right)]\delta_{q}^{m}.
\end{eqnarray}
\end{subequations}Clearly only certain combinations give rise to
nontrivial elements. For example, at ghost number three one has
\begin{eqnarray}
\left(\lambda\gamma^{m}\theta\right)\left(\lambda\gamma^{p}\theta\right)\left(\lambda\gamma^{q}\theta\right)\left(\theta\gamma_{npq}\theta\right) & = & \left(\frac{1}{10}\right)\delta_{n}^{m}\left(\lambda\gamma^{p}\theta\right)\left(\lambda\gamma^{q}\theta\right)\left(\lambda\gamma^{r}\theta\right)\left(\theta\gamma_{pqr}\theta\right)\nonumber \\
 &  & +\frac{1}{20}[Q,\left(\theta\gamma_{npq}\theta\right)\left(\lambda\gamma^{mpr}\theta\right)\left(\theta\gamma^{qst}\theta\right)\left(\lambda\gamma_{t}\theta\right)\eta_{rs}]\nonumber \\
 &  & +\frac{3}{40}[Q,\left(\theta\gamma^{mpq}\theta\right)\left(\lambda\gamma_{q}\theta\right)\left(\theta\gamma_{npr}\theta\right)\left(\lambda\gamma^{r}\theta\right)],
\end{eqnarray}
and the traceless composition from the left hand side of the equation
is BRST-exact.

As mentioned before, physical states in the pure spinor formalism
are defined to be in the ghost number one cohomology. The massless
vertex, $U$, is described by a superfield $A_{\alpha}$ built from
the zero modes of $X^{m}$ and $\theta^{\alpha}$:
\begin{equation}
U=\lambda^{\alpha}A_{\alpha}\left(X,\theta\right).\label{eq:unintegratedmassless}
\end{equation}
Observe that $\left\{ Q,U\right\} =\lambda^{\alpha}\lambda^{\beta}D_{\alpha}A_{\beta}$,
which can be Fierz decomposed to
\begin{eqnarray}
\left\{ Q,U\right\}  & = & \frac{1}{16}\left(\lambda\gamma^{m}\lambda\right)\left(D\gamma_{m}A\right)+\frac{1}{3!\cdot16}\left(\lambda\gamma^{mnp}\lambda\right)\left(D\gamma_{mnp}A\right)\nonumber \\
 &  & +\frac{1}{5!\cdot32}\left(\lambda\gamma^{mnpqr}\lambda\right)\left(D\gamma_{mnpqr}A\right)
\end{eqnarray}
The first term on the right hand side is proportional to the pure
spinor constraint while the second vanishes because $\gamma_{\alpha\beta}^{mnp}$
is antisymmetric in the spinor indices. BRST-closedness of the vertex
implies $D\gamma^{mnpqr}A=0$, which is the equation of motion for
the superfield $A_{\alpha}$ describing a massless vector boson, $a_{m}$,
and its superpartner, $\xi^{\alpha}$ \cite{Siegel:1978yi}. In the
gauge $\theta^{\alpha}A_{\alpha}=0$, the superfield can be expanded
as
\begin{eqnarray}
A_{\alpha} & = & \xi^{\beta}\left(\gamma_{m}\theta\right)_{\beta}\left(\gamma^{m}\theta\right)_{\alpha}-\frac{1}{8}\partial_{m}\xi^{\beta}\left(\gamma_{n}\theta\right)_{\beta}\left(\theta\gamma^{mnp}\theta\right)\left(\gamma_{p}\theta\right)_{\alpha}\nonumber \\
 &  & +a_{m}\left(\gamma^{m}\theta\right)_{\alpha}+\frac{1}{4}\partial_{n}a_{m}\left(\theta\gamma^{mnp}\theta\right)\left(\gamma_{p}\theta\right)_{\alpha}+\mathcal{O}\left(\theta^{5}\right).\label{eq:gaugesuperfield}
\end{eqnarray}
The gauge transformations of $A_{\alpha}$ assume the form $\delta A_{\alpha}=D_{\alpha}\Lambda$,
as $U$ is defined up to BRST-exact terms, $\delta U=\left[Q,\Lambda\right]$.
The integrated version of the vertex $U$ is given by
\begin{equation}
V=\oint\left\{ \Pi^{m}A_{m}+\partial\theta^{\alpha}A_{\alpha}+d_{\alpha}W^{\alpha}+N_{mn}F^{mn}\right\} ,\label{eq:integratedmassless}
\end{equation}
where
\begin{eqnarray}
A_{m} & = & \frac{1}{8}\left(D\gamma_{m}A\right),\label{eq:defvectorsuperfield}\\
W^{\alpha} & = & \frac{1}{10}\left[\left(\gamma^{m}D\right)^{\alpha}A_{m}-\partial_{m}\left(\gamma^{m}A\right)^{\alpha}\right],\\
F_{mn} & = & \frac{1}{2}\left(\partial_{m}A_{n}-\partial_{n}A_{m}\right),
\end{eqnarray}
and $N^{mn}=-\frac{1}{2}\omega\gamma^{mn}\lambda$ is the pure spinor
Lorentz current, satisfying
\begin{equation}
N^{mn}\left(z\right)\lambda^{\alpha}\left(y\right)\sim\frac{1}{2}\frac{\left(\gamma^{mn}\lambda\right)^{\alpha}}{\left(z-y\right)}.
\end{equation}
BRST-closedness of $V$ again relies on the pure spinor constraint
and it is easy to show that $\left[Q,V\right]=\ointctrclockwise\partial U$.

The extension of the covariant analysis to massive states ends up
introducing a lot of auxiliary superfields with unclear field content
(currently only the first massive level has a known covariant superfield
description \cite{Berkovits:2002qx}). Previous analyses of the pure
spinor cohomology relied on nontrivial operations on the BRST charge
$Q$ (infinity set of ghosts, similarity transformations, etc...),
which made the superfield character of the vertices very obscure.
Inspired by the DDF description of the bosonic string cohomology\footnote{The reader not familiar with the DDF operators in bosonic string is
advised to follow the quick review presented in the appendix.}, there is now a very transparent way of building the physical vertices
in an $SO\left(8\right)$-covariant way, which will be reviewed below.

\subsection{DDF-like operators\label{sub:DDF-like-operators}}

\

The DDF operators are built on the light-cone frame and it will be
useful to establish the $SO\left(8\right)$ notation beforehand.

Any $SO\left(1,9\right)$ vector, $K^{m}$, will be decomposed as
$\sqrt{2}K^{\pm}=\left(K^{0}\pm K^{9}\right)$, with transversal components
represented by $K^{i}$, with $i=1,\ldots,8$. In this notation, the
metric $\eta^{mn}$ is such that $\eta^{+-}=-1$, $\eta^{++}=\eta^{--}=\eta^{\pm i}=0$
and $\eta^{ij}$ is the flat $SO\left(8\right)$ vector metric. For
a rank-$2$ antisymmetric tensor $K^{mn}$, the $SO\left(8\right)$
components will be represented as 
\[
\left\{ K^{ij},K^{i}=K^{-i},\bar{K}^{i}=K^{+i},K=K^{+-}\right\} .
\]

Given a spinor $\xi^{\alpha}$, one can denote its $SO\left(8\right)$
components as $\xi_{a}$ and $\bar{\xi}_{\dot{a}}$, where $a,\dot{a}=1,\ldots,8$
are the $SO\left(8\right)$ spinorial indices, representing different
chiralities. Note that upper and lower indices in the $SO\left(8\right)$
language do not distinguish chiralities, \emph{i.e.}, one can define
a spinorial metric, $\eta_{ab}$ ($\eta_{\dot{a}\dot{b}}$), and its
inverse, $\eta^{ab}$ ($\eta^{\dot{a}\dot{b}}$), such that $\eta_{ac}\eta^{cb}=\delta_{a}^{b}$
($\eta_{\dot{a}\dot{c}}\eta^{\dot{c}\dot{b}}=\delta_{\dot{a}}^{\dot{b}}$),
which are responsible for lowering or raising spinorial indices, acting
as charge conjugation. 

The matrices $\gamma^{m}$ are conveniently written in terms of the
$8$-dimensional equivalent of the Pauli matrices, $\sigma_{a\dot{a}}^{i}$,
which satisfy the following properties\begin{subequations}\label{eq:pauliproperties}
\begin{eqnarray}
\left(\sigma_{a\dot{a}}^{i}\sigma_{b\dot{b}}^{j}+\sigma_{a\dot{b}}^{i}\sigma_{b\dot{a}}^{j}\right)\eta^{\dot{a}\dot{b}} & = & 2\eta^{ij}\eta_{ab},\\
\left(\sigma_{a\dot{a}}^{i}\sigma_{b\dot{b}}^{j}+\sigma_{a\dot{b}}^{i}\sigma_{b\dot{a}}^{j}\right)\eta^{ab} & = & 2\eta^{ij}\eta_{\dot{a}\dot{b}},\\
\left(\sigma_{a\dot{a}}^{i}\sigma_{b\dot{b}}^{j}+\sigma_{a\dot{b}}^{i}\sigma_{b\dot{a}}^{j}\right)\eta_{ij} & = & 2\eta_{ab}\eta_{\dot{a}\dot{b}}.
\end{eqnarray}
\end{subequations}The non-vanishing components of $\gamma_{\alpha\beta}^{m}$
and $\left(\gamma^{m}\right)^{\alpha\beta}$ are\begin{subequations}\label{eq:gammadecomposition}
\begin{equation}
\begin{array}{cc}
\gamma_{\alpha\beta}^{i}\equiv\sigma_{a\dot{a}}^{i}, & \left(\gamma^{i}\right)^{\alpha\beta}\equiv\sigma_{b\dot{b}}^{i}\eta^{ab}\eta^{\dot{a}\dot{b}},\\
\gamma_{\alpha\beta}^{+}\equiv-\sqrt{2}\eta_{ab}, & \left(\gamma^{+}\right)^{\alpha\beta}\equiv\sqrt{2}\eta^{\dot{a}\dot{b}},\\
\gamma_{\alpha\beta}^{-}\equiv-\sqrt{2}\eta_{\dot{a}\dot{b}}, & \left(\gamma^{-}\right)^{\alpha\beta}\equiv\sqrt{2}\eta^{ab},
\end{array}
\end{equation}
\end{subequations} and the usual anticommutation relation $\left\{ \gamma^{m},\gamma^{n}\right\} =2\eta^{mn}$
follows from \eqref{eq:pauliproperties}. From now on the light-cone
coordinates will be used, unless explicitly said otherwise. All the
$SO\left(8\right)$ metrics will be chosen to be equal to the identity
and no distinction will be made between upper and lower indices.

Several combinations of $\theta$'s and $\lambda$'s will appear,
so a short notation will help simplifying the results. The pure spinor
constraint is rewritten as 
\begin{equation}
\lambda_{a}\sigma_{a\dot{a}}^{i}\bar{\lambda}_{\dot{a}}=\lambda_{a}\lambda_{a}=\bar{\lambda}_{\dot{a}}\bar{\lambda}_{\dot{a}}=0,
\end{equation}
and the following definitions will be recurrent, relating some of
the $SO\left(1,9\right)$ bispinors to their $SO\left(8\right)$ decompositions:\begin{subequations}\label{eq:thetalambdacombinations}
\begin{equation}
\begin{array}{cc}
\begin{array}{rcl}
\theta^{ij} & \equiv & -\frac{1}{\sqrt{2}}\left(\theta\gamma^{+ij}\theta\right)\\
 & = & \theta_{a}\sigma_{ab}^{ij}\theta_{b},
\end{array} & \begin{array}{rcl}
\bar{\theta}^{ij} & \equiv & -\frac{1}{\sqrt{2}}\left(\theta\gamma^{-ij}\theta\right)\\
 & = & \bar{\theta}_{\dot{a}}\sigma_{\dot{a}\dot{b}}^{ij}\bar{\theta}_{\dot{b}},
\end{array}\end{array}
\end{equation}
\begin{equation}
\begin{array}{cc}
\begin{array}{rcl}
\Lambda^{ij} & \equiv & -\frac{1}{\sqrt{2}}\left(\lambda\gamma^{+ij}\theta\right)\\
 & = & \lambda_{a}\sigma_{ab}^{ij}\theta_{b},
\end{array} & \begin{array}{rcl}
\bar{\Lambda}^{ij} & \equiv & -\frac{1}{\sqrt{2}}\left(\theta\gamma^{-ij}\theta\right)\\
 & = & \bar{\lambda}_{\dot{a}}\sigma_{\dot{a}\dot{b}}^{ij}\bar{\theta}_{\dot{b}},
\end{array}\end{array}
\end{equation}
\begin{equation}
\begin{array}{cc}
\begin{array}{rcl}
\Lambda^{i} & \equiv & \frac{1}{2}\left(\lambda\gamma^{-}\gamma^{i+}\theta\right)\\
 & = & \bar{\lambda}_{\dot{a}}\sigma_{a\dot{a}}^{i}\theta_{a},
\end{array} & \begin{array}{rcl}
\bar{\Lambda}^{i} & \equiv & \frac{1}{2}\left(\lambda\gamma^{+}\gamma^{i-}\theta\right)\\
 & = & \lambda_{a}\sigma_{a\dot{a}}^{i}\bar{\theta}_{\dot{a}},
\end{array}\end{array}
\end{equation}
\begin{equation}
\begin{array}{cc}
\begin{array}{rcl}
\Lambda & \equiv & -\frac{1}{\sqrt{2}}\left(\lambda\gamma^{+}\theta\right)\\
 & = & \lambda_{a}\theta_{a},
\end{array} & \begin{array}{rcl}
\bar{\Lambda} & \equiv & -\frac{1}{\sqrt{2}}\left(\lambda\gamma^{-}\theta\right)\\
 & = & \bar{\lambda}_{\dot{a}}\bar{\theta}_{\dot{a}}.
\end{array}\end{array}
\end{equation}
\end{subequations}Here, $\sigma^{ij}\equiv\frac{1}{2}\left(\sigma^{i}\sigma^{j}-\sigma^{j}\sigma^{i}\right)$.

Having fixed the notation, the first step in determining the set of
operators to proceed to the DDF construction is to find a convenient
gauge for the superfield $A_{\alpha}$ of \eqref{eq:gaugesuperfield}
in a given light-cone frame. Working with momentum eigenfunctions
and choosing the frame where $k^{-}=k\sqrt{2}$ and $k^{+}=k^{i}=0$,
a quick analysis already determines the physical polarizations to
be $a^{i}$ and $\bar{\xi}_{\dot{a}}$. The component $a^{+}$
is removed by the condition $a_{m}k^{m}=0$ while $a^{-}$ is pure
gauge. For its superpartner, the equation of motion $k_{m}\left(\gamma^{m}\xi\right)_{\alpha}=0$
implies $\xi_{a}=0$, since $\left(\gamma^{+}\right)^{\alpha\beta}$
projects onto one of the $SO\left(8\right)$ chiralities. It turns
out that there is a gauge where $A_{\alpha}$ gets a very simple form,\begin{subequations}\label{eq:P-spinorsuperfields}
\begin{eqnarray}
A_{a}\left(k\right) & = & 0,\\
\bar{A}_{\dot{a}}\left(k\right) & = & e^{-ik\sqrt{2}X^{+}}\left\{ \delta_{il}-\frac{ik}{3!}\theta_{il}-\frac{k^{2}}{5!}\theta_{ij}\theta_{jl}+\frac{ik^{3}}{7!}\theta_{ij}\theta_{jk}\theta_{kl}\right\} a^{i}\left(\sigma^{l}\theta\right)_{\dot{a}}+\frac{i}{k}e^{-ik\sqrt{2}X^{+}}\bar{\xi}_{\dot{a}}\nonumber \\
 &  & +e^{-ik\sqrt{2}X^{+}}\left\{ \frac{1}{2!}\delta_{il}-\frac{ik}{4!}\theta_{il}-\frac{k^{2}}{6!}\theta_{ij}\theta_{jl}+\frac{ik^{3}}{8!}\theta_{ij}\theta_{jk}\theta_{kl}\right\} \left(\bar{\xi}\sigma^{i}\theta\right)\left(\sigma^{l}\theta\right)_{\dot{a}},\label{eq:spinorsuperfield}
\end{eqnarray}
\end{subequations}where $a^{i}$ and $\bar{\xi}_{\dot{a}}$
are the physical polarizations mentioned above. Observe that the unusual
singular term when $k\to0$ is necessary due to this particular gauge
choice, where all the dependence on the $\bar{\theta}_{\dot{a}}$
was removed. It is straightforward to show the action of the supersymmetric
derivative $D_{\alpha}$:\begin{subequations}\label{eq:eomSO8superfields}
\begin{eqnarray}
D_{a}\bar{A}_{\dot{a}}\left(k\right) & = & \sigma_{a\dot{a}}^{i}A_{i}\left(k\right),\\
\bar{D}_{\dot{a}}\bar{A}_{\dot{b}}\left(k\right) & = & 0,\\
D_{a}A_{i}\left(k\right) & = & ik\sigma_{a\dot{a}}^{i}\bar{A}_{\dot{a}}\left(k\right),\\
\bar{D}_{\dot{a}}A_{i}\left(k\right) & = & 0,
\end{eqnarray}
\end{subequations}where
\begin{eqnarray}
A_{i}\left(k\right) & = & e^{-ik\sqrt{2}X^{+}}\left\{ \delta_{ij}+\frac{ik}{2!}\theta_{ij}-\frac{k^{2}}{4!}\theta_{ik}\theta_{kj}-\frac{ik^{3}}{6!}\theta_{ik}\theta_{kl}\theta_{lj}+\frac{k^{4}}{8!}\theta_{ik}\theta_{kl}\theta_{lm}\theta_{mj}\right\} a^{j}\nonumber \\
 &  & +e^{-ik\sqrt{2}X^{+}}\left\{ \delta_{ij}+\frac{ik}{3!}\theta_{ij}-\frac{k^{2}}{5!}\theta_{ik}\theta_{kj}-\frac{ik^{3}}{7!}\theta_{ik}\theta_{kl}\theta_{lj}\right\} \left(\bar{\xi}\sigma^{j}\theta\right)\label{eq:vectorsuperfield}
\end{eqnarray}
represents the non-vanishing components of the superfield $A_{m}$
introduced in \eqref{eq:defvectorsuperfield} \cite{Brink:1983pf}.

The next step is to translate the above results to the massless pure
spinor cohomology. Instead of restricting the discussion to the open
string, it is more enriching to view them as coming from the holomorphic
sector of the closed string. The worldsheet scalars $X^{m}$ are the
only possible source of problems in this transition and will be written
as 
\begin{equation}
X^{m}\left(z,\bar{z}\right)=X_{L}^{m}\left(z\right)+X_{R}^{m}\left(\bar{z}\right).
\end{equation}
The subtleties coming from this holomorphic splitting will not play
any role in the construction of the physical spectrum and will be
ignored throughout this work.

Inserting the superfields of \eqref{eq:P-spinorsuperfields} in $U=\lambda^{\alpha}A_{\alpha}$,
one obtains
\begin{equation}
U=a_{i}\bar{U}_{i}+\bar{\xi}_{\dot{a}}\bar{Y}_{\dot{a}},\label{eq:defunintegratedpolarizations}
\end{equation}
where
\begin{eqnarray}
\bar{U}_{i}\left(z;k\right) & \equiv & e^{-ik\sqrt{2}X_{L}^{+}}\left\{ \Lambda_{i}-\frac{ik}{3!}\theta_{ij}\Lambda_{j}-\frac{k^{2}}{5!}\theta_{ij}\theta_{jk}\Lambda_{k}+\frac{ik^{3}}{7!}\theta_{ij}\theta_{jk}\theta_{kl}\Lambda_{l}\right\} ,\label{eq:Uiunintegrated}\\
\bar{Y}_{\dot{a}}\left(z;k\right) & \equiv & e^{-ik\sqrt{2}X_{L}^{+}}\left(\theta\sigma^{i}\right)_{\dot{a}}\left\{ \frac{1}{2!}\Lambda_{i}-\frac{ik}{4!}\theta_{ij}\Lambda_{j}-\frac{k^{2}}{6!}\theta_{ij}\theta_{jk}\Lambda_{k}+\frac{ik^{3}}{8!}\theta_{ij}\theta_{jk}\theta_{kl}\Lambda_{l}\right\} \nonumber \\
 &  & +\left(\frac{i}{k}\right)e^{-ik\sqrt{2}X_{L}^{+}}\bar{\lambda}_{\dot{a}},\label{eq:Yaunintegrated}
\end{eqnarray}
corresponding to the gauge fixed unintegrated massless vertices of
the $SO\left(8\right)$ vector and spinor polarizations\footnote{To match the notation of previous works, the vertices here differ
from the ones in \cite{Jusinskas:2014vqa} by imaginary factors.}. 

Both $\bar{U}_{i}$ and $\bar{Y}_{\dot{a}}$ transform nicely
under the action of the supersymmetry charge \eqref{eq:SUSYcharge},\begin{subequations}\label{eq:SUSYphysical}
\begin{eqnarray}
\{\bar{Q}_{\dot{a}},\bar{U}_{i}\} & = & 0,\\
{}[\bar{Q}_{\dot{a}},\bar{Y}_{\dot{b}}] & = & 0,\\
\{Q_{a},\bar{U}_{i}\} & = & -ik\sigma_{a\dot{a}}^{i}\bar{Y}_{\dot{a}},\\
{}[Q_{a},\bar{Y}_{\dot{a}}] & = & \sigma_{a\dot{a}}^{i}\bar{U}_{i},
\end{eqnarray}
\end{subequations}and BRST-closedness follow from the equations in
\eqref{eq:eomSO8superfields} and the pure spinor constraint.

The integrated vertex, denoted by $V_{\textrm{L.C.}}\left(k;a_{i},\bar{\xi}_{\dot{a}}\right)$,
comes from a simple insertion of the gauge fixed superfield and its
auxiliaries in \eqref{eq:integratedmassless},
\begin{equation}
V_{\textrm{L.C.}}\left(k;a_{i},\bar{\xi}_{\dot{a}}\right)=\oint\left\{ \left(\Pi_{i}-i\sqrt{2}k\bar{N}_{i}\right)A_{i}+\left(\partial\bar{\theta}_{\dot{a}}+ik\bar{d}_{\dot{a}}\right)\bar{A}_{\dot{a}}\right\} ,\label{eq:SO8integrated}
\end{equation}
with $\bar{N}^{i}$ denoting the components $N^{+i}$ of the
Lorentz ghost current. 

In the DDF construction, the integrated massless vertices constitute
a creation/annihilation algebra acting on a determined fundamental
state. A direct computation shows that the pure spinor vertices of
\eqref{eq:SO8integrated} satisfy the following commutation relation:
\begin{eqnarray}
\left[V_{\textrm{L.C.}}(k),V_{\textrm{L.C.}}(p)\right] & = & -\oint\left\{ A_{i}\left(p\right)\partial A_{i}\left(k\right)+ip\bar{A}_{\dot{a}}\left(p\right)\partial\bar{A}_{\dot{a}}\left(k\right)\right\} \nonumber \\
 &  & +ip\oint\bar{A}_{\dot{a}}\left(p\right)\partial\theta_{a}D_{a}\bar{A}_{\dot{a}}\left(k\right)\label{eq:VVcomputation}
\end{eqnarray}
Although far from obvious, the right hand side can be written in a
very simple way due to \eqref{eq:eomSO8superfields}. It might be
helpful to point out that $\partial=\Pi^{+}\partial_{+}+\partial\theta^{a}D_{a}$
whenever acting on superfields that depend only on $X_{L}^{+}$ and
$\theta^{a}$. Observe that
\begin{multline}
A_{i}\left(p\right)\partial A_{i}\left(k\right)+ip\bar{A}_{\dot{a}}\left(p\right)\partial\bar{A}_{\dot{a}}\left(k\right)-ip\bar{A}_{\dot{a}}\left(p\right)\partial\theta_{a}D_{a}\bar{A}_{\dot{a}}\left(k\right)=\\
=\frac{k}{k+p}\partial\left(A_{i}\left(p\right)A_{i}\left(k\right)+ip\bar{A}_{\dot{a}}\left(p\right)\bar{A}_{\dot{a}}\left(k\right)\right),\label{eq:totalderivativeproduct}
\end{multline}
so the integrand of \eqref{eq:VVcomputation} is a total derivative
for $(k+p)\neq0$. Another interesting consequence of \eqref{eq:totalderivativeproduct}
is that the expression inside the parentheses on the right hand side
is a constant for $(k+p)=0$, which can be shown to be: 
\begin{equation}
A_{i}\left(k;a_{i},\bar{\xi}_{\dot{a}}\right)A_{i}\left(-k;b_{i},\bar{\chi}_{\dot{a}}\right)+ik\bar{A}_{\dot{a}}\left(k;a_{i},\bar{\xi}_{\dot{a}}\right)\bar{A}_{\dot{a}}\left(-k;b_{i},\bar{\chi}_{\dot{a}}\right)=a_{i}b_{i}+\left(\frac{i}{k}\right)\bar{\xi}_{\dot{a}}\bar{\chi}_{\dot{a}}.\label{eq:superfieldproductconstant}
\end{equation}

Therefore,
\begin{eqnarray}
\left[V_{\textrm{L.C.}}(k),V_{\textrm{L.C.}}(p)\right] & = & \delta_{k+p}\oint\left\{ A_{i}\left(k\right)\partial A_{i}\left(-k\right)+ik\bar{A}_{\dot{a}}\left(k\right)\partial\bar{A}_{\dot{a}}\left(-k\right)\right\} \nonumber \\
 &  & +\delta_{k+p}\oint\left\{ ik\partial\theta_{a}D_{a}\bar{A}_{\dot{a}}\left(k\right)\bar{A}_{\dot{a}}\left(-k\right)\right\} \nonumber \\
 & = & \delta_{k+p}ik\sqrt{2}\oint\left\{ A_{i}\left(k\right)A_{i}\left(-k\right)+ik\bar{A}_{\dot{a}}\left(k\right)\bar{A}_{\dot{a}}\left(-k\right)\right\} \partial X^{+},
\end{eqnarray}
where in the last line surface contributions were again discarded.
Using the result \eqref{eq:superfieldproductconstant}, the commutator
takes the final form,
\begin{equation}
\left[V_{\textrm{L.C.}}(k;a_{i},\bar{\xi}_{\dot{a}}),V_{\textrm{L.C.}}(p;b_{i},\bar{\chi}_{\dot{a}})\right]=\delta_{k+p}\sqrt{2}\left\{ ka_{i}b_{i}+i\bar{\xi}_{\dot{a}}\bar{\chi}_{\dot{a}}\right\} P^{+},\label{eq:creation-annihilationFULL}
\end{equation}
constituting a supersymmetric creation/annihilation algebra whenever
acting on states with $P^{+}\neq0$.

Given the vertex \eqref{eq:defunintegratedpolarizations}, sometimes
it is easier to view supersymmetry as a passive transformation on
the polarizations $a_{i}$ and $\bar{\xi}_{\dot{a}}$ instead
of an active transformation on the basis $\bar{U}_{i}\left(k\right)$
and $\bar{Y}_{\dot{a}}\left(k\right)$:\begin{subequations}
\begin{equation}
\begin{array}{cc}
[Q_{a},a_{i}]=-\sigma_{a\dot{a}}^{i}\bar{\xi}_{\dot{a}}, & \{Q_{a},\bar{\xi}_{\dot{a}}\}=-ik\sigma_{a\dot{a}}^{i}a_{i}.\end{array}
\end{equation}
\end{subequations}When looking at the creation/annihilation algebra
of \eqref{eq:creation-annihilationFULL}, supersymmetry is consistent
with the combination $\delta_{k+p}\left(ka_{i}b_{i}+i\bar{\xi}_{\dot{a}}\bar{\chi}_{\dot{a}}\right)$.
Observe that
\begin{equation}
\frac{1}{2i}\left\{ Q_{a},\left[a_{i}(\sigma^{i}\bar{\chi})_{b}-b_{i}(\sigma^{i}\bar{\xi})_{b}\right]\right\} =\left(ka_{i}b_{i}+i\bar{\xi}_{\dot{a}}\bar{\chi}_{\dot{a}}\right)\eta_{ab}.\label{eq:SUSYpolarizations}
\end{equation}
In spite of this result, the right hand side of \eqref{eq:creation-annihilationFULL}
is the most general expression compatible with supersymmetry, since
$P^{+}$ itself is supersymmetric. In the next section a more general
construction will be introduced for the antifields, which includes
the structure of \eqref{eq:SUSYpolarizations}.

Another interesting feature of the vertex \eqref{eq:SO8integrated}
is its action on the operator $\left(\lambda\gamma^{-}\theta\right)=-\sqrt{2}\bar{\Lambda}$
defined in \eqref{eq:thetalambdacombinations}:
\begin{eqnarray}
\left[V_{\textrm{L.C.}}\left(k;a_{i},\bar{\xi}_{\dot{a}}\right),-\frac{\bar{\Lambda}}{2}\right] & = & \frac{ik}{2}\bar{\Lambda}_{i}A_{i}\left(k\right)+\frac{ik}{2}\bar{\lambda}_{\dot{a}}\bar{A}_{\dot{a}}\left(k\right)\nonumber \\
 & = & ik\bar{\lambda}_{\dot{a}}\bar{A}_{\dot{a}}\left(k\right)-\frac{ik}{2}\left\{ Q,\bar{\theta}_{\dot{a}}\bar{A}_{\dot{a}}\left(k\right)\right\} .\label{eq:masslessfromzeromomentum}
\end{eqnarray}
The unintegrated massless vertex, up to a gauge transformation, can
be thought of as coming from the action of the DDF operator on an
``unpolarised'' state of the zero-momentum cohomology.

The last step to build the physical spectrum through the algebra \eqref{eq:creation-annihilationFULL}
is to define the ground state which they should act upon. They are
of course build from the $P^{+}\neq0$ analogous of \eqref{eq:P-spinorsuperfields}:\begin{subequations}\label{eq:P+spinorsuperfields}
\begin{eqnarray}
A_{a}\left(k\right) & = & e^{-ik\sqrt{2}X_{L}^{-}}\left\{ \delta_{il}-\frac{ik}{3!}\bar{\theta}_{il}-\frac{k^{2}}{5!}\bar{\theta}_{ij}\bar{\theta}_{jl}+\frac{ik^{3}}{7!}\bar{\theta}_{ij}\bar{\theta}_{jk}\bar{\theta}_{kl}\right\} a^{i}\left(\sigma^{l}\bar{\theta}\right)_{a}+\frac{i}{k}e^{-ik\sqrt{2}X_{L}^{-}}\xi_{a}\nonumber \\
 &  & +e^{-ik\sqrt{2}X_{L}^{-}}\left\{ \frac{1}{2!}\delta_{il}-\frac{ik}{4!}\bar{\theta}_{il}-\frac{k^{2}}{6!}\bar{\theta}_{ij}\bar{\theta}_{jl}+\frac{ik^{3}}{8!}\bar{\theta}_{ij}\bar{\theta}_{jk}\bar{\theta}_{kl}\right\} \left(\xi\sigma^{i}\bar{\theta}\right)\left(\sigma^{l}\bar{\theta}\right)_{a},\\
\bar{A}_{\dot{a}}\left(k\right) & = & 0.
\end{eqnarray}
\end{subequations}Denoting the fundamental state by $\left|0,k\right\rangle $,
the state-operator map gives
\begin{eqnarray}
\left|0,k\right\rangle  & = & \lim_{z\to0}\lambda^{a}A_{a}\left(k\right)\left|0\right\rangle \nonumber \\
 & \equiv & a_{i}\left|i,k\right\rangle +\xi_{a}\left|a,k\right\rangle ,\label{eq:fieldground}
\end{eqnarray}
and the massive spectrum is obtained through the action of the operators
$V_{\textrm{L.C.}}$. The implementation is detailedly presented in
\cite{Jusinskas:2014vqa} and will not be repeated here.

Now the ghost number two cohomology will be discussed.

\section{Antifields vertex operators\label{sec:Antifields}}

\

In bosonic string theory, the antifields have an odd feature that
distinguishes them from the physical states (see appendix \ref{sec:bosonic}).
This is not the case in the pure spinor superstring and in this perspective
the definition of a physical state has to be better understood.

At the massless level, a generic element of the ghost number two cohomology
can be cast as
\begin{equation}
U^{*}=\lambda^{\alpha}\lambda^{\beta}A_{\alpha\beta},\label{eq:masslessantifieldvertex}
\end{equation}
where $A_{\alpha\beta}=A_{\alpha\beta}\left(X,\theta\right)$ is an
$SO\left(1,9\right)$ superfield constructed out of the zero modes
of $X^{m}$ and $\theta^{\alpha}$. Observe that
\begin{equation}
\left\{ Q,U^{*}\right\} =\lambda^{\gamma}\lambda^{\alpha}\lambda^{\beta}D_{\gamma}A_{\alpha\beta}\label{eq:BRSTantifield}
\end{equation}
and the superfield equation of motion follows from the BRST-closedness
of $U^{*}$:
\begin{equation}
D_{((\gamma}A_{\alpha\beta))}=0.\label{eq:antifieldeom}
\end{equation}
The double parentheses represent a symmetrized gamma-traceless operation
on the spinor indices, a consequence of the pure spinor condition. 

The field content of $A_{\alpha\beta}$ is composed by an anticommuting
vector field, $a_{m}^{*}$, and its superpartner, $\xi_{\alpha}^{*}$,
\begin{eqnarray}
U^{*} & = & \left(\lambda\gamma_{n}\theta\right)\left(\lambda\gamma_{p}\theta\right)\left(\theta\gamma^{np}\xi^{*}\right)+a_{n}^{*}\left(\lambda\gamma_{p}\theta\right)\left(\lambda\gamma_{q}\theta\right)\left(\theta\gamma^{npq}\theta\right)\nonumber \\
 &  & +\left(\frac{1}{8}\right)\partial_{m}\xi_{\alpha}^{*}\theta^{\alpha}\left(\lambda\gamma_{n}\theta\right)\left(\lambda\gamma_{p}\theta\right)\left(\theta\gamma^{mnp}\theta\right)\nonumber \\
 &  & -\left(\frac{1}{8}\right)\partial_{m}\xi_{\alpha}^{*}\left(\gamma_{np}\theta\right)^{\alpha}\left(\lambda\gamma^{n}\theta\right)\left(\lambda\gamma_{q}\theta\right)\left(\theta\gamma^{mpq}\theta\right)\nonumber \\
 &  & +\left(\frac{1}{40}\right)\partial_{m}a_{n}^{*}\left(\theta\gamma^{npr}\theta\right)\left(\lambda\gamma^{mqs}\theta\right)\left(\theta\gamma_{rst}\theta\right)\left(\lambda\gamma^{t}\theta\right)\eta_{pq}\nonumber \\
 &  & +\left(\frac{3}{80}\right)\partial_{m}a_{n}^{*}\left(\theta\gamma^{mpr}\theta\right)\left(\lambda\gamma_{r}\theta\right)\left(\theta\gamma^{nqs}\theta\right)\left(\lambda\gamma_{s}\theta\right)\eta_{pq}+\mathcal{O}\left(\theta^{7}\right),\label{eq:antifieldexpansion}
\end{eqnarray}
and their equations of motion follow from \eqref{eq:BRSTantifield}:
\begin{eqnarray}
\left\{ Q,U^{*}\right\}  & = & \left(\frac{1}{20}\right)\left(\partial\cdot a^{*}\right)\left(\lambda\gamma^{m}\theta\right)\left(\lambda\gamma^{n}\theta\right)\left(\lambda\gamma^{p}\theta\right)\left(\theta\gamma_{mnp}\theta\right)+\mathcal{O}\left(\theta^{6}\right),\\
 & \Rightarrow & \partial^{m}a_{m}^{*}=0.\nonumber 
\end{eqnarray}
This is expected for the gauge boson antifield. Since fields and antifields
are dual with respect to their equations of motion and gauge transformations,
it is no surprise that $\left\{ Q,U^{*}\right\} =0$ does not imply
an equation for $\xi_{\alpha}^{*}$, for there is no gauge freedom
associated to the gauge boson superpartner $\xi^{\alpha}$. On the
other hand, $\xi_{\alpha}^{*}$ has a nontrivial gauge transformation.
Of course $U^{*}$ is defined up to BRST-exact terms, which take the
form $\lambda^{\alpha}\lambda^{\beta}D_{\alpha}\Omega_{\beta}$ and
$\delta A_{\alpha\beta}=D_{((\alpha}\Omega_{\beta))}$ describes the
antifield gauge transformations. They can be individually expressed
as\begin{subequations}\label{eq:gaugeantifields}
\begin{eqnarray}
\delta a_{m}^{*} & = & \partial^{n}\left(\partial_{m}b_{n}-\partial_{n}b_{m}\right),\\
\delta\xi_{\alpha}^{*} & = & \gamma_{\alpha\beta}^{m}\partial_{m}\chi^{\beta},
\end{eqnarray}
\end{subequations}with gauge parameters $b_{m}$ and $\chi^{\alpha}$
\cite{Berkovits:2001rb}.

In this section, the cohomology at ghost number two will be further
analysed, including the massive levels. It will be shown that the
DDF-like extension to this sector displays some clear parallels with
the bosonic string case.

\subsection{Cohomology ring\label{sub:Cohomology-ring}}

\

A natural question about the ghost number two cohomology concerns
the composition of two elements from the ghost number one cohomology\footnote{This subject was previously studied in the amplitudes context \cite{Mafra:2010jq}.}.
To illustrate this construction consider the ordered product of two
massless superfields $A_{\alpha}$ and $A'_{\alpha}$ with momenta
$p^{m}$ and $q^{m}$, respectively:
\begin{equation}
\lambda^{\alpha}\lambda^{\beta}A_{\alpha\beta}\left(p,q;y\right)\equiv\lim_{z\to y}:\lambda^{\alpha}A_{\alpha}\left(p;z\right)\lambda^{\beta}A'_{\beta}\left(q;y\right):\label{eq:vertexsquared}
\end{equation}
Assuming $p^{m}$ parallel to $q^{m}$, there is clearly no ordering
issue and \eqref{eq:vertexsquared} is a massless BRST-closed operator. 

The existence of this vertex is expected, although it is not obvious
whether it is in the cohomology. In fact, it can be shown that \eqref{eq:vertexsquared}
is BRST-exact when $\left(p^{m}+q^{m}\right)\neq0$. When $\left(p^{m}+q^{m}\right)=0$,
it is a combination of two elements of the zero-momentum cohomology,
\emph{cf.} subsection \ref{sub:Masslesscohomology}. In order to check
this, \eqref{eq:vertexsquared} can be Lorentz rotated so that $p^{m}=p^{-}\equiv\sqrt{2}p$
and $q^{m}=q^{-}\equiv\sqrt{2}q$. After a gauge transformation, it
can be rewritten as
\begin{equation}
\bar{U}^{\left(2\right)}\left(p,q\right)=\bar{\lambda}_{\dot{a}}\bar{A}_{\dot{a}}\left(p\right)\bar{\lambda}_{\dot{b}}\bar{A}_{\dot{b}}\left(q\right).
\end{equation}
In this way, the $\theta$ expansion of $\bar{U}^{\left(2\right)}$
is easily obtained from the $SO\left(8\right)$ superfields discussed
in the previous section. The explicit construction for the vector
polarisations, for example, is given by
\begin{eqnarray}
\bar{U}_{i}\left(p\right)\bar{U}_{j}\left(q\right) & = & e^{-i\left(p+q\right)\sqrt{2}X_{L}^{+}}\left\{ \vphantom{\left(\frac{p^{2}q}{6!}\right)}\Lambda_{i}\Lambda_{j}-i\left(\frac{q}{3!}\right)\Lambda_{i}\theta_{jk}\Lambda_{k}-i\left(\frac{p}{3!}\right)\theta_{ik}\Lambda_{k}\Lambda_{j}\right.\nonumber \\
 &  & -\left(\frac{q^{2}}{5!}\right)\Lambda_{i}\theta_{jk}\theta_{kl}\Lambda_{l}-\left(\frac{p^{2}}{5!}\right)\theta_{ik}\theta_{kl}\Lambda_{l}\Lambda_{j}-\left(\frac{pq}{3!3!}\right)\theta_{ik}\Lambda_{k}\theta_{jl}\Lambda_{l}\nonumber \\
 &  & +i\left(\frac{q^{3}}{7!}\right)\Lambda_{i}\theta_{jk}\theta_{kl}\theta_{lm}\Lambda_{m}+i\left(\frac{p^{3}}{7!}\right)\theta_{ik}\theta_{kl}\theta_{lm}\Lambda_{m}\Lambda_{j}\nonumber \\
 &  & +\left.i\left(\frac{pq^{2}}{6!}\right)\theta_{ik}\Lambda_{k}\theta_{jl}\theta_{lm}\Lambda_{m}+i\left(\frac{p^{2}q}{6!}\right)\theta_{ik}\theta_{kl}\Lambda_{l}\theta_{jm}\Lambda_{m}\right\} .
\end{eqnarray}
This product is analysed in the appendix \ref{sec:computations},
equation \eqref{eq:UiUjappendix}. It can be cast in a very simple
form,
\begin{equation}
\bar{U}_{i}\left(p\right)\bar{U}_{j}\left(q\right)\approx i\,e^{-i\left(p+q\right)\sqrt{2}X_{L}^{+}}\left(\frac{p-q}{4}\right)\eta^{ij}\Lambda\bar{\Lambda}_{k}\left(\theta\sigma_{k}\bar{\theta}\right),\label{eq:UiUj}
\end{equation}
where the symbol $\approx$ means equal up to BRST-exact terms and
will be recurrent in the upcoming results. For $\left(p+q\right)\neq0$,
the right hand side can also be written as a BRST-exact expression:
\begin{equation}
e^{-i\left(p+q\right)\sqrt{2}X_{L}^{+}}\Lambda\bar{\Lambda}_{k}\left(\theta\sigma_{k}\bar{\theta}\right)=\left(\frac{-2}{p+q}\right)\left\{ Q,\left[\frac{\bar{\Lambda}}{p+q}+\frac{i}{2}\bar{\Lambda}_{k}\left(\theta\sigma_{k}\bar{\theta}\right)\right]e^{-i\left(p+q\right)\sqrt{2}X_{L}^{+}}\right\} .
\end{equation}
Therefore,
\begin{eqnarray}
\bar{U}_{i}\left(p\right)\bar{U}_{j}\left(q\right) & \approx & \begin{cases}
0 & \left(p+q\right)\neq0,\\
i\left(\frac{p}{2}\right)\eta^{ij}\Lambda\bar{\Lambda}_{k}\left(\theta\sigma_{k}\bar{\theta}\right) & \left(p+q\right)=0,
\end{cases}
\end{eqnarray}

When it comes to $\bar{U}_{i}\left(p\right)\bar{Y}_{\dot{a}}\left(q\right)$,
the procedure is very similar:
\begin{eqnarray}
\bar{U}_{i}\left(p\right)\bar{Y}_{\dot{a}}\left(q\right) & = & e^{-i\left(p+q\right)\sqrt{2}X_{L}^{+}}\left\{ \vphantom{\frac{p^{3}}{7!q}}\frac{i}{q}\Lambda_{i}\bar{\lambda}_{\dot{a}}-\frac{1}{2!}\Lambda_{i}\Lambda_{j}\left(\theta\sigma^{j}\right)_{\dot{a}}+\frac{p}{3!q}\theta_{ij}\Lambda_{j}\bar{\lambda}_{\dot{a}}\right.\nonumber \\
 &  & +\frac{iq}{4!}\Lambda_{i}\theta_{jk}\Lambda_{k}\left(\theta\sigma^{j}\right)_{\dot{a}}+\frac{ip}{2!3!}\theta_{ij}\Lambda_{j}\Lambda_{k}\left(\theta\sigma^{k}\right)_{\dot{a}}-\frac{ip^{2}}{5!q}\theta_{ij}\theta_{jk}\Lambda_{k}\bar{\lambda}_{\dot{a}}\nonumber \\
 &  & +\frac{q^{2}}{6!}\Lambda_{i}\theta_{jk}\theta_{kl}\Lambda_{l}\left(\theta\sigma^{j}\right)_{\dot{a}}+\frac{p^{2}}{2!5!}\theta_{ij}\theta_{jk}\Lambda_{k}\Lambda_{l}\left(\theta\sigma^{l}\right)_{\dot{a}}\nonumber \\
 &  & \left.+\frac{pq}{3!4!}\theta_{ij}\Lambda_{j}\theta_{kl}\Lambda_{l}\left(\theta\sigma^{k}\right)_{\dot{a}}-\frac{p^{3}}{7!q}\theta_{ij}\theta_{jk}\theta_{kl}\Lambda_{l}\bar{\lambda}_{\dot{a}}\right\} .
\end{eqnarray}
After the identification of the BRST-exact terms (also left to the
appendix, equation \eqref{eq:UiYaappendix}), the above expression
can be written as
\begin{equation}
\bar{U}_{i}\left(p\right)\bar{Y}_{\dot{a}}\left(q\right)\approx-\frac{1}{2!}e^{-i\left(p+q\right)\sqrt{2}X_{L}^{+}}\sigma_{a\dot{a}}^{i}\Lambda\left[\frac{1}{2}\bar{\Lambda}_{j}\left(\sigma^{j}\bar{\theta}\right)_{a}-\bar{\Lambda}\theta_{a}\right].\label{eq:UiYa}
\end{equation}
Observe that
\begin{equation}
e^{-ip\sqrt{2}X_{L}^{+}}\Lambda\left[\frac{1}{2}\bar{\Lambda}_{j}\left(\sigma^{j}\bar{\theta}\right)_{a}-\Lambda\bar{\Lambda}\theta_{a}\right]=-\frac{i}{p}\left[Q,e^{-ip\sqrt{2}X_{L}^{+}}\left(\frac{1}{2}\bar{\Lambda}_{j}\left(\sigma^{j}\bar{\theta}\right)_{a}-\Lambda\bar{\Lambda}\theta_{a}\right)\right],
\end{equation}
so when $(p+q)\neq0$, $\bar{U}_{i}\left(p\right)\bar{Y}_{\dot{a}}\left(q\right)$
is BRST-exact: 
\begin{eqnarray}
\bar{U}_{i}\left(p\right)\bar{Y}_{\dot{a}}\left(q\right) & \approx & \begin{cases}
0 & \left(p+q\right)\neq0,\\
\frac{1}{2!}\sigma_{a\dot{a}}^{i}\Lambda\left(\bar{\Lambda}\theta_{a}-\frac{1}{2}\bar{\Lambda}_{j}\left(\sigma^{j}\bar{\theta}\right)_{a}\right) & \left(p+q\right)=0.
\end{cases}
\end{eqnarray}

The same analysis can be done for the product $\bar{Y}_{\dot{a}}\left(p\right)\bar{Y}_{\dot{b}}\left(q\right)$,
so the conclusion is that $\bar{U}^{\left(2\right)}\left(p,q\right)$
is BRST-exact unless $\left(p^{m}+q^{m}\right)=0$. In that case it
is given in terms of the zero-momentum cohomology:
\begin{eqnarray}
\bar{U}^{\left(2\right)}\left(p,q\right) & \approx & \delta_{p+q}\left(\frac{1}{2}\right)\Lambda\left(\frac{1}{2}\bar{\Lambda}_{j}\left(\sigma^{j}\bar{\theta}\right)_{a}-\bar{\Lambda}\theta_{a}\right)\left\{ a_{i}\left(\sigma^{i}\bar{\chi}\right)_{a}-b_{i}\left(\sigma^{i}\bar{\xi}\right)_{a}\right\} \nonumber \\
 &  & +\delta_{p+q}\left(\frac{i}{2}\right)\Lambda\bar{\Lambda}_{i}\left(\theta\sigma_{i}\bar{\theta}\right)\left\{ pa_{j}b_{j}+i\bar{\xi}_{\dot{a}}\bar{\chi}_{\dot{a}}\right\} .
\end{eqnarray}
Here, $\left(a_{i},\bar{\xi}_{\dot{a}}\right)$ and $\left(b_{i},\bar{\chi}_{\dot{a}}\right)$
are the polarisations of $\bar{A}_{\dot{a}}\left(p\right)$ and
$\bar{A}_{\dot{a}}\left(q\right)$ respectively.

It is also possible to build an integrated version for $\bar{U}^{\left(2\right)}\left(p,q\right)$.
Denoting it by $\bar{V}^{\left(2\right)}\left(p,q\right)$, the
expression
\begin{eqnarray}
\bar{V}^{\left(2\right)}\left(p,q\right) & = & \ointctrclockwise\left\{ \left[\left(\Pi_{i}-i\sqrt{2}p\bar{N}_{i}\right)A_{i}\left(p\right)+\left(\partial\bar{\theta}_{\dot{a}}+ip\bar{d}_{\dot{a}}\right)\bar{A}_{\dot{a}}\left(p\right)\right]\bar{\lambda}_{\dot{b}}\bar{A}_{\dot{b}}\left(q\right)\right\} \nonumber \\
 &  & -\ointctrclockwise\left\{ \bar{\lambda}_{\dot{a}}\bar{A}_{\dot{a}}\left(p\right)\left[\left(\Pi_{i}-i\sqrt{2}q\bar{N}_{i}\right)A_{i}\left(q\right)+\left(\partial\bar{\theta}_{\dot{b}}+iq\bar{d}_{\dot{b}}\right)\bar{A}_{\dot{b}}\left(q\right)\right]\right\} \label{eq:integratedmasslesssquared}
\end{eqnarray}
is BRST-closed since
\begin{eqnarray}
\left[Q,\bar{V}^{\left(2\right)}\right] & = & \ointctrclockwise\left\{ \partial\left[\bar{\lambda}_{\dot{a}}\bar{A}_{\dot{a}}\left(p\right)\right]\bar{\lambda}_{\dot{b}}\bar{A}_{\dot{b}}\left(q\right)+\bar{\lambda}_{\dot{a}}\bar{A}_{\dot{a}}\left(p\right)\partial\left[\bar{\lambda}_{\dot{b}}\bar{A}_{\dot{b}}\left(q\right)\right]\right\} \nonumber \\
 & = & \ointctrclockwise\:\partial\bar{U}^{\left(2\right)}\left(p,q\right).
\end{eqnarray}

It turns out that the above construction is relevant for the extension
of the DDF operators to the ghost number two cohomology. For this
reason, it will be put in a more symmetrical way. Note that
\begin{equation}
\Lambda\bar{\Lambda}_{i}\left(\theta\sigma_{i}\bar{\theta}\right)=-\frac{1}{12}\theta_{ij}\Lambda_{i}\Lambda_{j}+\left\{ Q,\left[\frac{1}{2}\Lambda\left(\theta\sigma_{i}\bar{\theta}\right)\left(\theta\sigma_{i}\bar{\theta}\right)-\frac{1}{12}\theta_{ij}\left(\theta\sigma_{i}\bar{\theta}\right)\Lambda_{j}\right]\right\} .
\end{equation}
Therefore, after a gauge transformation
\begin{eqnarray}
\bar{U}^{\left(2\right)}\left(p,k\right) & \approx & \delta_{p+q}\left(\frac{1}{2}\right)\Lambda\left(\frac{1}{2}\bar{\Lambda}_{j}\left(\sigma^{j}\bar{\theta}\right)_{a}-\bar{\Lambda}\theta_{a}\right)\left\{ a_{i}\left(\sigma^{i}\bar{\chi}\right)_{a}-b_{i}\left(\sigma^{i}\bar{\xi}\right)_{a}\right\} \nonumber \\
 &  & -\delta_{p+q}\left(\frac{i}{24}\right)\theta_{ik}\Lambda_{i}\Lambda_{k}\left\{ pa_{j}b_{j}+i\bar{\xi}_{\dot{a}}\bar{\chi}_{\dot{a}}\right\} .
\end{eqnarray}
The integrated vertices associated to the right hand side of the above
equation will be denoted by $W_{a}^{*}\left(0\right)$ and $c_{0}^{+}$,
respectively, and are given by
\begin{eqnarray}
W_{a}^{*}\left(0\right) & \equiv & -\ointctrclockwise\left\{ \Lambda\left(d_{a}+\Pi_{i}\left(\sigma^{i}\bar{\theta}\right)_{a}+\sqrt{2}\partial X^{-}\theta_{a}-\frac{1}{2}\left(\bar{\theta}\sigma^{i}\partial\theta\right)\left(\sigma^{i}\bar{\theta}\right)_{a}\right)\right\} \nonumber \\
 &  & +\sqrt{2}\ointctrclockwise\left\{ \partial X^{+}\left(\frac{1}{2}\left(\sigma^{i}\bar{\theta}\right)_{a}\bar{\Lambda}_{i}+\bar{\Lambda}\theta_{a}\right)\right\} ,\\
c_{0}^{+} & \equiv & \frac{1}{6}\ointctrclockwise\left\{ \Pi_{i}\theta_{ij}\Lambda_{j}-\frac{3}{\sqrt{2}}\bar{N}_{i}\Lambda_{i}-\frac{3}{2}\left(\theta\sigma_{i}\bar{d}\right)\Lambda_{i}-\frac{1}{2}\left(\theta\sigma_{i}\partial\bar{\theta}\right)\theta_{ij}\Lambda_{j}\right\} ,\label{eq:c0+}
\end{eqnarray}
such that
\begin{eqnarray}
\bar{V}^{\left(2\right)}\left(p,q\right) & \approx & \frac{1}{2}\delta_{p+q}W_{a}^{*}\left(0\right)\left\{ a_{i}\left(\sigma^{i}\bar{\chi}\right)_{a}-b_{i}\left(\sigma^{i}\bar{\xi}\right)_{a}\right\} \nonumber \\
 &  & -i\delta_{p+q}c_{0}^{+}\left\{ pa_{j}b_{j}+i\bar{\xi}_{\dot{a}}\bar{\chi}_{\dot{a}}\right\} ,
\end{eqnarray}
and
\begin{eqnarray}
\left[Q,W_{a}^{*}\left(0\right)\right] & = & \ointctrclockwise\partial\left\{ \Lambda\left(\frac{1}{2}\bar{\Lambda}_{i}\left(\sigma^{i}\bar{\theta}\right)_{a}-\bar{\Lambda}\theta_{a}\right)\right\} ,\\
\left\{ Q,c_{0}^{+}\right\}  & = & \frac{1}{24}\ointctrclockwise\partial\left(\Lambda_{i}\Lambda_{k}\theta_{ik}\right).
\end{eqnarray}
The notation $W_{a}^{*}\left(0\right)$ will become clear soon. It
represents the zero-momentum limit of the antifield DDF operator. 

Clearly all the results derived here can be extended to the frame
where $P^{+}\neq0$. In this case, the analogous operators are defined
to be
\begin{eqnarray}
\bar{W}_{\dot{a}}^{*}\left(0\right) & \equiv & -\ointctrclockwise\left\{ \bar{\Lambda}\left(\bar{d}_{\dot{a}}+\Pi_{i}\left(\sigma^{i}\theta\right)_{\dot{a}}+\sqrt{2}\partial X^{+}\bar{\theta}_{\dot{a}}+\frac{1}{2}\left(\theta\sigma^{i}\partial\bar{\theta}\right)\left(\sigma^{i}\theta\right)_{\dot{a}}\right)\right\} \nonumber \\
 &  & +\ointctrclockwise\left\{ \sqrt{2}\partial X^{-}\left(\frac{1}{2}\left(\sigma^{i}\theta\right)_{\dot{a}}\Lambda_{i}+\Lambda\bar{\theta}_{\dot{a}}\right)\right\} ,\\
c_{0}^{-} & \equiv & \frac{1}{6}\ointctrclockwise\left\{ \Pi_{i}\bar{\theta}_{ij}\bar{\Lambda}_{j}-\frac{3}{\sqrt{2}}N_{i}\bar{\Lambda}_{i}+\frac{3}{2}\left(d\sigma_{i}\bar{\theta}\right)\bar{\Lambda}_{i}+\frac{1}{2}\left(\partial\theta\sigma_{i}\bar{\theta}\right)\bar{\theta}_{ij}\bar{\Lambda}_{j}\right\} .\label{eq:c0-}
\end{eqnarray}

Having a simple interpretation in terms of the integrated zero-momentum
vertices, the operators $c_{0}^{\pm}$ will be shown to play a similar
role to the zero mode of the bosonic string $c$ ghost, \emph{cf.}
equation \eqref{eq:bosonicC0} of the appendix. In order to understand
this relation, the BRST cohomology at ghost number two has to be further
discussed.

\subsection{Extended DDF construction}

\

Given the superfield $A_{\alpha\beta}$ of \eqref{eq:antifieldeom},
it might be possible to find a gauge transformation similar to what
was done for the physical states in the DDF description, where the
dependence on half of the $\theta$'s was removed, leaving only the
physical polarisations in a particular frame.

Instead of following this procedure and determining a convenient choice
for the gauge parameters in \eqref{eq:gaugeantifields}, a more direct
approach will be considered with the action of the operators $c_{0}^{\pm}$
on the gauge fixed massless states $\bar{\lambda}_{\dot{a}}\bar{A}_{\dot{a}}$,
\emph{cf.} equation \eqref{eq:spinorsuperfield}. Since they have
ghost number one and $\left\{ Q,c_{0}^{\pm}\right\} =0$, the resulting
operator shall have ghost number two and be BRST-closed. It is straightforward
to compute the anticommutators to obtain:
\begin{eqnarray}
\left\{ c_{0}^{+},\bar{\lambda}_{\dot{a}}\bar{A}_{\dot{a}}\left(k\right)\right\}  & = & -\frac{1}{2}\Lambda\bar{\lambda}_{\dot{a}}\bar{A}_{\dot{a}},\\
\left\{ c_{0}^{-},\bar{\lambda}_{\dot{a}}\bar{A}_{\dot{a}}\left(k\right)\right\}  & = & \bar{\Lambda}\bar{\Lambda}_{i}A_{i}\nonumber \\
 & = & \left\{ Q,\bar{\Lambda}\left(\bar{\theta}_{\dot{a}}\bar{A}_{\dot{a}}\right)\right\} +\bar{\Lambda}\left(\bar{\lambda}_{\dot{a}}\bar{A}_{\dot{a}}\right)\nonumber \\
 & \approx & \bar{\Lambda}\left(\bar{\lambda}_{\dot{a}}\bar{A}_{\dot{a}}\right).\label{eq:c0overmassless}
\end{eqnarray}

The first operator, $\Lambda\bar{\lambda}_{\dot{a}}\bar{A}_{\dot{a}}$,
can be easily shown to be BRST-exact (see appendix \ref{sec:computations}).
However this is not the case for $\bar{\Lambda}\left(\bar{\lambda}_{\dot{a}}\bar{A}_{\dot{a}}\right)$,
which is the analogous DDF gauge fixed operator for the massless ghost
number two cohomology. For completeness, observe that the action of
the operators $W_{a}^{*}\left(0\right)$ and $\bar{W}_{\dot{a}}^{*}\left(0\right)$
is completely neglectable, as they satisfy
\begin{eqnarray}
\{Q_{a},c_{0}^{+}\} & \approx & W_{a}^{*}\left(0\right),\\
\{\bar{Q}_{\dot{a}},c_{0}^{-}\} & \approx & \bar{W}_{\dot{a}}^{*}\left(0\right).
\end{eqnarray}
The Jacobi identity in its turn implies that both $[W_{a}^{*}\left(0\right),\bar{\lambda}_{\dot{a}}\bar{A}_{\dot{a}}\left(k\right)]$
and $[\bar{W}_{\dot{a}}^{*}\left(0\right),\bar{\lambda}_{\dot{c}}\bar{A}_{\dot{c}}\left(k\right)]$
are BRST-exact.

Defining
\begin{equation}
\bar{\Lambda}\left(\bar{\lambda}_{\dot{a}}\bar{A}_{\dot{a}}\right)\equiv a_{i}^{*}\bar{U}_{i}^{*}+\bar{\xi}_{\dot{a}}^{*}\bar{Y}_{\dot{a}}^{*},\label{eq:antifieldpolarization}
\end{equation}
the superfield expansion for each polarisation is given by\begin{subequations}\label{eq:SO8antifields}
\begin{eqnarray}
\bar{U}_{i}^{*}\left(z;k\right) & \equiv & \bar{\Lambda}\left\{ \Lambda_{i}-\frac{ik}{3!}\theta_{ij}\Lambda_{j}-\frac{k^{2}}{5!}\theta_{ij}\theta_{jk}\Lambda_{k}+\frac{ik^{3}}{7!}\theta_{ij}\theta_{jk}\theta_{kl}\Lambda_{l}\right\} e^{-ik\sqrt{2}X_{L}^{+}},\\
\bar{Y}_{\dot{a}}^{*}\left(z;k\right) & \equiv & \bar{\Lambda}\left(\theta\sigma_{i}\right)_{\dot{a}}\left\{ \frac{1}{2!}\Lambda_{i}-\frac{ik}{4!}\theta_{ij}\Lambda_{j}-\frac{k^{2}}{6!}\theta_{ij}\theta_{jk}\Lambda_{k}+\frac{ik^{3}}{8!}\theta_{ij}\theta_{jk}\theta_{kl}\Lambda_{l}\right\} e^{-ik\sqrt{2}X_{L}^{+}}\nonumber \\
 &  & +\bar{\Lambda}\left(\frac{i}{k}\right)\bar{\lambda}_{\dot{a}}e^{-ik\sqrt{2}X_{L}^{+}}.
\end{eqnarray}
\end{subequations}

It is worth to take a look at the statistics of the polarisations.
Now the $SO\left(8\right)$ vector polarisation, denoted by $a_{i}^{*}$,
should have fermionic statistics, while the polarisation $\bar{\xi}_{\dot{a}}^{*}$
is an $SO\left(8\right)$ bosonic spinor. In the following, however,
\emph{the physical polarisation statistics will be kept} (bosonic
vector and fermionic spinor), as the known properties of the superfields
$A_{i}$ and $\bar{A}_{\dot{a}}$ will be used constantly, avoiding
possible misunderstandings.

Maybe the best way to convince oneself that the vertices \eqref{eq:SO8antifields}
truly describe the antifields is to show that they are dual to the
physical massless states. Consider the two-point amplitude
\begin{equation}
f\left[(a_{i},\bar{\xi}_{\dot{a}}),(a_{i}^{*},\bar{\xi}_{\dot{a}}^{*}),k,p\right]=\left\langle \bar{\lambda}_{\dot{a}}\bar{A}_{\dot{a}}\left(k\right)\cdot\bar{\Lambda}\,\bar{\lambda}_{\dot{b}}\bar{A}_{\dot{b}}\left(p\right)\right\rangle ,
\end{equation}
where $\bar{\lambda}_{\dot{a}}\bar{A}_{\dot{a}}\left(k\right)$
describes the fields with polarisations $(a_{i},\bar{\xi}_{\dot{a}})$
and $\bar{\Lambda}\,\bar{\lambda}_{\dot{b}}\bar{A}_{\dot{b}}\left(p\right)$
the proposed antifield vertex with polarisations $(a_{i}^{*},\bar{\xi}_{\dot{a}}^{*})$.
According to the analysis of subsection \ref{sub:Cohomology-ring},
the product $U^{\left(2\right)}\left(k,p\right)=\bar{\lambda}_{\dot{a}}\bar{A}_{\dot{a}}\left(k\right)\cdot\bar{\lambda}_{\dot{b}}\bar{A}_{\dot{b}}\left(p\right)$
is BRST-exact unless $k+p=0$, that is
\begin{eqnarray}
U^{\left(2\right)}\left(k,p\right) & = & \delta_{k+p}\left(\frac{1}{2}\right)\Lambda\left(\frac{1}{2}\bar{\Lambda}_{j}\left(\sigma^{j}\bar{\theta}\right)_{a}-\bar{\Lambda}\theta_{a}\right)\left(a_{i}(\sigma^{i}\bar{\xi}^{*})_{a}-a_{i}^{*}(\sigma^{i}\bar{\xi})_{a}\right)\nonumber \\
 &  & -\delta_{k+p}\left(\frac{i}{24}\right)\theta_{ik}\Lambda_{i}\Lambda_{k}\left(ka_{j}a_{j}^{*}+i\bar{\xi}_{\dot{a}}\bar{\xi}_{\dot{a}}^{*}\right)+\left\{ Q,Z^{\left(1\right)}\right\} ,
\end{eqnarray}
where $Z^{\left(1\right)}$ is a ghost number one $SO\left(8\right)$
superfield unimportant to the present analysis, as it decouples from
the amplitude computation:
\begin{eqnarray}
f\left[(a_{i},\bar{\xi}_{\dot{a}}),(a_{i}^{*},\bar{\xi}_{\dot{a}}^{*}),k,p\right] & = & -\left\langle \bar{\Lambda}U^{\left(2\right)}\left(k,p\right)\right\rangle \nonumber \\
 & = & \delta_{k+p}\left(\frac{1}{2}\right)\left\langle -\bar{\Lambda}\Lambda\left(\frac{1}{2}\bar{\Lambda}_{j}\left(\sigma^{j}\bar{\theta}\right)_{a}+\bar{\Lambda}\theta_{a}\right)\right\rangle \left(a_{i}(\sigma^{i}\bar{\xi}^{*})_{a}-a_{i}^{*}(\sigma^{i}\bar{\xi})_{a}\right)\nonumber \\
 &  & +\delta_{k+p}\left(\frac{i}{24}\right)\left(ka_{j}a_{j}^{*}+i\bar{\xi}_{\dot{a}}\bar{\xi}_{\dot{a}}^{*}\right)\left\langle \bar{\Lambda}\theta_{ik}\Lambda_{i}\Lambda_{k}\right\rangle +\left\langle \left\{ Q,\bar{\Lambda}Z^{\left(1\right)}\right\} \right\rangle \nonumber \\
 & = & \delta_{k+p}\left(\frac{i}{24}\right)\left(ka_{j}a_{j}^{*}+i\bar{\xi}_{\dot{a}}\bar{\xi}_{\dot{a}}^{*}\right)\left\langle \bar{\Lambda}\theta_{ik}\Lambda_{i}\Lambda_{k}\right\rangle .
\end{eqnarray}
Observe that
\begin{equation}
\bar{\Lambda}\Lambda\left(-\frac{1}{2}\bar{\Lambda}_{j}\left(\sigma^{j}\bar{\theta}\right)_{a}+\bar{\Lambda}\theta_{a}\right)=\frac{1}{2}\left[Q,\bar{\Lambda}_{i}\left(\theta\sigma^{i}\bar{\theta}\right)\left(-\frac{1}{2}\bar{\Lambda}_{j}\left(\sigma^{j}\bar{\theta}\right)_{a}+\bar{\Lambda}\theta_{a}\right)\right],
\end{equation}
explaining the vanishing of the mixed polarisations term $\left(a_{i}(\sigma^{i}\bar{\xi}^{*})_{a}-a_{i}^{*}(\sigma^{i}\bar{\xi})_{a}\right)$. 

Now it remains to show that $\bar{\Lambda}\theta_{ik}\Lambda_{i}\Lambda_{k}$
is proportional to the pure spinor integration measure, given by
\[
\left\langle \left(\lambda\gamma^{m}\theta\right)\left(\lambda\gamma^{n}\theta\right)\left(\lambda\gamma^{p}\theta\right)\left(\theta\gamma_{mnp}\theta\right)\right\rangle .
\]
This result is discussed in the appendix, after equation \eqref{eq:SO8measure},
and it can be shown that:
\begin{eqnarray}
\bar{\Lambda}\Lambda_{i}\Lambda_{j}\theta_{ji} & = & \frac{1}{60}\left(\lambda\gamma^{m}\theta\right)\left(\lambda\gamma^{n}\theta\right)\left(\lambda\gamma^{p}\theta\right)\left(\theta\gamma_{mnp}\theta\right)\nonumber \\
 &  & -\frac{1}{6}\left[Q,\Lambda\bar{\Lambda}_{i}\left(\theta\sigma_{j}\bar{\theta}\right)\bar{\theta}_{ji}\right]\nonumber \\
 &  & -\frac{3}{4}\left[Q,\Lambda\bar{\Lambda}\theta_{ij}\bar{\theta}_{ji}\right]+\frac{5}{6}\left[Q,\bar{\Lambda}\Lambda_{i}\left(\bar{\theta}\sigma_{j}\theta\right)\theta_{ji}\right].\label{eq:so8measure}
\end{eqnarray}
Therefore
\begin{equation}
f\left[(a_{i},\bar{\xi}_{\dot{a}}),(a_{i}^{*},\bar{\xi}_{\dot{a}}^{*}),k,p\right]\propto\delta_{k+p}\left(ka_{j}a_{j}^{*}+i\bar{\xi}_{\dot{a}}\bar{\xi}_{\dot{a}}^{*}\right),
\end{equation}
as expected from the field-antifield 2-point amplitude. This shows
that $\bar{\Lambda}\,\bar{\lambda}_{\dot{b}}\bar{A}_{\dot{b}}\left(p\right)$
is indeed a proper antifield vertex.

From the above construction, the duality between fields and antifields
is explicit. The supersymmetry transformations are still very simple
but with an extra BRST-exact ingredient:\begin{subequations}\label{eq:SUSYantifields}
\begin{eqnarray}
[\bar{Q}_{\dot{a}},\bar{U}_{i}^{*}] & = & [Q,\bar{\theta}_{\dot{a}}\bar{U}_{i}]\\
 & \approx & 0,\nonumber \\
\{\bar{Q}_{\dot{a}},\bar{Y}_{\dot{b}}^{*}\} & = & \{Q,\bar{\theta}_{\dot{a}}\bar{Y}_{\dot{b}}\}\\
 & \approx & 0,\nonumber \\
{}[Q_{a},\bar{U}_{i}^{*}] & = & ik\sigma_{a\dot{a}}^{i}\bar{Y}_{\dot{a}},\\
\{Q_{a},\bar{Y}_{\dot{a}}^{*}\} & = & -\sigma_{a\dot{a}}^{i}\bar{U}_{i}.
\end{eqnarray}
\end{subequations}

The integrated vertex associated to \eqref{eq:SO8antifields}, denoted
by $V_{\textrm{L.C.}}^{*}$, can be easily guessed by observing the
role of the operator $\bar{\Lambda}$. Note that
\begin{equation}
\left\{ Q,\bar{\Lambda}\left[\left(\Pi_{i}-i\sqrt{2}k\bar{N}_{i}\right)A_{i}+\left(\partial\bar{\theta}_{\dot{a}}+ik\bar{d}_{\dot{a}}\right)\bar{A}_{\dot{a}}\right]\right\} =-\bar{\Lambda}\partial\left(\bar{\lambda}_{\dot{a}}\bar{A}_{\dot{a}}\right).
\end{equation}
Knowing that $\partial\bar{\Lambda}=-\sqrt{2}\left[Q,\partial X^{-}\right]$,
the obvious proposal for $V_{\textrm{L.C.}}^{*}$ is
\begin{equation}
V_{\textrm{L.C.}}^{*}\left(k;a_{i}^{*},\bar{\xi}_{\dot{a}}^{*}\right)=\ointctrclockwise\left\{ \bar{\Lambda}\left(\Pi_{i}-i\sqrt{2}k\bar{N}_{i}\right)A_{i}+\bar{\Lambda}\left(\partial\bar{\theta}_{\dot{a}}+ik\bar{d}_{\dot{a}}\right)\bar{A}_{\dot{a}}+\sqrt{2}\partial X^{-}\bar{\lambda}_{\dot{a}}\bar{A}_{\dot{a}}\right\} ,\label{eq:SO8integratedsantifields}
\end{equation}
which is BRST-closed by construction:
\begin{eqnarray}
\left\{ Q,V_{\textrm{L.C.}}^{*}\right\}  & = & -\ointctrclockwise\,\partial\left(\bar{\Lambda}\,\bar{\lambda}_{\dot{a}}\bar{A}_{\dot{a}}\right)\nonumber \\
 & = & 0,
\end{eqnarray}
It is important to note here that $V_{\textrm{L.C.}}^{*}$ has to
be appropriately ordered, since it contains products of operators
that diverge when approach each other, \emph{e.g.} $\partial X^{-}$
and $\bar{A}_{\dot{a}}$. The prescription used here is the usual
normal ordering where $:A\left(z\right)B\left(y\right):$ means the
absence of contractions between two generic operators $A$ and $B$.
From now on, this will be implicit in order to leave the notation
simpler.

As a consistency check, it is possible to show that
\begin{eqnarray}
\left[c_{0}^{+},V_{\textrm{L.C.}}\left(k;a_{i},\bar{\xi}_{\dot{a}}\right)\right] & \approx & +\oint\left\{ \frac{\Lambda}{2}\left[\Pi_{i}A_{i}-i\sqrt{2}k\bar{N}_{i}A_{i}+\partial\bar{\theta}_{\dot{a}}\bar{A}_{\dot{a}}+ik\bar{d}_{\dot{a}}\bar{A}_{\dot{a}}\right]\right\} \nonumber \\
 &  & +\frac{\sqrt{2}}{2}\oint\left\{ \partial X^{+}\left(\bar{\lambda}_{\dot{a}}\bar{A}_{\dot{a}}\right)\right\} \label{eq:cV+}\\
\left[c_{0}^{-},V_{\textrm{L.C.}}\left(k;a_{i},\bar{\xi}_{\dot{a}}\right)\right] & \approx & -\,V_{\textrm{L.C.}}^{*}\left(k;a_{i},\bar{\xi}_{\dot{a}}\right)\label{eq:cV-}
\end{eqnarray}
The vertex \eqref{eq:cV+} is the integrated form of $\Lambda\left(\bar{\lambda}_{\dot{a}}\bar{A}_{\dot{a}}\right)$,
which is BRST-exact. And the vertex \eqref{eq:cV-} agrees with the
proposed one in \eqref{eq:SO8integratedsantifields}, up to gauge
transformations. 

The natural step now is to understand the algebra of $V_{\textrm{L.C.}}^{*}$.
It is clear that it does not constitute a creation-annihilation algebra
as the operators are now charged under the ghost number current. On
the other hand, it gives rise to some interesting features.

Similarly to \eqref{eq:creation-annihilationFULL}, there might be
some subtleties when determining the algebra for $\left(k+p\right)=0$.
The details are discussed in the the appendix, equation \eqref{eq:VV*appendix}.
Computing the commutator between $V_{\textrm{L.C.}}$ and $V_{\textrm{L.C.}}^{*}$,
one obtains
\begin{eqnarray}
\left[V_{\textrm{L.C.}}(k;a_{i},\bar{\xi}_{\dot{a}}),V_{\textrm{L.C.}}^{*}(p;a_{i}^{*},\bar{\xi}_{\dot{a}}^{*})\right] & \approx & -ik\delta_{p+k}\left\{ a_{i}(\sigma^{i}\bar{\xi}^{*})_{a}-a_{i}^{*}(\sigma^{i}\bar{\xi})_{a}\right\} W_{a}^{*}\left(0\right)\nonumber \\
 &  & +2k\delta_{p+k}\left\{ ka_{j}a_{j}^{*}+i\bar{\xi}_{\dot{a}}\bar{\xi}_{\dot{a}}^{*}\right\} c_{0}^{+},\label{eq:VV*algebra}
\end{eqnarray}
and the operators $c_{0}^{+}$ and $W_{a}^{*}\left(0\right)$ naturally
appear in the extension of the algebra.

Going further and analysing the anticommutator of $V_{\textrm{L.C.}}^{*}$
with itself, a similar result is found (equation \eqref{eq:V*V*appendix}
of the appendix). The anticommutator assumes an elegant form,
\begin{equation}
\left\{ V_{\textrm{L.C.}}^{*}(k;a_{i}^{*},\bar{\xi}_{\dot{a}}^{*}),V_{\textrm{L.C.}}^{*}(p;b_{i}^{*},\bar{\chi}_{\dot{a}}^{*})\right\} \approx-4k\delta_{k+p}\left\{ ka_{j}^{*}b_{j}^{*}+i\bar{\xi}_{\dot{a}}^{*}\bar{\chi}_{\dot{a}}^{*}\right\} M,\label{eq:V*V*algebra}
\end{equation}
where
\begin{eqnarray}
M & \equiv & \frac{1}{6}\ointctrclockwise\left\{ \bar{\Lambda}\left(\Pi_{i}\theta_{ij}\Lambda_{j}-\frac{1}{2}\left(\theta\sigma_{i}\partial\bar{\theta}\right)\theta_{ij}\Lambda_{j}\right)+\frac{\sqrt{2}}{4}\partial X^{-}\theta_{ij}\Lambda_{i}\Lambda_{j}\right\} \nonumber \\
 &  & -\frac{1}{4}\ointctrclockwise\left\{ \bar{\Lambda}\left(\sqrt{2}\Lambda_{i}\bar{N}_{i}+\Lambda_{i}\left(\theta\sigma_{i}\bar{d}\right)\right)\right\} \label{eq:integratedmeasure}
\end{eqnarray}
is the integrated version of the pure spinor integration measure.
Note that
\begin{equation}
\left[Q,M\right]=\frac{1}{24}\ointctrclockwise\partial\left(\bar{\Lambda}\Lambda_{i}\Lambda_{j}\theta_{ji}\right),
\end{equation}
which is in accordance with the measure displayed in \eqref{eq:so8measure}.
All these results have a clear analogous in the bosonic string, \emph{cf.}
equations \ref{eq:VV*bosonicalgebra} and \ref{eq:V*V*bosonicalgebra}
of the appendix.

\subsection{The antifield spectrum}

\

Before generalising the construction of the spectrum to the antifields,
it is worth to understand better the properties of the massless vertices
introduced above.

First of all, this sector has an analogous property to \eqref{eq:masslessfromzeromomentum},
meaning that the massless unintegrated vertices of \eqref{eq:SO8antifields}
can be obtained from the action \eqref{eq:SO8integratedsantifields}
on the zero-momentum state $\bar{\Lambda}$:
\begin{equation}
\left[V_{\textrm{L.C.}}^{*}(k;a_{i}^{*},\bar{\xi}_{\dot{a}}^{*}),-\frac{\bar{\Lambda}}{2}\right]=ik\bar{\Lambda}\,\bar{\lambda}_{\dot{a}}\bar{A}_{\dot{a}}+\frac{ik}{2}\left[Q,\bar{\Lambda}\,\bar{\theta}_{\dot{a}}\bar{A}_{\dot{a}}\right].
\end{equation}

The zero-momentum limits of $\bar{\Lambda}\,\bar{\lambda}_{\dot{a}}\bar{A}_{\dot{a}}$
and $V_{\textrm{L.C.}}^{*}(k;a_{i}^{*},\bar{\xi}_{\dot{a}}^{*})$
present a subtlety for the $SO\left(8\right)$ vector polarisation.
Observe that
\begin{eqnarray}
\lim_{k\to0}\bar{U}_{i}^{*}\left(k\right) & = & \bar{\Lambda}\Lambda_{i}\nonumber \\
 & = & \frac{1}{2}\{Q,\bar{\theta}_{ij}\Lambda_{j}\}
\end{eqnarray}
is BRST-exact. A closer look clarifies this issue and the expected
zero-momentum state lies in fact at the next order in $k$. The easiest
way to solve this issue is to rescale the polarisation $a_{i}^{*}\to k^{-1}a_{i}^{*}$,
so that
\begin{eqnarray}
\lim_{k\to0}\bar{\Lambda}\,\bar{\lambda}_{\dot{a}}\bar{A}_{\dot{a}}(k;\frac{a_{i}^{*}}{k},\bar{\xi}_{\dot{a}}^{*}) & = & -\frac{i}{3!}a_{i}^{*}\left(3\Lambda\bar{\theta}_{ij}\Lambda_{j}+\bar{\Lambda}\theta_{ij}\Lambda_{j}\right)\nonumber \\
 &  & +\bar{\xi}_{\dot{a}}^{*}\bar{\Lambda}\left(-\frac{1}{2!}\Lambda_{i}\left(\theta\sigma_{i}\right)_{\dot{a}}+\Lambda\bar{\theta}_{\dot{a}}\right)\nonumber \\
 &  & +\left\{ Q,\lim_{k\to0}\left(\frac{1}{2k}\bar{\theta}_{ij}\Lambda_{j}e^{-ik\sqrt{2}X_{L}^{+}}\right)\right\} \nonumber \\
 &  & -\left\{ Q,\lim_{k\to0}\left(\frac{i}{k}\bar{\Lambda}\bar{\theta}_{\dot{a}}e^{-ik\sqrt{2}X_{L}^{+}}\right)\right\} .\label{eq:unantizeromomentum}
\end{eqnarray}
Note that this is just a feature of the procedure used here to determine
the antifield vertices, \emph{i.e.} the action of the $c_{0}^{-}$
operator, and does not mean that the vertices obtained in this way
are ill defined\footnote{\label{fn:extramomentum}In fact, this can be understood in the Lorentz
group analysis. From the light-cone point of view, there is a $U\left(1\right)$
charge associated to the Lorentz generator $L^{+-}$. For example,
the massless vertices $\bar{U}_{i}$ and $\bar{Y}_{\dot{a}}$
have charge $0$ and $\frac{1}{2}$, respectively. On the other hand,
the associated antifields $\bar{U}_{i}^{*}$ and $\bar{Y}_{\dot{a}}^{*}$
have charge $-1$ and $-\frac{1}{2}$. This is so because the operator
$c_{0}^{-}$ has charge $-1$ with respect to $L^{+-}$. While this
works well for the pair $\{\bar{Y}_{\dot{a}},\bar{Y}_{\dot{a}}^{*}\}$
(since the spinor field and antifield have opposite charge), the vector
pair $\{\bar{U}_{i},\bar{U}_{i}^{*}\}$ develop this asymmetry
under the action of $c_{0}^{-}$. This is the origin of the odd limit
discussed above.}.

The same analysis applies to the zero-momentum limit of $V_{\textrm{L.C.}}^{*}$.
Defining
\[
V_{\textrm{L.C.}}^{*}(k;\frac{a_{i}^{*}}{k},\bar{\xi}_{\dot{a}}^{*})\equiv a_{i}^{*}\bar{V}_{i}^{*}\left(k\right)+\bar{\xi}_{\dot{a}}^{*}\bar{W}_{\dot{a}}^{*}\left(k\right),
\]
it can be shown that
\begin{eqnarray}
\lim_{k\to0}V_{\textrm{L.C.}}^{*}(k;\frac{a_{i}^{*}}{k},\bar{\xi}_{\dot{a}}^{*}) & = & a_{i}^{*}\bar{V}_{i}^{*}\left(0\right)+\bar{\xi}_{\dot{a}}^{*}\bar{W}_{\dot{a}}^{*}\left(0\right)\nonumber \\
 &  & +\left(i\bar{\xi}_{\dot{a}}^{*}\sqrt{2}\right)\lim_{k\to0}\left\{ Q,\frac{1}{k}\ointctrclockwise\left(\partial X^{-}\bar{\theta}_{\dot{a}}e^{-ik\sqrt{2}X_{L}^{+}}\right)\right\} .\nonumber \\
 &  & +a_{i}^{*}\lim_{k\to0}\left[Q,\frac{1}{k}\ointctrclockwise\left\{ \left(\bar{\theta}\sigma^{i}d\right)+\frac{1}{2}\bar{\theta}_{ij}\Pi_{j}+\sqrt{2}N^{i}\right\} e^{-ik\sqrt{2}X^{+}}\right]\nonumber \\
 &  & +\left(\frac{1}{2}a_{i}^{*}\right)\lim_{k\to0}\left[Q,\frac{1}{k}\ointctrclockwise\left\{ \partial\bar{\theta}_{ij}\left(\theta\sigma^{j}\bar{\theta}\right)+\bar{\theta}_{ij}\left(\partial\theta\sigma^{j}\bar{\theta}\right)\right\} e^{-ik\sqrt{2}X^{+}}\right]\nonumber \\
 &  & -\left(\sqrt{2}a_{i}^{*}\right)\lim_{k\to0}\left[Q,\frac{1}{k}\ointctrclockwise\left\{ \Pi^{-}\left(\theta\sigma^{i}\bar{\theta}\right)\right\} e^{-ik\sqrt{2}X^{+}}\right],\label{eq:inantizeromomentum}
\end{eqnarray}
where\begin{subequations}
\begin{eqnarray}
\bar{V}_{i}^{*}\left(0\right) & = & -i\ointctrclockwise\bar{\Lambda}\left\{ \left(\theta\sigma^{i}\bar{d}\right)+\frac{1}{2}\theta_{ij}\Pi_{j}+\sqrt{2}\,\bar{N}^{i}-\frac{1}{3!}\theta_{ij}\left(\theta\sigma^{j}\partial\bar{\theta}\right)\right\} \nonumber \\
 &  & -i\ointctrclockwise\Lambda\left\{ \left(\bar{\theta}\sigma^{i}d\right)+\frac{1}{2}\bar{\theta}_{ij}\Pi_{j}+\sqrt{2}N^{i}+\frac{1}{2}\bar{\theta}_{ij}\left(\partial\theta\sigma^{j}\bar{\theta}\right)\right\} \nonumber \\
 &  & -i\ointctrclockwise\Lambda\left\{ \frac{1}{2}\partial\bar{\theta}_{ij}\left(\theta\sigma^{j}\bar{\theta}\right)-\sqrt{2}\Pi^{-}\left(\theta\sigma^{i}\bar{\theta}\right)\right\} \nonumber \\
 &  & -i\sqrt{2}\ointctrclockwise\left\{ \partial X^{+}\left(\bar{\Lambda}_{ij}\left(\theta\sigma^{j}\bar{\theta}\right)+\frac{1}{2}\bar{\theta}_{ij}\bar{\Lambda}_{j}\right)+\frac{1}{3!}\partial X^{-}\theta_{ij}\Lambda_{j}\right\} \\
\bar{W}_{\dot{a}}^{*}\left(0\right) & = & -\ointctrclockwise\left\{ \bar{\Lambda}\left(\bar{d}_{\dot{a}}+\Pi_{i}\left(\sigma^{i}\theta\right)_{\dot{a}}+\sqrt{2}\partial X^{+}\bar{\theta}_{\dot{a}}+\frac{1}{2}\left(\theta\sigma^{i}\partial\bar{\theta}\right)\left(\sigma^{i}\theta\right)_{\dot{a}}\right)\right\} \nonumber \\
 &  & +\ointctrclockwise\left\{ \sqrt{2}\partial X^{-}\left(\frac{1}{2}\left(\sigma^{i}\theta\right)_{\dot{a}}\Lambda_{i}+\Lambda\bar{\theta}_{\dot{a}}\right)\right\} 
\end{eqnarray}
\end{subequations}The existence of singular terms in the zero-momentum
limit of \eqref{eq:unantizeromomentum} and \eqref{eq:inantizeromomentum}
is due to a singular gauge choice and could be of course removed by
a gauge transformation. 

Concerning the antifield spectrum, one starts defining the ground
state similarly to the ghost number one case, 
\begin{eqnarray}
\left|0,k\right\rangle ^{*} & = & \lim_{z\to0}\Lambda\lambda^{a}A_{a}\left(k\right)\left|0\right\rangle \nonumber \\
 & \equiv & a_{i}\left|i,k\right\rangle ^{*}+\xi_{a}\left|a,k\right\rangle ^{*},\label{eq:antifieldground}
\end{eqnarray}
\emph{cf.} equation \eqref{eq:P+spinorsuperfields}, such that
\begin{equation}
\Lambda\lambda^{a}A_{a}\left(k\right)\approx\left\{ c_{0}^{+},\lambda_{a}A_{a}\left(k\right)\right\} .
\end{equation}
The excited states are built by the action of the creation operators
$V_{\textrm{L.C.}}$ of \eqref{eq:SO8integrated} on \eqref{eq:antifieldground}.
This is exactly the same as presented in \cite{Jusinskas:2014vqa}
for the physical states, the only difference being the ground state,
which has now ghost number two.

In this way, each physical state has a correspondent antifield. This
map can be made more precise with the action of the $c_{0}^{+}$ operator.
Since $\left[c_{0}^{+},V_{\textrm{L.C.}}\right]\approx0$, given any
DDF state in the physical spectrum ($P^{-}\neq0$) of the form
\begin{equation}
\left|\psi\right\rangle =\prod_{k}\sum_{n}C{}_{k,n}V_{\textrm{L.C.}}^{n}\left(k\right)\left(a_{i}\left|i\right\rangle +\xi_{a}\left|a\right\rangle \right),\label{eq:DDF=0000231}
\end{equation}
one can define the antifield by
\begin{eqnarray}
\left|\psi\right\rangle ^{*} & \equiv & c_{0}^{+}\left|\psi\right\rangle \nonumber \\
 & \approx & \prod_{k}\sum_{n}C{}_{k,n}V_{\textrm{L.C.}}^{n}\left(k\right)\left(a_{i}\left|i\right\rangle ^{*}+\xi_{a}\left|a\right\rangle ^{*}\right).\label{eq:DDF=0000232}
\end{eqnarray}
It should be kept in mind that the $V_{\textrm{L.C.}}$'s in the above
construction have independent polarisations among each other, in such
a way that any element of the cohomology can be described by either
\eqref{eq:DDF=0000231} or \eqref{eq:DDF=0000232} for a given set
of polarisations $\left\{ a_{i}^{1},a_{i}^{2},\ldots,\bar{\xi}_{\dot{a}}^{1},\bar{\xi}_{\dot{a}}^{2},\ldots,\right\} $,
up to Lorentz transformations and gauge transformations.

Observe that both $\left|\psi\right\rangle $ and $\left|\psi\right\rangle ^{*}$
depend only on half of the $\lambda^{\alpha}$ components. The components
$\bar{\lambda}_{\dot{a}}$ clearly decouple from the ground states
\eqref{eq:fieldground} and \eqref{eq:antifieldground} while the
ghost contributions from the creation operators are all encoded in
$\bar{N}_{i}=-\frac{1}{\sqrt{2}}(\lambda\sigma_{i}\bar{\omega})$.
If the frame $P^{+}\neq0$ is chosen for the ground state instead,
the DDF spectrum will depend only on $\bar{\lambda}_{\dot{a}}$.

Another interesting property of the unintegrated vertices $\bar{U}_{i}^{*}$
and $\bar{Y}_{\dot{a}}^{*}$ is that they can be written as\emph{
singular} BRST-exact states, in a direct analogy with \eqref{eq:singgaugebos}
in the bosonic string. The key ingredient here is
\begin{equation}
\bar{\Lambda}e^{-i\epsilon\sqrt{2}X_{L}^{-}}=-\frac{i}{\epsilon}\left[Q,e^{-i\epsilon\sqrt{2}X_{L}^{-}}\right],
\end{equation}
so that any infinitesimal massive deformation of the form $:\bar{\Lambda}\,\bar{\lambda}_{\dot{a}}\bar{A}_{\dot{a}}\left(k\right)e^{-i\epsilon\sqrt{2}X_{L}^{-}}:$
is a BRST-exact state:
\begin{equation}
:\left(\bar{\Lambda}\,\bar{\lambda}_{\dot{a}}\bar{A}_{\dot{a}}\left(k\right)e^{-i\epsilon\sqrt{2}X_{L}^{-}}\right):=-\frac{i}{\epsilon}\left[Q,:\left(\bar{\lambda}_{\dot{a}}\bar{A}_{\dot{a}}\left(k\right)e^{-i\epsilon\sqrt{2}X_{L}^{-}}\right):\right].\label{eq:antfieldpuregaugeP+}
\end{equation}
This corresponds to a well known property of the massless antifields.
Analysing, for example, the gauge transformations of \eqref{eq:gaugeantifields},
it is easy to show that both $a_{m}^{*}$ and $\xi_{\alpha}^{*}$
are pure gauge if $k^{m}k_{m}\neq0$, \emph{i.e.} if they are massive.

For the $P^{-}\neq0$ sector, the same property holds
\begin{equation}
:\left(\Lambda\,\lambda_{a}A_{a}\left(k\right)e^{-i\epsilon\sqrt{2}X_{L}^{+}}\right):=-\frac{i}{\epsilon}\left[Q,:\left(\lambda_{a}A_{a}\left(k\right)e^{-i\epsilon\sqrt{2}X_{L}^{+}}\right):\right],\label{eq:antfieldpuregaugeP-}
\end{equation}
which is trivially extended to the massive spectrum. Since $\left[Q,V_{\textrm{L.C.}}\right]=0$,
any DDF antifield of the form $\left|\psi\right\rangle ^{*}$ is a
\emph{singular} BRST-exact state. Although hidden in the covariant
description, this result is equivalent to the statement that BRST-closedness
does not impose the mass-shell condition on ghost number two vertex
operators.

It is useful to point out that the action of $c_{0}^{\pm}$ is meaningful
only in the ghost number one cohomology. One can try, for example,
to build a ghost number three state by the successive action of $c_{0}^{-}$'s:
\[
\left|\psi\right\rangle ^{**}=\left(c_{0}^{-}\right)^{2}\left|\psi\right\rangle .
\]
It can be shown, however, that\begin{subequations}\label{eq:c0nilpotent}
\begin{eqnarray}
\left\{ c_{0}^{-},c_{0}^{-}\right\}  & \approx & 0,\\
\left\{ c_{0}^{+},c_{0}^{+}\right\}  & \approx & 0.
\end{eqnarray}
\end{subequations}In other words, they are nilpotent within the BRST
cohomology. The proof is left to the appendix, equation \eqref{eq:c0squared}.
The remaining option is the linear combination
\begin{equation}
\left|\psi\right\rangle ^{**}=c_{0}^{-}c_{0}^{+}\left|\psi\right\rangle +\alpha\,c_{0}^{+}c_{0}^{-}\left|\psi\right\rangle ,
\end{equation}
where $\alpha$ is an arbitrary constant. Taking the ground state
as a reference, one can define
\begin{equation}
\left|0,k\right\rangle ^{**}\equiv c_{0}^{-}\left|0,k\right\rangle ^{*},
\end{equation}
with $\left|0,k\right\rangle ^{*}\approx c_{0}^{+}\left|0,k\right\rangle $.
Since $c_{0}^{-}\left|0,k\right\rangle \approx0$, $\left|0,k\right\rangle ^{**}$
can be cast as
\begin{equation}
\left|0,k\right\rangle ^{**}\approx\left\{ c_{0}^{-},c_{0}^{+}\right\} \left|0,k\right\rangle .
\end{equation}

The anticommutator $\left\{ c_{0}^{-},c_{0}^{+}\right\} $ is easily
obtained from \eqref{eq:VV*algebra}. Observe that
\begin{eqnarray}
\left\{ c_{0}^{-},\left[V_{\textrm{L.C.}}(k;a_{i},\bar{\xi}_{\dot{a}}),V_{\textrm{L.C.}}^{*}(-k;a_{i}^{*},\bar{\xi}_{\dot{a}}^{*})\right]\right\}  & \approx & +ik\delta_{p+k}\left(a_{i}(\sigma^{i}\bar{\xi}^{*})_{a}-a_{i}^{*}(\sigma^{i}\bar{\xi})_{a}\right)\left[c_{0}^{-},W_{a}^{*}\left(0\right)\right]\nonumber \\
 &  & +2k\delta_{p+k}\left(ka_{j}a_{j}^{*}+i\bar{\xi}_{\dot{a}}\bar{\xi}_{\dot{a}}^{*}\right)\left\{ c_{0}^{-},c_{0}^{+}\right\} .
\end{eqnarray}
Using the results of \eqref{eq:cV-} and \eqref{eq:c0nilpotent},
the Jacobi identity on the left hand side of the above equation implies
that
\begin{eqnarray}
\left\{ V_{\textrm{L.C.}}^{*}(k;a_{i},\bar{\xi}_{\dot{a}}),V_{\textrm{L.C.}}^{*}(-k;a_{i}^{*},\bar{\xi}_{\dot{a}}^{*})\right\}  & \approx & -ik\delta_{p+k}\left(a_{i}(\sigma^{i}\bar{\xi}^{*})_{a}-a_{i}^{*}(\sigma^{i}\bar{\xi})_{a}\right)\left[c_{0}^{-},W_{a}^{*}\left(0\right)\right]\nonumber \\
 &  & -2k\delta_{p+k}\left(ka_{j}a_{j}^{*}+i\bar{\xi}_{\dot{a}}\bar{\xi}_{\dot{a}}^{*}\right)\left\{ c_{0}^{-},c_{0}^{+}\right\} ,
\end{eqnarray}
which was already computed in \eqref{eq:V*V*algebra}. Comparing both
sides of the equation, one obtains\begin{subequations} 
\begin{eqnarray}
\left\{ c_{0}^{-},c_{0}^{+}\right\}  & \approx & 2M,\\
\left[c_{0}^{-},W_{a}^{*}\left(0\right)\right] & \approx & 0,
\end{eqnarray}
\end{subequations}where $M$ is defined in \eqref{eq:integratedmeasure}.

$\left|0,k\right\rangle ^{**}$ is then proportional to $M\left|0,k\right\rangle $.
A direct computation shows that $\left\{ M,\lambda_{a}A_{a}\left(k\right)\right\} $
is trivial, which in its turn implies that
\begin{equation}
\left|0,k\right\rangle ^{**}\approx0.
\end{equation}
A similar analysis can be made for states of the form $V_{\textrm{L.C.}}^{*}\left(k\right)\left|0\right\rangle ^{*}$
but the conclusion is the same: given the DDF structure discussed
here, it is impossible to build any higher ghost number state with
nonzero momentum, in accordance with the known statements about the
pure spinor cohomology.

Next section will discuss the role of the $b$ ghost in the structures
presented so far, showing that the fields and antifields are indeed
in one-to-one correspondence.

\section{Siegel gauge and the physical state condition\label{sec:DDFandb}}

\

The DDF perspective on the pure spinor cohomology shows that there
is an essential difference between states at ghost number one and
two. Although the physical states are mirrored by the antifield spectrum,
it was shown that the latter has a singular kinematic condition much
like their correspondent in bosonic string theory, as follows from
the discussion after equation \eqref{eq:antfieldpuregaugeP-}.

For the bosonic string, there is a way to make this distinction very
precise, which is currently known as the Siegel gauge. Physical states
are defined to be in the cohomology of the BRST charge plus an extra
condition: they have to be annihilated by the $b$ ghost zero mode,
$b_{0}$. This is a simple requirement, as one will be selecting only
the states which have no $c$ ghost zero mode, $c_{0}$, and they
all fall in the ghost number one spectrum. The physical state condition
can be understood as a consequence of the unitarity of the scattering
amplitudes, that can be shown to imply the projection onto the subspace
of states annihilated by $b_{0}$. In fact, it is easy to show that
the physical states are $b_{0}$-exact because the cohomology of $b_{0}$
is trivial due to the relation
\begin{equation}
\left\{ b_{0},c_{0}\right\} =1.
\end{equation}

However, the pure spinor formalism does not have a fundamental $\left(b,c\right)$
system. The $b$ ghost is a composite operator and the $c$ ghost
is simply absent. In order to understand the implications of the Siegel
gauge in the cohomology, it is useful to recall first some basic properties
of the pure spinor $b$ ghost.

\subsection{Quick review on the $b$ ghost}

\

In general terms, the fundamental property of a $b$ ghost is 

\begin{equation}
\left\{ Q,b\right\} =T,\label{eq:QbT}
\end{equation}
which is ultimately related to the BRST invariance of loop amplitudes
because of the connection with the energy-momentum tensor, $T$. When
expanded in Laurent modes, two equations are particularly interesting:
\begin{eqnarray}
\{Q,b_{0}\} & = & L_{0},\\
\{Q,b_{-1}\} & = & L_{-1}.
\end{eqnarray}
Observe that any BRST-closed operator $U$ with conformal weight $h$
can be written as a BRST-exact operator for $h\neq0$,
\begin{equation}
U=\frac{1}{h}\left\{ Q,\left[b_{0},U\right]\right\} ,\label{eq:nonscalarBRST-trivial}
\end{equation}
and that there is a simple recipe for constructing the integrated
vertex associated to $U$, defined by 
\begin{equation}
V\equiv\ointctrclockwise\left[b_{-1},U\right],
\end{equation}
and satisfying $\left\{ Q,V\right\} =\ointctrclockwise\partial U$.

The first proposal of a $b$ ghost like field in the pure spinor formalism
was presented in \cite{Berkovits:2004px}, with a complicated set
of picture raised operators. This is so because there is no natural
ghost number $-1$ field, as the pure spinor conjugate, $\omega_{\alpha}$,
has a gauge freedom associated to the pure spinor constraint and always
appear in gauge invariant combinations such as the ghost number current,
$J$, the Lorentz ghost current, $N^{mn}$, and the energy-momentum
tensor, $T_{\lambda}=-\omega_{\alpha}\partial\lambda^{\alpha}$. The
simplest way to overcome this difficulty is to introduce a constant
spinor $C_{\alpha}$, such that the product $C_{\alpha}\lambda^{\alpha}$
is nonzero (different patches of the pure spinor variable require
different $C$'s to ensure this condition). The noncovariant $b$
ghost is defined to be
\begin{equation}
b_{nc}\equiv\frac{C_{\alpha}G^{\alpha}}{C_{\beta}\lambda^{\beta}},
\end{equation}
with 
\begin{equation}
G^{\alpha}=\frac{1}{2}\Pi^{m}\left(\gamma_{m}d\right)_{\alpha}-\frac{1}{4}N_{mn}\left(\gamma^{mn}\partial\theta\right)^{\alpha}-\frac{1}{4}J\partial\theta^{\alpha}-\frac{7}{2}\partial^{2}\theta^{\alpha}.\label{eq:Galpha}
\end{equation}
Note that $G^{\alpha}$ satisfies
\begin{equation}
\left\{ Q,G^{\alpha}\right\} =\lambda^{\alpha}\left(T_{\textrm{matter}}+T_{\lambda}\right),
\end{equation}
such that
\begin{equation}
\left\{ Q,b_{nc}\right\} =T,
\end{equation}
where $T$ is the total energy-momentum tensor.

A covariant version of the $b$ ghost was presented later in \cite{Berkovits:2005bt},
with the introduction of the so-called non-minimal formalism. The
non-minimal variables consist of two conjugate pairs, $(\hat{\omega}^{\alpha},\hat{\lambda}_{\alpha})$
and $(s^{\alpha},r_{\alpha})$, such that the BRST charge is modified
to
\begin{equation}
Q_{nm}=\ointctrclockwise\left\{ \lambda^{\alpha}d_{\alpha}+\hat{\omega}^{\alpha}r_{\alpha}\right\} .
\end{equation}
$\hat{\lambda}_{\alpha}$ is also a pure spinor, \emph{i.e.} $(\hat{\lambda}\gamma^{m}\hat{\lambda})=0$,
while $r_{\alpha}$ is constrained through $(\hat{\lambda}\gamma^{m}r)=0$.
They decouple from the spectrum (quartet argument) and $Q_{nm}$ has
the same cohomology of $Q$. The $b$ ghost assumes a more robust
form, enabling a systematic investigation of its properties. For the
purposes of this work, it is sufficient to state that $b$ is defined
by
\begin{equation}
b=-s^{\alpha}\partial\hat{\lambda}_{\alpha}+\frac{\left(\hat{\lambda}_{\alpha}G^{\alpha}\right)}{\left(\hat{\lambda}\lambda\right)}-2!\frac{\left(\hat{\lambda}_{\alpha}r_{\beta}H^{\alpha\beta}\right)}{\left(\hat{\lambda}\lambda\right)^{2}}-3!\frac{\left(\hat{\lambda}_{\alpha}r_{\beta}r_{\gamma}K^{\alpha\beta\gamma}\right)}{\left(\hat{\lambda}\lambda\right)^{3}}+4!\frac{\left(\hat{\lambda}_{\alpha}r_{\beta}r_{\gamma}r_{\lambda}L^{\alpha\beta\gamma\lambda}\right)}{\left(\hat{\lambda}\lambda\right)^{4}}\label{eq:quantumb}
\end{equation}
with\begin{subequations}
\begin{eqnarray}
H^{\alpha\beta} & = & \frac{1}{4\cdot96}\gamma_{mnp}^{\alpha\beta}\left(d\gamma^{mnp}d+24N^{mn}\Pi^{p}\right),\\
K^{\alpha\beta\gamma} & = & -\frac{1}{96}N_{mn}\gamma_{mnp}^{[\alpha\beta}\left(\gamma^{p}d\right)^{\gamma]},\\
L^{\alpha\beta\gamma\lambda} & = & -\frac{3}{\left(96\right)^{2}}N^{mn}N^{rs}\eta^{pq}\gamma_{mnp}^{[\alpha\beta}\gamma_{qrs}^{\gamma]\lambda}.
\end{eqnarray}
\end{subequations}Here $[\alpha\beta\gamma]$ means antisymmetrisation
of the spinor indices.

The relevant properties for the analysis of the physical state condition
are nilpotency \cite{Chandia:2010ix,Jusinskas:2013yca} and non-uniqueness
of $b$ . The former can be stated as
\begin{equation}
b\left(z\right)b\left(y\right)\sim0,\label{eq:bbOPE}
\end{equation}
and naturally brings questions about its cohomology, which is of course
connected to the Siegel gauge discussion. While for the bosonic string
\eqref{eq:bbOPE} is trivially satisfied, the composite character
of the pure spinor $b$ ghost makes it far from obvious and $b_{0}$
is likely to have a nontrivial cohomology. The absence of a $c$ ghost
makes this subject even more intriguing.

Concerning non-uniqueness, different operators $b$ and $b'$ satisfying
\eqref{eq:QbT} have to differ by a BRST-exact term (it follows from
the discussion around equation \eqref{eq:nonscalarBRST-trivial}).
This is a useful property, as different forms of the $b$ ghost might
be suitable in different contexts. Although nilpotency is not assured
by these deformations, it can be stated for a general class of BRST-exact
terms \cite{Jusinskas:2013sha}.

These results will be used in the investigation of the Siegel gauge
in the pure spinor cohomology that follows.

\subsection{The physical state condition}

\

According to equation \eqref{eq:nonscalarBRST-trivial}, any operator
$\mathcal{O}$ in the cohomology of $Q$ has to be a worldsheet scalar.
This implies, in particular, that the (anti)commutator $[b_{0},\mathcal{O}]$
has to be BRST-invariant. In fact, $[b_{0},\mathcal{O}]$ can either
be (1) vanishing, (2) BRST-exact, or (3) also an element of the BRST
cohomology. Given that $b_{0}$ itself is defined up to BRST-exact
terms, the conditions (1) and (2) should be physically equivalent.
In this sense, the so-called Siegel gauge (1) is a stronger condition
than what is required from the consistency of the above analysis.

Due to the composite nature of pure spinor $b$ ghost, the implementation
of the Siegel gauge is not trivial. For the massless spectrum, the
first discussions on the subject were presented in \cite{Grassi:2009fe,Aisaka:2009yp}.
While the work of Grassi and Vanhove discussed the Siegel gauge by
explicitly computing the action of the noncovariant $b$, Aisaka and
Berkovits assumed nilpotency of the non-minimal $b$ ghost and built
the massless ghost number one vertex $\tilde{U}$ as a $b_{0}$-exact
state coming from the antifield $A_{\alpha\beta}$, \emph{cf.} equation
\eqref{eq:masslessantifieldvertex}:
\[
\tilde{U}=\left[b_{0},\lambda^{\alpha}\lambda^{\beta}A_{\alpha\beta}\right].
\]
It is clear that $\tilde{U}=U+\left[Q_{nm},\Lambda\right]$ and the
condition $\{b_{0},\tilde{U}\}=0$ is satisfied only in a gauge in
which the non-minimal variables are present. Both approaches are simple
enough when the massless spectrum is concerned but can hardly be extended
to the massive levels. The explicit action of $b_{0}$, even in its
simplest form, would be a cumbersome computation and determining the
BRST-exact pieces to understand the physical implications would be
far from trivial. Besides, the cohomology of $b_{0}$ is not known
and the construction of $b_{0}$-exact states is not assured, as there
might be ghost number two states in the cohomology of $b_{0}$.

However, since the DDF spectrum relies on massless vertex operators,
the analysis of the Siegel gauge can be performed in a straightforward
way. One can start by examining the double poles of the operator $G^{\alpha}$
in the OPE's with the massless vertices of \eqref{eq:defunintegratedpolarizations}
and \eqref{eq:antifieldpolarization}:\begin{subequations}\label{eq:GstateOPEs}
\begin{eqnarray}
z\bar{G}_{\dot{a}}\left(z\right)\:\bar{\lambda}_{\dot{b}}\bar{A}_{\dot{b}}\left(k\right) & \sim & \textrm{regular},\\
zG_{a}\left(z\right)\:\bar{\lambda}_{\dot{b}}\bar{A}_{\dot{b}}\left(k\right) & \sim & \textrm{regular},\\
z\bar{G}_{\dot{a}}\left(z\right)\:\bar{\Lambda}\,\bar{\lambda}_{\dot{b}}\bar{A}_{\dot{b}}\left(k\right) & \sim & \frac{1}{z}\left(2ik\right)\bar{\lambda}_{\dot{a}}\bar{\lambda}_{\dot{b}}\bar{A}_{\dot{b}}\left(k\right),\\
zG_{a}\left(z\right)\:\bar{\Lambda}\,\bar{\lambda}_{\dot{b}}\bar{A}_{\dot{b}}\left(k\right) & \sim & \textrm{regular}.
\end{eqnarray}
\end{subequations}Note the appearance of the momentum factor $k$,
in accordance with the previous section (see footnote \ref{fn:extramomentum}).
From \eqref{eq:GstateOPEs} and the discussion on the non-uniqueness
of $b$, it follows that\begin{subequations}\label{eq:b0actionDDFU}
\begin{eqnarray}
\left\{ b_{0},\bar{\lambda}_{\dot{a}}\bar{A}_{\dot{a}}\left(k\right)\right\}  & \approx & 0,\\
\left[b_{0},\bar{\Lambda}\,\bar{\lambda}_{\dot{a}}\bar{A}_{\dot{a}}\left(k\right)\right] & \approx & 2ik\,\bar{\lambda}_{\dot{a}}\bar{A}_{\dot{a}}\left(k\right).
\end{eqnarray}
\end{subequations}These equations should hold for any well-defined
version of the $b$ ghost, whether in the minimal or non-minimal formalism\footnote{One can define non-covariant versions of the $b$ ghost as 
\[
\begin{array}{cc}
b_{-}\equiv\frac{\bar{C}_{\dot{a}}\bar{G}_{\dot{a}}}{\bar{C}_{\dot{b}}\bar{\lambda}_{\dot{b}}}, & b_{+}\equiv\frac{C_{a}G_{a}}{\vphantom{\bar{C}_{\dot{a}}}C_{b}\lambda_{b}},\end{array}
\]
which clearly satisfy $\{Q,b_{\pm}\}=T$. With this definition, the
equations in \eqref{eq:b0actionDDFU} are true for $b_{-}$ while
for $b_{+}$ they do not work at all. The key word here is \emph{well-defined}.
When one chooses to work with the DDF spectrum in the $P^{-}\neq0$
frame, for example, the components $\lambda_{a}$ decouple and the
states depend only on $\bar{\lambda}_{\dot{a}}$. This effectively
means $\lambda_{a}=0$ and then $b_{+}$ is a singular operator in
this subspace. This is of course just a hint on the solution of this
puzzle and a more formal understanding has yet to be achieved. }. The extension to the integrated vertices \eqref{eq:SO8integrated}
and \eqref{eq:SO8integratedsantifields} is straightforward, as they
can be built from the unintegrated vertices with the action of the
mode $b_{-1}$, up to gauge transformations:\begin{subequations}\label{eq:b0actionDDFV}
\begin{eqnarray}
\left[b_{0},V_{\textrm{L.C.}}\left(k;a_{i},\bar{\xi}_{\dot{a}}\right)\right] & \approx & 0,\\
\left\{ b_{0},V_{\textrm{L.C.}}^{*}\left(k;a_{i}^{*},\bar{\xi}_{\dot{a}}^{*}\right)\right\}  & \approx & -2ik\,V_{\textrm{L.C.}}\left(k;a_{i}^{*},\bar{\xi}_{\dot{a}}^{*}\right).
\end{eqnarray}
\end{subequations}

It is in principle possible to build a creation/annihilation algebra
already in the Siegel gauge through the $b_{0}$-exact construction
just described. However, the $b$ ghost will clearly spoil the structure
depicted in \eqref{eq:creation-annihilationFULL} with the introduction
of BRST-exact pieces. Besides, massive states might acquire higher
and higher powers of inverse $\lambda^{\alpha}$, \emph{e.g.} $(\hat{\lambda}\lambda)^{-1}$
in the non-minimal formalism. This is a potential problem due to existence
of the operator
\begin{eqnarray}
\xi & = & \frac{\hat{\lambda}\cdot\theta}{\hat{\lambda}\cdot\lambda-r\cdot\theta}\nonumber \\
 & = & \frac{\hat{\lambda}\cdot\theta}{\hat{\lambda}\cdot\lambda}\sum_{n=0}^{11}\left(\frac{r\cdot\theta}{\hat{\lambda}\cdot\lambda}\right)^{n},
\end{eqnarray}
which trivialises the BRST cohomology, for $\{Q,\xi\}=1$. Also, high
inverse powers of $\lambda^{\alpha}\hat{\lambda}_{\alpha}$ are hard
to deal with when regularising the scattering amplitudes \cite{Berkovits:2006vi}.
Because of these subtleties, the $b_{0}$-exact creation operators
are not advantageous.

In the full analysis, when arbitrary states of the DDF spectrum are
considered, the general conclusion is\begin{subequations}\label{eq:b0actionDDFVspectrum}
\begin{eqnarray}
b_{0}\left|\psi\right\rangle  & \approx & 0,\\
b_{0}\left|\psi\right\rangle ^{*} & \approx & \left|\psi\right\rangle ,
\end{eqnarray}
\end{subequations}where $\left|\mathcal{\psi}\right\rangle $ and
$\left|\psi\right\rangle ^{*}$ were defined in \eqref{eq:DDF=0000231}
and \eqref{eq:DDF=0000232}. In other words, if the BRST cohomology
is spanned by the DDF spectrum, there is no BRST-closed ghost number
two element in the $b_{0}$ cohomology. Also, any BRST-closed ghost
number one vertex can be written as $b_{0}$-exact, up to a gauge
transformation. 

Therefore, the physical state condition for the pure spinor formalism
can be compactly written as
\begin{equation}
b_{0}\left|\mathcal{O}\right\rangle \approx0,\label{eq:physicalstate}
\end{equation}
which is compatible also with the unitarity analysis of the amplitudes,
in a direct analogy with the bosonic string. Here, $\left|\mathcal{O}\right\rangle $
is a state defined by the state-operator map of a generic operator
$\mathcal{O}$ in the BRST cohomology.

Given the roles of the operators $b_{0}$ and $c_{0}^{\pm}$ as mapping
vertex operators of different ghost numbers, there should be a relation
among them that exposes this inversion character. The results stated
above can be made a bit more precise by investigating this relation,
as will be shown next.

\subsection{The DDF conjugates of $b_{0}$ }

\

As mentioned in the \nameref{sec:Introduction}, the bosonic string
analogous of the quantity $\{b_{0},c_{0}^{\pm}\}$ is very clear.
For the pure spinor superstring, however, this quantity is much harder
to obtain by brute force, due to the complicated nature of the operators
involved. At this point, the extended DDF algebra comes in handy,
allowing an indirect computation of the anticommutator.

Consider the quantity defined by
\begin{equation}
I\equiv\left\{ b_{0},\left[V_{\textrm{L.C.}}\left(k;a_{i},\bar{\xi}_{\dot{a}}\right),V_{\textrm{L.C.}}^{*}\left(p;a_{i}^{*},\bar{\xi}_{\dot{a}}^{*}\right)\right]\right\} .\label{eq:b0c0auxiliar}
\end{equation}
The Jacobi identity implies that
\begin{eqnarray}
I & = & \left\{ \left[b_{0},V_{\textrm{L.C.}}\left(k;a_{i},\bar{\xi}_{\dot{a}}\right)\right],V_{\textrm{L.C.}}^{*}\left(p;a_{i}^{*},\bar{\xi}_{\dot{a}}^{*}\right)\right\} \nonumber \\
 &  & +\left[V_{\textrm{L.C.}}\left(k;a_{i},\bar{\xi}_{\dot{a}}\right),\left\{ b_{0},V_{\textrm{L.C.}}^{*}\left(p;a_{i}^{*},\bar{\xi}_{\dot{a}}^{*}\right)\right\} \right].
\end{eqnarray}
The inside (anti)commutators can be replaced according to equation
\eqref{eq:b0actionDDFV}, so the result is
\begin{eqnarray}
I & \approx & 2ip\left[V_{\textrm{L.C.}}\left(p;a_{i}^{*},\bar{\xi}_{\dot{a}}^{*}\right),V_{\textrm{L.C.}}\left(k;a_{i},\bar{\xi}_{\dot{a}}\right)\right]\nonumber \\
 & \approx & 2\sqrt{2}ik\delta_{k+p}\left(ka_{j}a_{j}^{*}+i\bar{\xi}_{\dot{a}}\bar{\xi}_{\dot{a}}^{*}\right)P^{+},\label{eq:b0c0auxiliarleft}
\end{eqnarray}
\emph{cf.} the creation/annihilation algebra of \eqref{eq:creation-annihilationFULL}.

On the other hand, using \eqref{eq:VV*algebra}, $I$ can be written
as
\begin{eqnarray}
I & = & ik\delta_{p+k}\left\{ a_{i}(\sigma^{i}\bar{\xi}^{*})_{a}-a_{i}^{*}(\sigma^{i}\bar{\xi})_{a}\right\} \left[b_{0},W_{a}^{*}\left(0\right)\right]\nonumber \\
 &  & +2k\delta_{p+k}\left(ka_{j}a_{j}^{*}+i\bar{\xi}_{\dot{a}}\bar{\xi}_{\dot{a}}^{*}\right)\left\{ b_{0},c_{0}^{+}\right\} .
\end{eqnarray}
According to equation \eqref{eq:b0actionDDFV}, $\left[b_{0},W_{a}^{*}\left(0\right)\right]\approx0$,
as $W_{a}^{*}\left(0\right)$ is a zero-momentum vertex, and the first
term on the right hand side is BRST-exact. Therefore,
\begin{equation}
I\approx2k\delta_{p+k}\left(ka_{j}a_{j}^{*}+i\bar{\xi}_{\dot{a}}\bar{\xi}_{\dot{a}}^{*}\right)\left\{ b_{0},c_{0}^{+}\right\} .\label{eq:b0c0auxiliarright}
\end{equation}

Comparing now the equations \eqref{eq:b0c0auxiliarleft} and \eqref{eq:b0c0auxiliarright},
one obtains
\begin{equation}
\left\{ b_{0},c_{0}^{+}\right\} \approx i\sqrt{2}P^{+}.\label{eq:b0c0+}
\end{equation}
The same procedure can be followed in the DDF frame $P^{-}\neq0$,
so that
\begin{equation}
\left\{ b_{0},c_{0}^{-}\right\} \approx i\sqrt{2}P^{-}.\label{eq:b0c0-}
\end{equation}

In spite of having a very simple form, equations \eqref{eq:b0c0+}
and \eqref{eq:b0c0-} differ from the bosonic case by an essential
term, as they were derived up to BRST-exact quantities. In their full
form, they can be cast as
\begin{equation}
i\sqrt{2}P^{\pm}=\left\{ b_{0},c_{0}^{\pm}\right\} +\left\{ Q,\xi^{\pm}\right\} ,\label{eq:b0c0full}
\end{equation}
where $\xi^{\pm}$ depends on several quantities that can be traced
back to the specific form of the chosen $b$ ghost. It should be emphasised
again that these results hold for both minimal and non-minimal formalisms,
although the latter allows a clearer interpretation due to the non-minimal
variables\footnote{In this case, the result \eqref{eq:b0c0full} is not so surprising,
because the non-minimal $b$ ghost and the BRST current are the fermionic
generators of a $\mathcal{N}=2$ topological algebra \cite{Berkovits:2005bt}.
Thanks to C. Maccaferri for this observation.}. 

As a consistency check, consider the action of equation \eqref{eq:b0c0full}
in certain sets of states:
\begin{itemize}
\item suppose there is a ghost number one state $\left|\phi\right\rangle $
annihilated by $b_{0}$ with $P^{+}=1$, then
\begin{eqnarray}
i\sqrt{2}\left|\phi\right\rangle  & = & b_{0}c_{0}^{+}\left|\phi\right\rangle +\left\{ Q,\xi^{+}\right\} \left|\phi\right\rangle .
\end{eqnarray}
Besides, if $\left|\phi\right\rangle $ is BRST-closed, then
\begin{equation}
i\sqrt{2}\left|\phi\right\rangle =b_{0}c_{0}^{+}\left|\phi\right\rangle +Q\xi^{+}\left|\phi\right\rangle .
\end{equation}
According to the spectrum analysis of section \ref{sec:Antifields},
$c_{0}^{+}\left|\phi\right\rangle $ should be a BRST-closed ghost
number two state. Therefore, if $\left|\phi\right\rangle $ is annihilated
by $b_{0}$, then it is $b_{0}$-exact up to a gauge transformation:
\begin{equation}
i\sqrt{2}\left|\phi\right\rangle \approx b_{0}\left|\phi\right\rangle ^{*}.
\end{equation}

\item now, suppose there is a ghost number two state $\left|\phi\right\rangle ^{*}$
in the same conditions (BRST-closed and annihilated by $b_{0}$),
then
\begin{equation}
i\sqrt{2}\left|\phi\right\rangle ^{*}=b_{0}c_{0}^{+}\left|\phi\right\rangle ^{*}+Q\xi^{+}\left|\phi\right\rangle ^{*}.
\end{equation}
As discussed for the antifields, $c_{0}^{+}\left|\phi\right\rangle ^{*}$
is BRST-exact and the above equation implies that $\left|\phi\right\rangle ^{*}$
itself is BRST-exact. In particular, if $\left|\phi\right\rangle ^{*}$
is an element of the BRST cohomology, this leads to a contradiction.
Therefore, it is not possible to have a ghost number two state in
the BRST cohomology annihilated by $b_{0}$.
\end{itemize}
These results are in agreement with the physical state condition of
\eqref{eq:physicalstate}.

\section{Summary and conclusions\label{sec:Summary}}

\

Given the success of the DDF construction in determining the physical
spectrum \cite{Berkovits:2000nn,Jusinskas:2014vqa,Berkovits:2014bra},
the starting point of the analysis introduced here for the ghost number
two cohomology was a convenient solution for the massless superfield
$A_{\alpha\beta}$ of \eqref{eq:masslessantifieldvertex}. It should
correspond to the ground state of the antifield spectrum. In the frame
with momentum $P^{-}\neq0$, the nonvanishing components can be cast
as
\begin{equation}
\bar{A}_{\dot{a}\dot{b}}=\frac{1}{2}(\bar{\theta}_{\dot{a}}\bar{A}_{\dot{b}}+\bar{\theta}_{\dot{b}}\bar{A}_{\dot{a}})-\frac{1}{8}\eta_{\dot{a}\dot{b}}(\bar{\theta}_{\dot{c}}\bar{A}_{\dot{c}}),\label{eq:antifieldDDF}
\end{equation}
where $\bar{A}_{\dot{a}}$ is displayed in \eqref{eq:spinorsuperfield}.
Unlike what happens for the massless field, there is a dependence
on $\bar{\theta}_{\dot{a}}$ that cannot be removed by a gauge
transformation. The equation of motion $D_{((\alpha}A_{\beta\gamma))}=0$
is satisfied, as can be readily seen from the BRST-closedness of $U^{*}=\bar{\lambda}_{\dot{a}}\bar{\lambda}_{\dot{b}}\bar{A}_{\dot{a}\dot{b}}$,
and the solution above is the antifield equivalent of the $SO\left(8\right)$-covariant
Yang-Mills superfields introduced in \cite{Brink:1983pf}. There is
of course an analogous solution in the frame $P^{+}\neq0$, with nonvanishing
components $A_{ab}$. 

It is interesting to point out that \eqref{eq:antifieldDDF} did not
come from an ingenious gauge choice but from the action of the operator
$c_{0}^{-}$ on the massless ghost number one states, \emph{cf.} equation
\eqref{eq:c0overmassless}. This operation conveniently enables the
construction of both integrated and unintegrated massless vertices
of the ghost number two cohomology. The former, for example, is given
by
\begin{equation}
V^{*}=\ointctrclockwise\left\{ \bar{\Lambda}\left(\Pi_{i}-i\sqrt{2}k\bar{N}_{i}\right)A_{i}+\bar{\Lambda}\left(\partial\bar{\theta}_{\dot{a}}+ik\bar{d}_{\dot{a}}\right)\bar{A}_{\dot{a}}+\sqrt{2}\partial X^{-}\bar{\lambda}_{\dot{a}}\bar{A}_{\dot{a}}\right\} 
\end{equation}
and can be shown to satisfy $\left\{ Q,V^{*}\right\} =-\ointctrclockwise\,\partial(\bar{\lambda}_{\dot{a}}\bar{\lambda}_{\dot{b}}\bar{A}_{\dot{a}\dot{b}})$.
At this level, supersymmetry of the polarisations is manifest up to
BRST-exact terms. Not surprisingly, the operators $c_{0}^{\pm}$,
together with their supersymmetric partners, naturally emerge in the
extension of the DDF algebra to the next ghost number level. 

As in the bosonic string, the antifield vertices have a very peculiar
kinematic property. Roughly speaking, the ghost number two vertices
correspond to \emph{singular} BRST-exact operators. For the massless
case, for example, one has
\begin{equation}
:\left(\bar{\lambda}_{\dot{a}}\bar{\lambda}_{\dot{b}}\bar{A}_{\dot{a}\dot{b}}e^{-i\epsilon\sqrt{2}X_{L}^{-}}\right):=-\frac{i}{\epsilon}\left[Q,:\left(\bar{\lambda}_{\dot{a}}\bar{A}_{\dot{a}}\left(k\right)e^{-i\epsilon\sqrt{2}X_{L}^{-}}\right):\right].
\end{equation}
The limit $\epsilon\to0$ is singular, so the vertex is in the cohomology.
However any massive ($\epsilon\neq0$) deformation is BRST-exact.
Another way of stating this property is saying that BRST-closedness
does not impose the mass-shell condition for the antifields. In the
covariant description of the vertex, this property is hidden. 

The doubling of the pure spinor cohomology can be simply described
as the action of the operators $c_{0}^{\pm}$, which play the role
of the $c$ ghost zero mode. In fact, $c_{0}^{+}$ and $c_{0}^{-}$
constitute a sort of DDF conjugates of the $b$ ghost zero mode, $b_{0}$.
This can be stated as
\begin{equation}
\left\{ b_{0},c_{0}^{\pm}\right\} =i\sqrt{2}P^{\pm}-\left\{ Q,\xi^{\pm}\right\} ,\label{eq:bcDDFconjugates}
\end{equation}
and the conjugate interpretation holds in the subspaces of momentum
$P^{+}\neq0$ and $P^{-}\neq0$. Recall that the $b$ ghost is not
a unique fundamental field in the pure spinor formalism and the operators
$\xi^{+}$ and $\xi^{-}$ clearly depend on its specific form. Although
the demonstration of the above relation relied on the extended DDF
algebra, its validity is not restricted to the BRST cohomology. This
illustrates the power of the DDF construction, since very general
statements can be made by knowing only a couple of properties of the
massless operators. 

Due to the BRST-exact piece in \eqref{eq:bcDDFconjugates}, the analysis
of the $b_{0}$ cohomology could be carried out only within the BRST-closed
operators. It was shown that every ghost number one state is $b_{0}$-exact
up to a gauge transformation. Besides, assuming that the antifield
spectrum is also spanned by the DDF operators, there is no ghost number
two state in the BRST cohomology annihilated by the $b$ ghost zero
mode. Among other implications, this result enables a formal definition
of a physical state condition in the pure spinor superstring,
\begin{equation}
b_{0}\left|\phi\right\rangle \approx0,\label{eq:physicalcond}
\end{equation}
where $\left|\phi\right\rangle $ is a generic BRST-closed state and
the equality holds up to BRST-exact terms. The condition \eqref{eq:physicalcond}
applies to the open string or to the holomorphic sector of the closed
string. In the last case, the antiholomorphic sector should be annihilated
by the corresponding $b$ ghost zero mode, denoted by $\bar{b}_{0}$.

It is worth noting that the symmetric way in which $Q$ and $b_{0}$
appear in \eqref{eq:bcDDFconjugates} might be interpreted as coming
from their dual roles in the topological string algebra of the non-minimal
formalism \cite{Berkovits:2005bt}. There is a very unusual $c$ ghost
like field introduced in \cite{Jusinskas:2013sha} which was shown
to satisfy
\begin{equation}
b\left(z\right)c\left(y\right)\sim\frac{1}{\left(z-y\right)},
\end{equation}
but its singular character prevents any well-defined attempt to trivialise
the $b$ ghost cohomology through this construction. In fact, this
proposal for the $c$ ghost seems to be completely unrelated to the
DDF motivated $c_{0}^{\pm}$ operators introduced in equations \eqref{eq:c0+}
and \eqref{eq:c0-}. Also, for any choice of $b$, the vanishing of
the term $\left\{ Q,\xi^{\pm}\right\} $ is very unlikely to happen
(even for the simplest case, $\xi^{\pm}$ is a cumbersome operator),
which is a strong indication of a nontrivial cohomology for $b_{0}$. 

The cohomology of the $b$ ghost remains a mystery in the pure spinor
formalism. A natural starting point would be to study the massless
subspace and maybe even develop a DDF-like approach to extend it to
higher mass states. There is no analogous feature in the other superstring
formalisms so it is hard to stablish a guiding direction at this point.
The interest in this subject resides mostly on the implementation
of the Siegel gauge in the superstring field theory. There, fields
and antifields should appear in a symmetrical way and a better understanding
of the Siegel gauge might help to clarify the second quantized version
of the pure spinor formalism. The bosonic closed string field, for
example, is annihilated by the holomorphic-antilomorphic combination
$(b_{0}-\bar{b}_{0})$, and it is not known how this condition extends
to the pure spinor case.

\

\textbf{Acknowledgements:} I would like to thank Thales Azevedo, Nathan
Berkovits, Carlo Maccaferri, Andrei Mikhailov and Martin Schnabl and
for useful comments and discussions, especially Thales Azevedo for
carefully reading the draft. This research has been supported by the
Grant Agency of the Czech Republic, under the grant P201/12/G028.

\appendix

\section{Some explicit computations\label{sec:computations}}

\

This appendix contains some of the results that were not completely
developed in sections \ref{sec:Antifields} and \ref{sec:DDFandb}.
It is a collection of demonstrations rather than a cohesive text.
There is no logical connection between each part and they are organised
according to their order of appearance in the paper.

\paragraph*{Composition of $\bar{U}_{i}\left(p\right)\bar{U}_{j}\left(q\right)$}

\

According to equation \eqref{eq:Uiunintegrated}, the product $\bar{U}_{i}\left(p\right)\bar{U}_{j}\left(q\right)$
can be cast as

\begin{eqnarray}
\bar{U}_{i}\left(p\right)\bar{U}_{j}\left(q\right) & = & e^{-i\left(p+q\right)\sqrt{2}X_{L}^{+}}\left\{ \vphantom{\left(\frac{p^{2}q}{6!}\right)}\Lambda_{i}\Lambda_{j}-i\left(\frac{q}{3!}\right)\Lambda_{i}\theta_{jk}\Lambda_{k}-i\left(\frac{p}{3!}\right)\theta_{ik}\Lambda_{k}\Lambda_{j}\right.\nonumber \\
 &  & -\left(\frac{q^{2}}{5!}\right)\Lambda_{i}\theta_{jk}\theta_{kl}\Lambda_{l}-\left(\frac{p^{2}}{5!}\right)\theta_{ik}\theta_{kl}\Lambda_{l}\Lambda_{j}-\left(\frac{pq}{3!3!}\right)\theta_{ik}\Lambda_{k}\theta_{jl}\Lambda_{l}\nonumber \\
 &  & +i\left(\frac{q^{3}}{7!}\right)\Lambda_{i}\theta_{jk}\theta_{kl}\theta_{lm}\Lambda_{m}+i\left(\frac{p^{3}}{7!}\right)\theta_{ik}\theta_{kl}\theta_{lm}\Lambda_{m}\Lambda_{j}\nonumber \\
 &  & +\left.i\left(\frac{pq^{2}}{6!}\right)\theta_{ik}\Lambda_{k}\theta_{jl}\theta_{lm}\Lambda_{m}+i\left(\frac{p^{2}q}{6!}\right)\theta_{ik}\theta_{kl}\Lambda_{l}\theta_{jm}\Lambda_{m}\right\} ,\label{eq:UiUjappendix}
\end{eqnarray}
where the higher powers of $\theta_{a}$ vanish because there are
only $8$ independent components and it is an anticommuting variable.

At this point, it is better to rewrite the expression inside the curly
brackets in a more suggestive way. Observe that
\begin{eqnarray}
\Lambda_{i}\Lambda_{j} & = & \frac{1}{2}\left\{ Q,\Lambda_{i}\left(\theta\sigma_{j}\bar{\theta}\right)\right\} -\frac{1}{2}\left\{ Q,\Lambda_{j}\left(\theta\sigma_{i}\bar{\theta}\right)\right\} -\frac{1}{4}\left\{ Q,\left(\Lambda\bar{\theta}_{ij}+\bar{\Lambda}\theta_{ij}\right)\right\} \nonumber \\
 &  & +\frac{1}{8}\left\{ Q,\left(\bar{\Lambda}_{ik}\theta_{kj}-\theta_{ik}\bar{\Lambda}_{kj}\right)\right\} ,
\end{eqnarray}
and
\begin{eqnarray}
\Lambda_{i}\theta_{jk}\Lambda_{k} & = & \frac{1}{2}\left\{ Q,\Lambda_{i}\left(\theta\sigma_{k}\bar{\theta}\right)\theta_{jk}\right\} -\frac{1}{2}\left\{ Q,\left(\theta\sigma_{i}\bar{\theta}\right)\Lambda_{k}\theta_{jk}\right\} -\frac{1}{4}\left\{ Q,\Lambda\bar{\theta}_{ik}\theta_{jk}\right\} \nonumber \\
 &  & +\frac{1}{8}\left\{ Q,\bar{\Lambda}\theta_{ik}\theta_{kj}\right\} +\frac{1}{8}\left\{ Q,\left(\Lambda_{il}\bar{\theta}_{lk}-\bar{\theta}_{il}\Lambda_{lk}\right)\theta_{jk}\right\} \nonumber \\
 &  & +\frac{1}{4}\left\{ Q,\theta_{ij}\left(\theta\sigma_{k}\bar{\theta}\right)\bar{\Lambda}_{k}\right\} -\left\{ Q,\left(\theta\sigma_{i}\bar{\theta}\right)\theta_{jk}\bar{\Lambda}_{k}\right\} \nonumber \\
 &  & +3\Lambda\left(\theta\sigma_{i}\bar{\theta}\right)\Lambda_{j}-3\Lambda\Lambda_{i}\left(\theta\sigma_{j}\bar{\theta}\right)+\frac{3}{2}\Lambda\bar{\Lambda}\theta_{ij}+\frac{3}{4}\Lambda\left(\bar{\theta}_{ik}\Lambda_{kj}-\Lambda_{ik}\bar{\theta}_{kj}\right)\nonumber \\
 &  & +\frac{3}{2}\eta^{ij}\Lambda\bar{\Lambda}_{k}\left(\theta\sigma_{k}\bar{\theta}\right).\label{eq:auxiliarc0squared}
\end{eqnarray}

Going further on higher orders of $\theta$, the BRST-exact structures
are recurrent and straightforward to determine: 
\begin{eqnarray}
\bar{U}_{i}\left(p\right)\bar{U}_{j}\left(q\right) & = & \frac{1}{2}\left\{ Q,\left[\Lambda_{i}\left(\theta\sigma_{j}\bar{\theta}\right)-\Lambda_{j}\left(\theta\sigma_{i}\bar{\theta}\right)\right]e^{-i\left(p+q\right)\sqrt{2}X_{L}^{+}}\right\} \nonumber \\
 &  & -\frac{1}{4}\left\{ Q,\left[\Lambda\bar{\theta}_{ij}+\bar{\Lambda}\theta_{ij}-\frac{1}{2}\left(\bar{\Lambda}_{ik}\theta_{kj}-\theta_{ik}\bar{\Lambda}_{kj}\right)\right]e^{-i\left(p+q\right)\sqrt{2}X_{L}^{+}}\right\} \nonumber \\
 &  & -i\left(\frac{q}{4!}\right)\left\{ Q,\left[2\Lambda_{i}\left(\theta\sigma_{k}\bar{\theta}\right)-2\left(\theta\sigma_{i}\bar{\theta}\right)\Lambda_{k}\right]\theta_{jk}e^{-i\left(p+q\right)\sqrt{2}X_{L}^{+}}\right\} \nonumber \\
 &  & +i\left(\frac{q}{4!}\right)\left\{ Q,\left[\Lambda\bar{\theta}_{ik}-\frac{1}{2}\left(-\bar{\theta}_{il}\Lambda_{lk}+\Lambda_{il}\bar{\theta}_{lk}\right)\right]\theta_{jk}e^{-i\left(p+q\right)\sqrt{2}X_{L}^{+}}\right\} \nonumber \\
 &  & +i\left(\frac{p}{4!}\right)\left\{ Q,\left[2\Lambda_{j}\left(\theta\sigma_{k}\bar{\theta}\right)-2\left(\theta\sigma_{j}\bar{\theta}\right)\Lambda_{k}\right]\theta_{ik}e^{-i\left(p+q\right)\sqrt{2}X_{L}^{+}}\right\} \nonumber \\
 &  & -i\left(\frac{p}{4!}\right)\left\{ Q,\left[\Lambda\bar{\theta}_{jk}-\frac{1}{2}\left(-\bar{\theta}_{jl}\Lambda_{lk}+\Lambda_{jl}\bar{\theta}_{lk}\right)\right]\theta_{ik}e^{-i\left(p+q\right)\sqrt{2}X_{L}^{+}}\right\} \nonumber \\
 &  & -i\left(\frac{q}{3!}\right)\left\{ Q,\left[\frac{1}{8}\bar{\Lambda}\theta_{ik}\theta_{kj}+\frac{1}{4}\theta_{ij}\left(\theta\sigma_{k}\bar{\theta}\right)\bar{\Lambda}_{k}-\left(\theta\sigma_{i}\bar{\theta}\right)\theta_{jk}\bar{\Lambda}_{k}\right]e^{-i\left(p+q\right)\sqrt{2}X_{L}^{+}}\right\} \nonumber \\
 &  & +i\left(\frac{p}{3!}\right)\left\{ Q,\left[\frac{1}{8}\bar{\Lambda}\theta_{jk}\theta_{ki}+\frac{1}{4}\theta_{ji}\left(\theta\sigma_{k}\bar{\theta}\right)\bar{\Lambda}_{k}-\left(\theta\sigma_{j}\bar{\theta}\right)\theta_{ik}\bar{\Lambda}_{k}\right]e^{-i\left(p+q\right)\sqrt{2}X_{L}^{+}}\right\} \nonumber \\
 &  & +i\,e^{-i\left(p+q\right)\sqrt{2}X_{L}^{+}}\left(\frac{p-q}{4}\right)\eta^{ij}\Lambda\bar{\Lambda}_{k}\left(\theta\sigma_{k}\bar{\theta}\right)+\mathcal{O}\left(\theta^{6}\right),
\end{eqnarray}
which gives the equation displayed in \eqref{eq:UiUj}.

\paragraph*{Composition of $\bar{U}_{i}\left(p\right)\bar{Y}_{\dot{a}}\left(q\right)$}

\

For the product $\bar{U}_{i}\left(p\right)\bar{Y}_{\dot{a}}\left(q\right)$,
the analysis is very similar. The explicit expression, according to
\eqref{eq:Uiunintegrated} and \eqref{eq:Yaunintegrated}, is
\begin{eqnarray}
\bar{U}_{i}\left(p\right)\bar{Y}_{\dot{a}}\left(q\right) & = & e^{-i\left(p+q\right)\sqrt{2}X_{L}^{+}}\left\{ \vphantom{\frac{p^{3}}{7!q}}\frac{i}{q}\Lambda_{i}\bar{\lambda}_{\dot{a}}-\frac{1}{2!}\Lambda_{i}\Lambda_{j}\left(\theta\sigma^{j}\right)_{\dot{a}}+\frac{p}{3!q}\theta_{ij}\Lambda_{j}\bar{\lambda}_{\dot{a}}\right.\nonumber \\
 &  & +\frac{iq}{4!}\Lambda_{i}\theta_{jk}\Lambda_{k}\left(\theta\sigma^{j}\right)_{\dot{a}}+\frac{ip}{2!3!}\theta_{ij}\Lambda_{j}\Lambda_{k}\left(\theta\sigma^{k}\right)_{\dot{a}}-\frac{ip^{2}}{5!q}\theta_{ij}\theta_{jk}\Lambda_{k}\bar{\lambda}_{\dot{a}}\nonumber \\
 &  & +\frac{q^{2}}{6!}\Lambda_{i}\theta_{jk}\theta_{kl}\Lambda_{l}\left(\theta\sigma^{j}\right)_{\dot{a}}+\frac{p^{2}}{2!5!}\theta_{ij}\theta_{jk}\Lambda_{k}\Lambda_{l}\left(\theta\sigma^{l}\right)_{\dot{a}}\nonumber \\
 &  & \left.+\frac{pq}{3!4!}\theta_{ij}\Lambda_{j}\theta_{kl}\Lambda_{l}\left(\theta\sigma^{k}\right)_{\dot{a}}-\frac{p^{3}}{7!q}\theta_{ij}\theta_{jk}\theta_{kl}\Lambda_{l}\bar{\lambda}_{\dot{a}}\right\} .\label{eq:UiYaappendix}
\end{eqnarray}

Here again, the BRST-exact structures are straightforward to obtain.
For example,
\begin{eqnarray}
\Lambda_{i}\bar{\lambda}_{\dot{a}} & = & -\left[Q,\Lambda_{i}\bar{\theta}_{\dot{a}}\right],\\
\Lambda_{i}\Lambda_{j}\left(\sigma^{j}\theta\right)_{\dot{a}} & = & +\frac{1}{4}\left[Q,\left(\bar{\Lambda}_{ik}\theta_{kj}-\theta_{ik}\bar{\Lambda}_{kj}\right)\left(\sigma^{j}\theta\right)_{\dot{a}}\right]-\left[Q,\theta_{ij}\bar{\Lambda}_{j}\bar{\theta}_{\dot{a}}\right]\nonumber \\
 &  & -\left[Q,\Lambda_{j}\left(\theta\sigma_{i}\bar{\theta}\right)\left(\sigma^{j}\theta\right)_{\dot{a}}\right]-\frac{1}{2}\left[Q,\bar{\Lambda}\theta_{ij}\left(\sigma^{j}\theta\right)_{\dot{a}}\right]\nonumber \\
 &  & -\frac{1}{6}\left[Q,\bar{\Lambda}\theta_{ij}\left(\sigma^{j}\theta\right)_{\dot{a}}\right]+\frac{1}{2}\left[Q,\left(\theta\sigma^{j}\bar{\theta}\right)\Lambda_{j}\left(\sigma^{i}\theta\right)_{\dot{a}}\right]\nonumber \\
 &  & +\sigma_{a\dot{a}}^{i}\Lambda\left[\frac{1}{2}\bar{\Lambda}_{j}\left(\sigma^{j}\bar{\theta}\right)_{a}-\Lambda\bar{\Lambda}\theta_{a}\right]-2\Lambda\Lambda_{i}\bar{\theta}_{\dot{a}},\\
\theta_{ij}\Lambda_{j}\bar{\lambda}_{\dot{a}} & = & -\left[Q,\theta_{ij}\Lambda_{j}\bar{\theta}_{\dot{a}}\right]+6\Lambda\Lambda_{i}\bar{\theta}_{\dot{a}}.
\end{eqnarray}

Gathering these results, it is possible to show that
\begin{eqnarray}
\bar{U}_{i}\left(p\right)\bar{Y}_{\dot{a}}\left(q\right) & = & -\frac{i}{q}\left[Q,\Lambda_{i}\bar{\theta}_{\dot{a}}e^{-i\left(p+q\right)\sqrt{2}X_{L}^{+}}\right]+\frac{1}{2}\left[Q,\theta_{ij}\bar{\Lambda}_{j}\bar{\theta}_{\dot{a}}e^{-i\left(p+q\right)\sqrt{2}X_{L}^{+}}\right]\nonumber \\
 &  & -\frac{1}{8}\left[Q,\left(\bar{\Lambda}_{ik}\theta_{kj}-\theta_{ik}\bar{\Lambda}_{kj}\right)\left(\sigma^{j}\theta\right)_{\dot{a}}e^{-i\left(p+q\right)\sqrt{2}X_{L}^{+}}\right]\nonumber \\
 &  & +\frac{1}{2}\left[Q,\Lambda_{j}\left(\theta\sigma_{i}\bar{\theta}\right)\left(\sigma^{j}\theta\right)_{\dot{a}}e^{-i\left(p+q\right)\sqrt{2}X_{L}^{+}}\right]\nonumber \\
 &  & +\frac{1}{4}\left[Q,\bar{\Lambda}\theta_{ij}\left(\sigma^{j}\theta\right)_{\dot{a}}e^{-i\left(p+q\right)\sqrt{2}X_{L}^{+}}\right]\nonumber \\
 &  & +\frac{1}{12}\left[Q,\bar{\Lambda}\theta_{ij}\left(\sigma^{j}\theta\right)_{\dot{a}}e^{-i\left(p+q\right)\sqrt{2}X_{L}^{+}}\right]\nonumber \\
 &  & -\frac{1}{4}\left[Q,\left(\theta\sigma^{j}\bar{\theta}\right)\Lambda_{j}\left(\sigma^{i}\theta\right)_{\dot{a}}e^{-i\left(p+q\right)\sqrt{2}X_{L}^{+}}\right]\nonumber \\
 &  & -\frac{p}{3!q}\left[Q,e^{-i\left(p+q\right)\sqrt{2}X_{L}^{+}}\theta_{ij}\Lambda_{j}\bar{\theta}_{\dot{a}}\right]\nonumber \\
 &  & -\frac{1}{2!}e^{-i\left(p+q\right)\sqrt{2}X_{L}^{+}}\sigma_{a\dot{a}}^{i}\Lambda\left[\frac{1}{2}\bar{\Lambda}_{j}\left(\sigma^{j}\bar{\theta}\right)_{a}-\bar{\Lambda}\theta_{a}\right]+\mathcal{O}\left(\theta^{5}\right).
\end{eqnarray}
The higher powers of $\theta_{a}$ can be worked out too, although
they are much more complex. The final result is displayed in equation
\eqref{eq:UiYa}.

\paragraph*{BRST-exactness of $\left\{ c_{0}^{+},\bar{\lambda}_{\dot{a}}\bar{A}_{\dot{a}}\left(k\right)\right\} $}

\

It was claimed in the text that $\Lambda\bar{\lambda}_{\dot{a}}\bar{A}_{\dot{a}}\left(k\right)$
is BRST-exact. Analysing first the vector polarisation, given by
\begin{equation}
\Lambda\bar{U}_{i}\left(z;k\right)=\Lambda\left\{ \Lambda_{i}-\frac{ik}{3!}\theta_{ij}\Lambda_{j}-\frac{k^{2}}{5!}\theta_{ij}\theta_{jk}\Lambda_{k}+\frac{ik^{3}}{7!}\theta_{ij}\theta_{jk}\theta_{kl}\Lambda_{l}\right\} e^{-ik\sqrt{2}X_{L}^{+}},
\end{equation}
the aim here is to identify the BRST-exact structures. To do that,
the following identities are very useful:\begin{subequations}
\begin{eqnarray}
\Lambda\Lambda_{i} & = & \frac{1!}{3!}\left\{ Q,\theta_{ij}\Lambda_{j}\right\} \\
\Lambda\theta_{ij}\Lambda_{j} & = & \frac{3!}{5!}\left\{ Q,\theta_{ij}\theta_{jk}\Lambda_{k}\right\} \\
\Lambda\theta_{ij}\theta_{jk}\Lambda_{k} & = & \frac{5!}{7!}\left\{ Q,\theta_{ij}\theta_{jk}\theta_{kl}\Lambda_{l}\right\} \\
\Lambda\theta_{ij}\theta_{jk}\theta_{kl}\Lambda_{l} & = & \frac{7!}{9!}\left\{ Q,\theta_{ij}\theta_{jk}\theta_{kl}\theta_{lm}\Lambda_{m}\right\} \nonumber \\
 & = & 0.
\end{eqnarray}
\end{subequations}The last equation vanishes because $\theta_{ij}\theta_{jk}\theta_{kl}\theta_{lm}\Lambda_{m}$
contains nine $\theta_{a}$'s. Inserting these equations in $\Lambda\bar{U}_{i}\left(z;k\right)$,
it is easy to show that 
\begin{eqnarray}
\Lambda\bar{U}_{i}\left(z;k\right) & = & \left(\frac{1}{3!}\left\{ Q,\theta_{ij}\Lambda_{j}\right\} -\frac{ik}{5!}\left\{ Q,\theta_{ij}\theta_{jk}\Lambda_{k}\right\} -\frac{k^{2}}{7!}\left\{ Q,\theta_{ij}\theta_{jk}\theta_{kl}\Lambda_{l}\right\} \right)e^{-ik\sqrt{2}X_{L}^{+}},\nonumber \\
 & = & \left\{ Q,\left(\frac{1}{3!}\theta_{ij}\Lambda_{j}-\frac{ik}{5!}\theta_{ij}\theta_{jk}\Lambda-\frac{k^{2}}{7!}\theta_{ij}\theta_{jk}\theta_{kl}\Lambda_{l}\right)e^{-ik\sqrt{2}X_{L}^{+}}\right\} \nonumber \\
 &  & +\Lambda\left(\frac{ik}{3!}\theta_{ij}\Lambda_{j}+\frac{k^{2}}{5!}\theta_{ij}\theta_{jk}\Lambda_{k}-i\frac{k^{3}}{7!}\theta_{ij}\theta_{jk}\theta_{kl}\Lambda_{l}\right)e^{-ik\sqrt{2}X_{L}^{+}}.
\end{eqnarray}
This procedure can be repeated, such that
\begin{eqnarray}
\Lambda\bar{U}_{i}\left(z;k\right) & = & \left\{ Q,\left[\frac{1}{3!}\theta_{ij}\Lambda_{j}-\frac{ik}{5!}\theta_{ij}\theta_{jk}\Lambda_{k}-\frac{k^{2}}{7!}\theta_{ij}\theta_{jk}\theta_{kl}\Lambda_{l}\right]e^{-ik\sqrt{2}X_{L}^{+}}\right\} \nonumber \\
 &  & +\left\{ \frac{ik}{5!}\left\{ Q,\theta_{ij}\theta_{jk}\Lambda_{k}\right\} +\frac{k^{2}}{7!}\left\{ Q,\theta_{ij}\theta_{jk}\theta_{kl}\Lambda_{l}\right\} \right\} e^{-ik\sqrt{2}X_{L}^{+}}\nonumber \\
 & = & \left\{ Q,\left(\frac{1}{3!}\theta_{ij}\Lambda_{j}-\frac{ik}{5!}\theta_{ij}\theta_{jk}\Lambda_{k}-\frac{k^{2}}{7!}\theta_{ij}\theta_{jk}\theta_{kl}\Lambda_{l}\right)e^{-ik\sqrt{2}X_{L}^{+}}\right\} \nonumber \\
 &  & +\left\{ Q,\left(\frac{ik}{5!}\theta_{ij}\theta_{jk}\Lambda_{k}+\frac{k^{2}}{7!}\theta_{ij}\theta_{jk}\theta_{kl}\Lambda_{l}\right)e^{-ik\sqrt{2}X_{L}^{+}}\right\} \nonumber \\
 &  & +\Lambda\left\{ -\frac{k^{2}}{5!}\theta_{ij}\theta_{jk}\Lambda_{k}+i\frac{k^{3}}{7!}\theta_{ij}\theta_{jk}\theta_{kl}\Lambda_{l}\right\} e^{-ik\sqrt{2}X_{L}^{+}}\nonumber \\
 & = & \left\{ Q,\left(\frac{1}{3!}\theta_{ij}\Lambda_{j}-\frac{ik}{5!}\theta_{ij}\theta_{jk}\Lambda_{k}-\frac{k^{2}}{7!}\theta_{ij}\theta_{jk}\theta_{kl}\Lambda_{l}\right)e^{-ik\sqrt{2}X_{L}^{+}}\right\} \nonumber \\
 &  & +\left\{ Q,\left(\frac{ik}{5!}\theta_{ij}\theta_{jk}\Lambda_{k}+\frac{k^{2}}{7!}\theta_{ij}\theta_{jk}\theta_{kl}\Lambda_{l}\right)e^{-ik\sqrt{2}X_{L}^{+}}\right\} \nonumber \\
 &  & -\frac{k^{2}}{7!}\left\{ Q,\theta_{ij}\theta_{jk}\theta_{kl}\Lambda_{l}\right\} e^{-ik\sqrt{2}X_{L}^{+}}\nonumber \\
 & = & \left\{ Q,\left(\frac{1}{3!}\theta_{ij}\Lambda_{j}-\frac{ik}{5!}\theta_{ij}\theta_{jk}\Lambda_{k}-\frac{k^{2}}{7!}\theta_{ij}\theta_{jk}\theta_{kl}\Lambda_{l}\right)e^{-ik\sqrt{2}X_{L}^{+}}\right\} \nonumber \\
 &  & +\left\{ Q,\left(\frac{ik}{5!}\theta_{ij}\theta_{jk}\Lambda_{k}+\frac{k^{2}}{7!}\theta_{ij}\theta_{jk}\theta_{kl}\Lambda_{l}\right)e^{-ik\sqrt{2}X_{L}^{+}}\right\} \nonumber \\
 &  & -\left\{ Q,\left(\frac{k^{2}}{7!}\theta_{ij}\theta_{jk}\theta_{kl}\Lambda_{l}\right)e^{-ik\sqrt{2}X_{L}^{+}}\right\} .
\end{eqnarray}
The final result is
\begin{equation}
\Lambda\bar{U}_{i}\left(z;k\right)=\left\{ Q,\left(\frac{1}{3!}\theta_{ij}\Lambda_{j}-\frac{k^{2}}{7!}\theta_{ij}\theta_{jk}\theta_{kl}\Lambda_{l}\right)e^{-ik\sqrt{2}X_{L}^{+}}\right\} .
\end{equation}

For the spinor polarisation, the supersymmetry charge $Q_{a}$ comes
in hand, as it commutes with $Q$. Because $\left\{ Q_{a},\bar{U}_{i}\right\} =-ik\sigma_{a\dot{a}}^{i}\bar{Y}_{\dot{a}}$,
\begin{eqnarray}
\left[Q_{a},\Lambda\bar{U}_{i}\right] & = & \lambda_{a}\bar{U}_{i}-ik\sigma_{a\dot{a}}^{i}\left(\Lambda\bar{Y}_{\dot{a}}\right)\nonumber \\
 & = & \left[Q,\theta_{a}\bar{U}_{i}\right]-ik\sigma_{a\dot{a}}^{i}\left(\Lambda\bar{Y}_{\dot{a}}\right),
\end{eqnarray}
which implies that
\begin{eqnarray}
ik\sigma_{a\dot{a}}^{i}\left(\Lambda\bar{Y}_{\dot{a}}\right) & = & \left[Q,\left\{ Q_{a},\left(\frac{1}{3!}\theta_{ij}\Lambda_{j}-\frac{k^{2}}{7!}\theta_{ij}\theta_{jk}\theta_{kl}\Lambda_{l}\right)e^{-ik\sqrt{2}X_{L}^{+}}\right\} \right]\nonumber \\
 &  & +\left[Q,\theta_{a}\bar{U}_{i}\right].
\end{eqnarray}
Therefore, $\Lambda\bar{\lambda}_{\dot{a}}\bar{A}_{\dot{a}}\left(k\right)=a_{i}\Lambda\bar{U}_{i}-\bar{\xi}_{\dot{a}}\Lambda\bar{Y}_{\dot{a}}$
is BRST-exact, as claimed.

\paragraph*{Commutator $\left[V_{\textrm{L.C.}}^{*},V_{\textrm{L.C.}}\right]$}

\

This commutator is straightforward to obtain from the $SO\left(8\right)$
decomposed OPE's:\begin{subequations}
\begin{eqnarray}
\bar{d}_{\dot{a}}\left(z\right)\Pi^{i}\left(y\right) & \sim & \frac{\left(\sigma^{i}\partial\theta\right)_{\dot{a}}}{\left(z-y\right)},\\
\bar{d}_{\dot{a}}\left(z\right)X_{L}^{+}\left(y\right) & \sim & \textrm{regular},\\
\bar{d}_{\dot{a}}\left(z\right)\bar{d}_{\dot{b}}\left(y\right) & \sim & -\frac{\sqrt{2}\eta_{\dot{a}\dot{b}}\Pi^{+}}{\left(z-y\right)},\\
\bar{N}_{i}\left(z\right)\bar{\lambda}_{\dot{a}}\left(y\right) & \sim & \frac{1}{\sqrt{2}}\frac{\sigma_{a\dot{a}}^{i}\lambda_{a}}{\left(z-y\right)}.
\end{eqnarray}
\end{subequations}

The less trivial part is to organise the result in a simple way. For
$\left(k+p\right)\neq0$, it can be cast as
\begin{eqnarray}
\left[V_{\textrm{L.C.}}^{*}(p),V_{\textrm{L.C.}}(k)\right] & = & +2ik\bar{V}^{\left(2\right)}\left(p,k\right)\nonumber \\
 &  & +ik\oint\partial\left\{ \bar{\lambda}_{\dot{a}}\bar{A}_{\dot{a}}\left(p\right)\bar{\theta}_{\dot{c}}\bar{A}_{\dot{c}}\left(k\right)\right\} \nonumber \\
 &  &+ \frac{k}{k+p}\oint\partial\left\{ \bar{\Lambda}\left[A_{i}\left(p\right)A_{i}\left(k\right)-ik\bar{A}_{\dot{a}}\left(p\right)\bar{A}_{\dot{a}}\left(k\right)\right]\right\} \nonumber \\
 &  & +\frac{\sqrt{2}k}{k+p}\left[Q,\oint\left\{ \partial X^{-}\left[A_{i}\left(p\right)A_{i}\left(k\right)-ik\bar{A}_{\dot{a}}\left(p\right)\bar{A}_{\dot{a}}\left(k\right)\right]\right\} \right]\nonumber \\
 &  & -ik\left[Q,\oint\left\{ \left(\Pi_{i}-i\sqrt{2}p\bar{N}_{i}\right)A_{i}\left(p\right)\bar{\theta}_{\dot{c}}A_{\dot{c}}\left(k\right)\right\} \right]\nonumber \\
 &  & -ik\left[Q,\oint\left\{ \left(\partial\bar{\theta}_{\dot{a}}+ip\bar{d}_{\dot{a}}\right)\bar{A}_{\dot{a}}\left(p\right)\bar{\theta}_{\dot{c}}A_{\dot{c}}\left(k\right)\right\} \right],\label{eq:VV*appendix}
\end{eqnarray}
where $\bar{V}^{\left(2\right)}\left(p,k\right)$ was defined
in \eqref{eq:integratedmasslesssquared}. For $\left(k+p\right)=0$,
it is a bit more subtle. It can be shown that
\begin{eqnarray}
\left[V_{\textrm{L.C.}}^{*}(-k),V_{\textrm{L.C.}}(k)\right] & = & +2ik\bar{V}^{\left(2\right)}\left(-k,k\right)\nonumber \\
 &  & -\ointctrclockwise\partial\left[\frac{1}{2}\bar{\Lambda}A_{i}\left(k\right)A_{i}\left(-k\right)-ik\bar{\lambda}_{\dot{a}}\bar{A}_{\dot{a}}\left(-k\right)\bar{\theta}_{\dot{c}}\bar{A}_{\dot{c}}\left(k\right)\right]\nonumber \\
 &  & +\frac{ik}{2}\oint\partial\left[\bar{\Lambda}\theta_{a}\left(\sigma^{i}\bar{A}\left(k\right)\right)_{a}A^{i}\left(-k\right)\right]\nonumber \\
 &  & +\frac{ik}{2}\oint\partial\left[\bar{\Lambda}\theta_{a}A_{i}\left(k\right)\left(\sigma^{i}\bar{A}\left(-k\right)\right)_{a}\right]\nonumber \\
 &  & -\frac{1}{\sqrt{2}}\left[Q,\ointctrclockwise\left\{ \partial X^{-}A_{i}\left(k\right)A_{i}\left(-k\right)\right\} \right]\nonumber \\
 &  & +\frac{ik}{\sqrt{2}}\left[Q,\ointctrclockwise\left\{ \partial X^{-}\theta_{a}\left(\sigma^{i}\bar{A}\left(k\right)\right)_{a}A^{i}\left(-k\right)\right\} \right]\nonumber \\
 &  & +\frac{ik}{\sqrt{2}}\left[Q,\ointctrclockwise\left\{ \partial X^{-}\theta_{a}A_{i}\left(k\right)\left(\sigma^{i}\bar{A}\left(-k\right)\right)_{a}\right\} \right]\nonumber \\
 &  & +2i\left(ka_{j}a_{j}^{*}+i\bar{\xi}_{\dot{a}}\bar{\xi}_{\dot{a}}^{*}\right)\left[Q,\oint\tilde{N}\right]\nonumber \\
 &  & -ik\left[Q,\oint\left\{ \left(\Pi_{i}+i\sqrt{2}k\bar{N}_{i}\right)A_{i}\left(-k\right)\bar{\theta}_{\dot{c}}A_{\dot{c}}\left(k\right)\right\} \right]\nonumber \\
 &  & -ik\left[Q,\oint\left\{ \left(\partial\bar{\theta}_{\dot{a}}-ik\bar{d}_{\dot{a}}\right)\bar{A}_{\dot{a}}\left(-k\right)\bar{\theta}_{\dot{c}}A_{\dot{c}}\left(k\right)\right\} \right],
\end{eqnarray}
where $\tilde{N}\equiv\left(N+\frac{1}{2}\theta_{a}p_{a}-\frac{1}{2}\bar{\theta}_{\dot{a}}\bar{p}_{\dot{a}}\right)$
and $N=N^{+-}$ is the Lorentz ghost current . Following the analysis
of the subsection \ref{sub:Cohomology-ring}, $\bar{V}^{\left(2\right)}\left(p,k\right)$
is BRST-exact for a non null resulting momentum and proportional to
$c_{0}^{+}$ when $p=-k$. This demonstrates the resulting commutator
displayed in \eqref{eq:VV*algebra}.

\paragraph*{Anticommutator $\left\{ V_{\textrm{L.C.}}^{*},V_{\textrm{L.C.}}^{*}\right\} $}

\

A direct computation gives
\begin{eqnarray}
\left\{ V_{\textrm{L.C.}}^{*}\left(k;a_{i}^{*},\bar{\xi}_{\dot{a}}^{*}\right),V_{\textrm{L.C.}}^{*}\left(p;b_{i}^{*},\bar{\chi}_{\dot{a}}^{*}\right)\right\}  & = & -2i\left(k-p\right)V^{\left(3\right)}\left(k,p\right)\nonumber \\
 &  & -ik\ointctrclockwise\partial\left[\bar{\Lambda}\,\bar{\theta}_{\dot{b}}\bar{A}_{\dot{b}}\left(k\right)\bar{\lambda}_{\dot{a}}\bar{A}_{\dot{a}}\left(p\right)\right]\nonumber \\
 &  & -ik\left\{ Q,\ointctrclockwise\left[\bar{\Lambda}\left(\bar{\theta}_{\dot{b}}\bar{A}_{\dot{b}}\left(k\right)\right)\left(\Pi_{i}-i\sqrt{2}p\bar{N}_{i}\right)A_{i}\left(p\right)\right]\right\} \nonumber \\
 &  & -ik\left\{ Q,\ointctrclockwise\left[\bar{\Lambda}\left(\bar{\theta}_{\dot{b}}\bar{A}_{\dot{b}}\left(k\right)\right)\left(\partial\bar{\theta}_{\dot{a}}+ip\bar{d}_{\dot{a}}\right)\bar{A}_{\dot{a}}\left(p\right)\right]\right\} \nonumber \\
 &  & -ip\left\{ Q,\ointctrclockwise\left[\bar{\Lambda}\left(\Pi_{i}-ik\sqrt{2}\bar{N}_{i}\right)A_{i}\left(k\right)\left(\bar{\theta}_{\dot{a}}\bar{A}_{\dot{a}}\left(p\right)\right)\right]\right\} \nonumber \\
 &  & -ip\left\{ Q,\ointctrclockwise\left[\bar{\Lambda}\left(\partial\bar{\theta}_{\dot{b}}+ik\bar{d}_{\dot{b}}\right)\bar{A}_{\dot{b}}\left(k\right)\left(\bar{\theta}_{\dot{a}}\bar{A}_{\dot{a}}\left(p\right)\right)\right]\right\} \nonumber \\
 &  & +\sqrt{2}\left\{ Q,\ointctrclockwise\left[\bar{\Lambda}\partial X^{-}A_{i}\left(k\right)A_{i}\left(p\right)\right]\right\} \nonumber \\
 &  & +i\sqrt{2}\left(k-p\right)\left\{ Q,\ointctrclockwise\left[\bar{\Lambda}\partial X^{-}\bar{A}_{\dot{a}}\left(k\right)\bar{A}_{\dot{a}}\left(p\right)\right]\right\} \label{eq:V*V*appendix}
\end{eqnarray}
where
\begin{eqnarray}
V^{\left(3\right)}\left(k,p\right) & \equiv & \ointctrclockwise\left\{ \bar{\Lambda}\left(\Pi_{i}-ik\sqrt{2}\bar{N}_{i}\right)A_{i}\left(k\right)\bar{\lambda}_{\dot{a}}\bar{A}_{\dot{a}}\left(p\right)\right\} \nonumber \\
 &  & +\ointctrclockwise\left\{ \bar{\Lambda}\left(\partial\bar{\theta}_{\dot{b}}+ik\bar{d}_{\dot{b}}\right)\bar{A}_{\dot{b}}\left(k\right)\bar{\lambda}_{\dot{a}}\bar{A}_{\dot{a}}\left(p\right)\right\} \nonumber \\
 &  & -\ointctrclockwise\left\{ \bar{\Lambda}\left(\Pi_{i}-ip\sqrt{2}\bar{N}_{i}\right)A_{i}\left(k\right)\bar{\lambda}_{\dot{b}}\bar{A}_{\dot{b}}\left(p\right)\right\} \nonumber \\
 &  & -\ointctrclockwise\left\{ \bar{\Lambda}\left(\partial\bar{\theta}_{\dot{a}}+ip\bar{d}_{\dot{a}}\right)\bar{A}_{\dot{b}}\left(k\right)\bar{\lambda}_{\dot{b}}\bar{A}_{\dot{b}}\left(p\right)\right\} \nonumber \\
 &  & +\ointctrclockwise\left\{ \sqrt{2}\partial X^{-}\bar{\lambda}_{\dot{b}}\bar{A}_{\dot{b}}\left(k\right)\bar{\lambda}_{\dot{a}}\bar{A}_{\dot{a}}\left(p\right)\right\} 
\end{eqnarray}
is the analogous of the operator $V^{\left(2\right)}\left(k,p\right)$
but now with a massless composition of one field, $\bar{\lambda}_{\dot{a}}\bar{A}_{\dot{a}}\left(p\right)$,
and one antifield, $\bar{\Lambda}\,\bar{\lambda}_{\dot{b}}\bar{A}_{\dot{b}}\left(k\right)$.
Observe that
\begin{eqnarray}
\left[Q,V^{\left(3\right)}\left(k,p\right)\right] & = & \ointctrclockwise\partial\left(\bar{\Lambda}\,\bar{\lambda}_{\dot{a}}\bar{A}_{\dot{a}}\left(p\right)\bar{\lambda}_{\dot{b}}\bar{A}_{\dot{b}}\left(k\right)\right).
\end{eqnarray}
Following the same steps as before, it is now trivial to show that
$V^{\left(3\right)}\left(k,p\right)$ is pure gauge for $\left(k+p\right)\neq0$.
As for $\left(k+p\right)=0$, $V^{\left(3\right)}\left(k,-k\right)$
is proportional to an $SO\left(8\right)$ version of the integrated
vertex associated to the pure spinor measure of integration, given
by $M$ in \eqref{eq:integratedmeasure}. This demonstrates the result
\eqref{eq:V*V*algebra}.

For completeness, the integrated form of the covariant measure can
be written in a very simple way,
\begin{equation}
M_{cov}=\ointctrclockwise\left\{ \left(8\partial X^{m}+\left(\theta\gamma^{m}\partial\theta\right)\right)\left(\lambda\gamma^{n}\theta\right)\left(\lambda\gamma^{p}\theta\right)\left(\theta\gamma_{mnp}\theta\right)\right\} ,
\end{equation}
satisfying
\begin{equation}
\left[Q,M_{cov}\right]=\ointctrclockwise\partial\left\{ \left(\lambda\gamma^{m}\theta\right)\left(\lambda\gamma^{n}\theta\right)\left(\lambda\gamma^{p}\theta\right)\left(\theta\gamma_{mnp}\theta\right)\right\} .
\end{equation}
Note that $\left[Q,\partial X^{m}\right]=\frac{1}{2}\partial\left(\lambda\gamma^{m}\theta\right)$
and 
\begin{equation}
\left(\lambda\gamma^{m}\theta\right)\left(\lambda\gamma^{n}\theta\right)\left(\lambda\gamma^{p}\partial\theta\right)\left(\theta\gamma_{mnp}\theta\right)=\left(\lambda\gamma^{m}\theta\right)\left(\lambda\gamma^{n}\theta\right)\left(\lambda\gamma^{p}\theta\right)\left(\theta\gamma_{mnp}\partial\theta\right),
\end{equation}
which follows from $\eta_{mn}\left(\gamma_{\alpha\beta}^{m}\gamma_{\gamma\lambda}^{n}+\gamma_{\alpha\gamma}^{m}\gamma_{\beta\lambda}^{n}+\gamma_{\alpha\lambda}^{m}\gamma_{\beta\gamma}^{n}\right)=0$
and the pure spinor constraint.

\paragraph*{$SO\left(8\right)$ decomposition of $\left(\lambda\gamma^{m}\theta\right)\left(\lambda\gamma^{n}\theta\right)\left(\lambda\gamma^{p}\theta\right)\left(\theta\gamma_{mnp}\theta\right)$}

\

After some algebraic manipulations, the pure spinor measure can be
cast in a simple $SO\left(8\right)$ version:
\begin{eqnarray}
\left(\lambda\gamma^{m}\theta\right)\left(\lambda\gamma^{n}\theta\right)\left(\lambda\gamma^{p}\theta\right)\left(\theta\gamma_{mnp}\theta\right) & = & 60\left(\bar{\Lambda}\Lambda\bar{\Lambda}_{ij}\theta_{ji}+\Lambda\bar{\Lambda}\Lambda_{ij}\bar{\theta}_{ji}\right)\nonumber \\
 &  & +10\left(\bar{\Lambda}\Lambda_{i}\Lambda_{j}\theta_{ji}+\Lambda\bar{\Lambda}_{i}\bar{\Lambda}_{j}\bar{\theta}_{ji}\right)\label{eq:SO8measure}
\end{eqnarray}

Considering now the BRST-exact quantities\begin{subequations}
\begin{eqnarray}
\left[Q,\Lambda\bar{\Lambda}_{i}\left(\theta\sigma_{j}\bar{\theta}\right)\bar{\theta}_{ji}\right] & = & \Lambda\bar{\Lambda}_{i}\bar{\Lambda}_{j}\bar{\theta}_{ji}-3\Lambda\bar{\Lambda}\Lambda_{ij}\bar{\theta}_{ji},\\
\left[Q,\Lambda\bar{\Lambda}\theta_{ij}\bar{\theta}_{ji}\right] & = & 2\Lambda\bar{\Lambda}\Lambda_{ij}\bar{\theta}_{ji}-2\bar{\Lambda}\Lambda\bar{\Lambda}_{ij}\theta_{ji},\\
\left[Q,\bar{\Lambda}\Lambda_{i}\left(\bar{\theta}\sigma_{j}\theta\right)\theta_{ji}\right] & = & \bar{\Lambda}\Lambda_{i}\Lambda_{j}\theta_{ji}-3\bar{\Lambda}\Lambda\bar{\Lambda}_{ij}\theta_{ji},
\end{eqnarray}
\end{subequations}the measure can be rewritten as
\begin{eqnarray}
\frac{1}{10}\left(\lambda\gamma^{m}\theta\right)\left(\lambda\gamma^{n}\theta\right)\left(\lambda\gamma^{p}\theta\right)\left(\theta\gamma_{mnp}\theta\right) & = & \left(1+C\right)\bar{\Lambda}\Lambda_{i}\Lambda_{j}\theta_{ji}\nonumber \\
+A\left[Q,\Lambda\bar{\Lambda}_{i}\left(\theta\sigma_{j}\bar{\theta}\right)\bar{\theta}_{ji}\right] &  & +\left(6-2B-3C\right)\bar{\Lambda}\Lambda\bar{\Lambda}_{ij}\theta_{ji}\nonumber \\
+B\left[Q,\Lambda\bar{\Lambda}\theta_{ij}\bar{\theta}_{ji}\right]+C\left[Q,\bar{\Lambda}\Lambda_{i}\left(\bar{\theta}\sigma_{j}\theta\right)\theta_{ji}\right] &  & +\left(6-3A+2B\right)\Lambda\bar{\Lambda}\Lambda_{ij}\bar{\theta}_{ji}\nonumber \\
 &  & +\left(1+A\right)\Lambda\bar{\Lambda}_{i}\bar{\Lambda}_{j}\bar{\theta}_{ji}.
\end{eqnarray}

A deeper analysis shows that this is the most general construction
for the measure in $SO\left(8\right)$. The particular case with $A=-1$,
$B=-\frac{9}{2}$ and $C=5$ is displayed in equation \eqref{eq:so8measure}.
In lower dimensions, the analysis is not so simple because there are
more independent contributions. In \cite{Azevedo:2014tva}, there
is a very complete discussion for the case $D=4$.

\paragraph*{Anticommutator $\left\{ c_{0}^{+},c_{0}^{+}\right\} $}

\

Given the operator $c_{0}^{+}$ in \eqref{eq:c0+}, its anticommutator
with itself is computed to be:
\begin{eqnarray}
\left\{ c_{0}^{+},c_{0}^{+}\right\}  & = & \frac{1}{6}\ointctrclockwise\left\{ \Pi_{i}\theta_{ij}\Lambda_{j}\Lambda-\frac{1}{2}\left(\theta\sigma_{i}\partial\bar{\theta}\right)\theta_{ij}\Lambda_{j}\Lambda\right\} \nonumber \\
 &  & -\frac{1}{4}\ointctrclockwise\left\{ \sqrt{2}\Lambda_{i}\bar{N}_{i}\Lambda+\left(\theta\sigma_{i}\bar{d}\right)\Lambda_{i}\Lambda\right\} +\ointctrclockwise\Lambda\partial\Lambda\nonumber \\
 &  & +\frac{\sqrt{2}}{16}\ointctrclockwise\partial X^{+}\Lambda_{i}\theta_{ij}\Lambda_{j}-\frac{5}{48}\ointctrclockwise\left(\theta\partial\theta\right)\Lambda_{i}\theta_{ij}\Lambda_{j}\nonumber \\
 &  & +\frac{2}{144}\ointctrclockwise\theta_{ik}\partial\Lambda_{k}\theta_{ij}\Lambda_{j}+\frac{5}{144}\ointctrclockwise\partial\theta_{ik}\Lambda_{k}\theta_{ij}\Lambda_{j}.\label{eq:c0squared}
\end{eqnarray}
The easiest way to see that $\left\{ c_{0}^{+},c_{0}^{+}\right\} $
is BRST-exact is recognising that it is the integrated version of
a BRST-exact expression. Observe that
\[
\left[Q,\left\{ c_{0}^{+},c_{0}^{+}\right\} \right]=\frac{1}{4}\ointctrclockwise\partial\left[\Lambda\Lambda_{i}\theta_{ij}\Lambda_{j}\right].
\]
Now using equation \eqref{eq:auxiliarc0squared}, the surface term
can be rewritten as
\begin{eqnarray}
\Lambda\Lambda_{i}\theta_{ij}\Lambda_{j} & = & \left[Q,\left(\Lambda_{i}+\bar{\Lambda}_{i}\right)\left(\theta\sigma_{j}\bar{\theta}\right)\theta_{ij}\Lambda\right]\nonumber \\
 &  & -\frac{1}{4}\left[Q,\Lambda\Lambda_{ik}\bar{\theta}_{kj}\theta_{ij}\right].
\end{eqnarray}
Therefore, $\left\{ c_{0}^{+},c_{0}^{+}\right\} $ is BRST-exact.
A similar procedure can be used to show that $\left\{ c_{0}^{-},c_{0}^{-}\right\} $
is also trivial.

\section{Fields and antifields in bosonic string theory\label{sec:bosonic}}

\

The bosonic string action after gauge fixing can be simply written
as
\begin{equation}
S_{B}=\frac{1}{2\pi}\int d^{2}z\left\{ \frac{1}{2}\partial X^{m}\bar{\partial}X_{m}+b\bar{\partial}c+\bar{b}\partial\bar{c}\right\} ,
\end{equation}
together with the (holomorphic) BRST charge
\begin{equation}
Q_{B}=\ointctrclockwise\left\{ cT_{m}+bc\partial c\right\} ,
\end{equation}
such that $T_{m}=-\frac{1}{2}\partial X^{m}\partial X_{m}$ is the
matter energy-momentum tensor and $Q_{B}^{2}=0$ for $D=26$. The
ghost energy-momentum tensor is $T_{gh}=-2b\partial c+c\partial b$.
Notice that
\begin{equation}
\left\{ Q_{B},b\right\} =T_{m}+T_{gh},
\end{equation}
which implies that the cohomology of $Q_{B}$ is nontrivial only for
null conformal weight states. In terms of the eigenstates of the momentum
operator $P^{m}=\frac{1}{2\pi}\ointctrclockwise\partial X^{m}$, it
is straightforward to determine the cohomology of the bosonic string.
The zero-momentum part is given by
\[
\left\{ \begin{array}{cccc}
\mathbbm{1}, & c\partial X^{m}, & c\partial c\partial X^{m}, & c\partial c\partial^{2}c\end{array}\right\} ,
\]
which is organised according to the ghost number current $J_{B}=\ointctrclockwise cb$.
The last state is the tree-level ghost measure of integration.

The ground states correspond to the tachyon and antitachyon vertices,
\begin{eqnarray}
U_{t} & = & c\,e^{ik\cdot X},\\
U_{t}^{*} & = & c\partial c\,e^{ik\cdot X}.
\end{eqnarray}

In bosonic string theory, the cohomology at ghost number two is in
one-to-one correspondence with the physical states, defined to be
at the ghost number one cohomology. In short terms, for each physical
state it is possible to obtain its dual (antifield) by the action
of the $c$ ghost zero mode, $c_{0}$. For example, the massless gauge
boson, described by the vertex
\begin{equation}
a_{m}\:c\:\partial X^{m}e^{ik\cdot x},\label{eq:masslessbos}
\end{equation}
immediately determines its antifield,
\begin{equation}
a_{m}^{*}\:c\:\partial c\:\partial X^{m}e^{ik\cdot x}.\label{eq:antimasslessbos}
\end{equation}
In the state-operator map, $\partial c$ is associated to $c_{0}$. 

BRST-closedness of \eqref{eq:masslessbos} with respect to $Q_{B}$
implies $a^{m}k_{m}=k^{m}k_{m}=0$. For the antifield, however, this
is a bit more subtle, as the massless condition is not imposed. It
is straightforward to rewrite it as a (singular) BRST-exact expression
\begin{equation}
a_{m}^{*}\:c\:\partial c\:\partial X^{m}e^{ik\cdot x}=-\left(\frac{2}{k^{2}}\right)a_{m}^{*}\left[Q_{B},c\:\partial X^{m}e^{ik\cdot x}\right].\label{eq:singgaugebos}
\end{equation}
In fact, this can be extended to the massive states as well and the
whole antifields spectrum presents an analogous expression. This is
an odd feature when amplitudes are concerned as it would induce the
presence of $\delta\left(k^{2}+m^{2}\right)$ insertions instead of
the usual poles/cuts structures. It can be shown, in fact, that unitarity
implies the decoupling of states that are not annihilated by the $b$
ghost zero-mode, $b_{0}$. That is why $b_{0}\left|\psi\right\rangle =0$
is called a physical state condition.

Determining the cohomology and all the gauge transformations is not
an easy task when the goal is to find the physical degrees of freedom.
In this sense, an explicit gauge fixing of the reparametrisation and
residual symmetries might be more interesting. This leads to the well
known light-cone gauge, where the excitation modes from the light-cone
directions $\left(X^{+},X^{-}\right)$ decouple from the spectrum
(in this case, there are no ghosts).

\paragraph*{DDF operators}

\

In \cite{Del Giudice:1971fp}, it was proposed a method of constructing
the physical vertex operators that can be used to match the the BRST-description
and the nice features of the light-cone spectrum, which is known as
DDF construction.

It is based on a particular choice of the Lorentz frame that enables
an explicit decoupling of the unphysical degrees of freedom. In particular,
the massless states are set to have momentum $k^{i}=0$ and $k^{-}\neq0$
(or $k^{+}\neq0$) non-null, so that the integrated vertex associated
to \eqref{eq:masslessbos} is given by
\begin{equation}
\bar{V}^{i}\left(k\right)=\ointctrclockwise\partial X^{i}e^{-ik\sqrt{2}X_{L}^{+}}.
\end{equation}
This corresponds to the physical polarisations, since $\bar{V}^{+}$
is pure gauge and $\bar{V}^{-}$ is not BRST-closed. Observe
that
\begin{equation}
\left[\bar{V}^{i}\left(k\right),\bar{V}^{j}\left(p\right)\right]=\sqrt{2}k\eta^{ij}\delta_{k+p}P^{+},
\end{equation}
which constitutes a creation/annihilation algebra whenever acting
on states with $P^{+}\neq0$. The reason for the $\sqrt{2}$ factor
is to make it compatible with the pure spinor description in the main
text. There, this is a convenient choice for the superfield expansions.

\paragraph*{Extended DDF algebra}

\

The integrated vertices associated to the zero-momentum states $c\partial c\partial X^{\pm}$
are given by
\begin{equation}
c_{0}^{\pm}\equiv-\sqrt{2}\ointctrclockwise\partial c\partial X^{\pm}.\label{eq:bosonicC0}
\end{equation}
The normalisation is chosen in order to make the comparison with the
results of section \ref{sec:Antifields} easier.

It is possible to generalise the DDF construction to the antifields
vertex operators. In the frame $P^{-}\neq0$, they are defined as
\begin{eqnarray}
\bar{V}_{i}^{*}\left(k\right) & \equiv & -\left[c_{0}^{-},\bar{V}_{i}\left(k\right)\right],\nonumber \\
 & = & -2ik\ointctrclockwise\partial c\partial X_{i}e^{-ik\sqrt{2}X_{L}^{+}},
\end{eqnarray}
and the creation/annihilation algebra is easily extended:
\begin{eqnarray}
[\bar{V}^{i}\left(k\right),\bar{V}_{j}^{*}\left(p\right)] & = & 2\sqrt{2}kp\delta_{j}^{i}\ointctrclockwise\partial c\partial X^{+}e^{-i\left(k+p\right)\sqrt{2}X_{L}^{+}},\nonumber \\
 & = & \delta_{j}^{i}\frac{\sqrt{2}kp}{\left(k+p\right)^{2}}\left[Q,\ointctrclockwise\partial X^{-}e^{-i\left(k+p\right)\sqrt{2}X_{L}^{+}}\right]\nonumber \\
 &  & +2k^{2}c_{0}^{+}\delta_{k+p}\delta_{j}^{i},\label{eq:VV*bosonicalgebra}\\
\{\bar{V}_{i}^{*}\left(k\right),\bar{V}_{j}^{*}\left(p\right)\} & = & 4\eta_{ij}kp\ointctrclockwise\left(\partial c\partial^{2}c\right)e^{-i\left(k+p\right)\sqrt{2}X_{L}^{+}},\nonumber \\
 & = & -i\left(\frac{2\sqrt{2}kp}{k+p}\right)\eta_{ij}\left\{ Q,\ointctrclockwise\left(\partial c\partial X^{-}\right)e^{-i\left(k+p\right)\sqrt{2}X_{L}^{+}}\right\} \nonumber \\
 &  & -4k^{2}\eta_{ij}M_{bos}.\label{eq:V*V*bosonicalgebra}
\end{eqnarray}
Here, $M_{bos}\equiv\ointctrclockwise\left(\partial c\partial^{2}c\right)$,
such that
\begin{equation}
\left[Q,M_{bos}\right]=\ointctrclockwise\partial\left(c\partial c\partial^{2}c\right),
\end{equation}
\emph{i.e.} $M_{bos}$ is the integrated vertex associated to the
ghost measure $c\partial c\partial^{2}c$.

\paragraph*{Spectrum}

\

To obtain the physical spectrum associated to the bosonic creation/annihilation
algebra, one has to define a ground state. This enables a one-to-one
map with the light-cone spectrum. The natural option is the tachyon
vertex
\begin{equation}
U_{t}\left(z;k\right)=c\,\exp\left\{ -ik\sqrt{2}X_{L}^{+}+\frac{i}{\sqrt{2}k}X_{L}^{-}\right\} ,
\end{equation}
with $k^{m}k_{m}=-m^{2}=2$. Choosing, for simplicity, $k=\frac{\sqrt{2}}{2}$,
the ground state will be defined by the state-operator map
\begin{equation}
\left|0\right\rangle _{t}\equiv\lim_{z\to0}U_{t}\left(z;k=\frac{\sqrt{2}}{2}\right)\left|0\right\rangle .
\end{equation}

Due to the OPE
\begin{equation}
e^{-ik\sqrt{2}X_{L}^{+}}\left(y\right)e^{iX_{L}^{-}}\left(z\right)\sim\left(y-z\right)^{\sqrt{2}k}:e^{i\left(X_{L}^{-}-\sqrt{2}kX_{L}^{+}\right)}:+\ldots\label{eq:OPEclosedX+X-}
\end{equation}
it is straightforward to show that $\bar{V}^{i}\left(k\right)\left|0\right\rangle _{t}=0$
for $k\geq0$. For $k<0$, this operation makes sense only for $k=-\frac{m}{\sqrt{2}}$,
with $m\in\mathbb{\mathbb{Z}}^{+}$. This is the only way the OPE
above will have integer poles. Defining
\begin{equation}
\bar{V}_{m}^{i}\equiv\ointctrclockwise\partial X^{i}e^{-imX_{L}^{+}},\label{eq:DDFbosinteger}
\end{equation}
states of the form
\begin{equation}
\prod_{i}\prod_{m>0}\sum_{n}C_{i,m}\left(n\right)\left(\bar{V}_{-m}^{i}\right)^{n}\left|0\right\rangle _{t},
\end{equation}
are BRST-closed by construction and span the (left-moving) light-cone
spectrum of the bosonic string, with mass
\begin{equation}
m_{\textrm{closed}}^{2}=2N-2,
\end{equation}
and $N\in\mathbb{\mathbb{Z}}^{*}$. The coefficients $C_{i,m}\left(n\right)$
are just numerical constants. Note that for the \emph{particular solution}
$X_{L}^{+}=\ln z$ and the Laurent expansion
\begin{equation}
\partial X^{i}=\sum_{n}\frac{\alpha_{n}^{i}}{z^{n+1}},
\end{equation}
the operator in \eqref{eq:DDFbosinteger} takes the form
\begin{equation}
\bar{V}_{m}^{i}=\alpha_{m}^{i},
\end{equation}
making the map between the light-cone gauge and the DDF operators
even more explicit.

The simplest (excited) DDF state is the massless vector, given by
\begin{equation}
\bar{V}_{-1}^{i}\left|0\right\rangle _{t}\Rightarrow c\,\partial X^{i}e^{iX_{L}^{-}}.
\end{equation}

The differences between open and closed string will not be discussed
here, but can be found in \cite{Jusinskas:2014vqa} for the pure spinor
case.

\end{document}